\RequirePackage{fix-cm}
\documentclass[smallextended]{svjour3}      
\smartqed  
\usepackage{graphicx}
\usepackage{enumitem}
\usepackage{amsmath}
\usepackage{multirow,tabularx}
\usepackage{cite}
\usepackage{hyperref}
\hypersetup{colorlinks,allcolors=black}
\usepackage[pagewise]{lineno}
\journalname{Indian Journal of Physics}
\begin{document}

\title{A comprehensive analysis using 9 dark matter halo models on the spiral galaxy NGC 4321 
}

\titlerunning{Dark matter analysis on galaxy M100}        

\author{Tan Wei Shen$^{1*}$ \and Zamri Zainal Abidin$^{1}$ \and
        Norsiah Hashim$^{2}$ \and 
}


\institute{Tan Wei Shen \at
              $^{1}$Radio Cosmology Research Lab, Physics Dept., Faculty of Science, University of Malaya, Kuala Lumpur, Malaysia \\
              \email{weishen06@hotmail.com}           
           \and
           Zamri Zainal Abidin \at
              $^{1}$Radio Cosmology Research Lab, Physics Dept., Faculty of Science, University of Malaya, Kuala Lumpur, Malaysia  \\
              \email{zzaa@um.edu.my}           
            \and
           Norsiah Hashim \at
              $^{2}$Mathematics Section, Centre for Foundation Studies in Science, University of Malaya, Kuala Lumpur, Malaysia \\
              \email{norsiahashim@um.edu.my}           
}

\date{Received: date / Accepted: date}

\maketitle

\begin{abstract}
This paper addressed the dark matter analysis on the spiral galaxy NGC 4321 (M100) by considering the nine different dark matter profiles, so far lacking in the scientific literature, i.e. Pseudoisothermal, Burkert, NFW, Moore, Einasto, core-modified, DC14, coreNFW and Lucky13 profiles. In this paper, we analyzed the rotation curve analysis on the galaxy NGC 4321 by using nonlinear fitting of star, gaseous and dark matter halo equations with selected VLA HI observation data. Among the nine dark matter profiles, four dark matter profiles (DC14, Lucky13, Burkert and Moore profiles) showed declining features and hence not suitable for this galaxy. This is concluded to be mainly due to the characteristics of those dark matter profiles and also the varying levels of problems within the inner region fittings. For the remaining five accepted dark matter profiles, we further conducted the analysis by using reduced chi-square test. Four out of the five accepted dark matter profiles lie within the range of 0.40 $<$ $\chi_{red}^2$ $<$ 1.70, except for the core-modified profile. In addition, Pseudoisothermal profile achieved the best fitting with $\chi_{red}^2$ nearest to 1, mainly due to its linearity in the inner region and flatness at large radii.
\end{abstract}

\section{Introduction}
\label{intro}
Dark matter is one of the most important investigations since there are still a lot of mysteries related to it. In galaxies, the distribution of the baryonic components cannot justify the observed profiles or the amplitudes of the measured circular velocities \cite{bosma198121}. Furthermore, the profiles of the rotation curve imply that the distribution of light does not match the distribution of mass within the galaxies \cite{persic1996universal}. The explanation of the lacking distribution leads to the suggestions of a presence of an additional invisible mass component \cite{rubin1980rotational}. This is usually solved by adding an extra mass component, i.e. the dark matter halo \cite{donato2009constant}.

Though initial evidence for dark matter came from rotation curves of galaxies, more compelling evidence for non-baryonic matter exists. In 1980, Vera Rubin and Kent Ford present the observations of a set of spiral galaxies that orbital velocities of stars in galaxies were unexpectedly high at large distances from the nucleus by using the new sophisticated optical spectrograph that they developed \cite{rubin1970rotation}. This unexpected result indicated that the falloff in luminous mass with distance from the centre is balanced by an increase in non-luminous mass. Although initially met with skepticism, Rubin's result of the existence of dark matter became scientifically accepted after the subsequent decades by studying more than 200 galaxies and enough documented data proved that the universe was virtually 90 percent undiscovered matter \cite{rubin1980rotational}.

NGC 4321 (also known as M100) is a grand design galaxy located in the Virgo cluster. NGC 4321 is an SAB(s)bc galaxy that has two symmetric well-defined spiral arms \cite{elmegreen1987arm}. M100 has receive much observational and theoretical attention, not only because it is one of the closest galaxies in Virgo cluster, but also because its bar of moderate strength gives rise to a particularly clear resonant circumnuclear structure \cite{knapen2000kinematics}. Its relative proximity and moderate inclination make it suitable to study the content, distribution and kinematics of the neutral hydrogen gas in both its molecular (CO) and atomic (HI) forms of its interstellar medium \cite{cepa1992star}.

Rotation of spiral galaxies is measured by spectroscopic observations of emission lines such as H$\alpha$, HI and CO lines. In these lines, the velocity dispersion is negligibly small compared to rotational velocity. This implies that the pressure term in the Virial theorem is also negligible so that the mass can be calculated in sufficient accuracy by the dynamical balance between the gravitational and centrifugal forces \cite{sofue2017rotation}. HI rotation curves are most often derived from velocity fields. A velocity field aims to give a compact and accurate shorthand description of the dynamics of a galaxy by assigning a typical velocity to every spatial position. That is, for every position, one uses the velocity that most accurately represents the circular motion of the bulk of the quiescent component of the gas as it moves around the center of the galaxy \cite{de2008high}. Fitting the inner part of the rotation curve is challenging. For the case of M100, the existence of intermediate bar \cite{garcia1994gas} and the known starburst activity in the center of this galaxy \cite{knapen1995central} may induced non circular motions.

A large number of HI observation studies in the broader literature have examined this galaxy, namely, HI rotation curves for this galaxy were derived by Vera Rubin \cite{bosma198121,rubin1980rotational}, HI observation by VLA and Nobeyama radio observatory to study large-scale star-formation processes \cite{knapen1993large}, HI rotation curves to see the behavior of the approaching and receding sides of NGC 4321 by Knapen \cite{knapen1993star} and study of environmental effects on HI gas properties of cluster galaxies \cite{chung2009vla}. Besides the usage of HI observational data to measure the galaxy's rotation curves, CO and H$\alpha$ have also been previously used for this purpose since 1980, namely, central CO rotation curves of this galaxy derived by Sofue \cite{sofue1999central}, H$\alpha$ rotation curves and Position-Velocity diagram by Daigle \cite{daigle2006halpha} and H$\alpha$ rotation curves of the inner disc by Morales \cite{castillo2007non} and CO rotation curve to study the dark matter in the central region by using ALMA \cite{ali2018dark}.

Previous CO rotation curve studies by Ali \cite{ali2018dark} have exclusively focused on the dark matter in the central region of this galaxy, which is defined to be up to the radius 0.7 kpc with NFW dark matter profile. The unexpected findings of the dark matter in the central region signal the need for additional studies to understand more deeply about the overall dark matter distribution in this galaxy. To fill this literature gap, this paper will employ HI observed data up to radius 10 kpc to identify the overall dark matter distribution by using a rotation curve with the nine different dark matter profiles. The nine dark matter profiles that we used are Pseudoisothermal, Burkert, NFW, Moore, Einasto, core-modified, DC14, coreNFW and Lucky13 dark matter profiles. More detailed information about the nine dark matter profiles will be discussed in Methodology section.

\section{Methodology}

This section is organized as follows: In subsection 2.1 we introduce the mass modeling of rotation curve technique and then each parameter in the subsequent subsection. The HI observed data processing is explained in subsection 2.2. In subsection 2.3, we present the observed data calculation by using the tilted-ring method. The rotation curve comparison with previous works is explained in subsection 2.4. In the last few subsections, we discuss the mass model parameter of the nine dark matter profiles, stellar, and gas in subsection 2.5, 2.6, and 2.7, respectively.

\subsection{\textit{Mass modeling rotation curve}}
The measurement of the rotation curves of disc galaxies is a powerful tool to investigate the nature of dark matter \cite{frusciante2012distribution}. Rotation curve has been used to assess the existence, the amount and the distribution of this dark component \cite{donato2009constant}. Most galaxies have rotation curves that show sharp rising or high velocity in the very centre, following by a slowly rising or constant velocity rotation in the outer parts. The flatness of the rotation curves in the outer part implies that galaxies contain large amounts of dark matter \cite{xin2013revised}. The contribution of each component to the rotation curve is computed by using Mathematica software. By summing in quadrature the contribution of the three components (disc, gas, dark matter) in all possible combinations and get the best-fitting to the observed data, we will obtain the most accurate value of star, gas and dark matter \cite{pato2015dark}. The rotation curve can be represented as a model, which is the sum of the contribution from the star, halo and gas components \cite{randriamampandry2014galaxy} as follows:
\begin{equation}
V_{rot}^{2}=V_{gas}^{2}+V_{star}^{2}+V_{DM}^{2}
\end{equation}

\textit{\subsection{VLA (Very Large Array) HI data}}

The observed data is NGC 4321 HI (neutral atomic hydrogen) 21 cm line from VLA archive data with project number AS0750\textunderscore D030325. The data reduction for inspection, flagging, bandpass calibration, gain calibration, continuum subtraction, cleaning, imaging, moment mapping and PV (Position-Velocity) Diagram generating is processed by using CASA (Common Astronomy Software Application) software. The distance adopted is 17.1 Mpc \cite{yuan1998spiral}. The overall HI observational parameters of VLA on galaxy NGC 4321 are detailed in Table 1.

\begin{table}
\centering
\scriptsize
\begin{tabular}{|l|l|}
\hline
Parameter                     & Value            \\ \hline \hline
Observation date              & 25th March 2003  \\ \hline
Total observation time        & 12910 seconds    \\ \hline
Total antennas used           & 27               \\ \hline
Antennas diameter (each)      & 25.0m            \\ \hline
Total data recorded           & 855738           \\ \hline
Polarizations                 & LL and RR        \\ \hline
Configurations                & D                \\ \hline
Band                          & L                \\ \hline
RA                            & 12:22:54         \\ \hline
Dec                           & +15:49:20        \\ \hline
Systemic velocity             & 1575 $kms^{-1}$ \\ \hline
Position Angle                & -26$^{\circ}$             \\ \hline
Rest Frequency                & 1420.41 MHz      \\ \hline
Total bandwidth observation   & 1538.1 kHz       \\ \hline
Divided channel               & 63               \\ \hline
Each channel frequency        & 24.41 kHz        \\ \hline
Restoring beam (major, minor) & (52.95", 47.66") \\ \hline
Velocity resolution           & 10 $kms^{-1}$          \\ \hline
Primary calibrator            & 1331+305         \\ \hline
Secondary calibrator          & 1221+282         \\ \hline
\end{tabular}
\newline
\small
\newline
Table 1: HI observational parameter of VLA on galaxy NGC 4321
\end{table}
Figure 1 shows the velocity channel map of HI in the central region of NGC 4321 with restoring beam of $52.95" \times 47.66"$ was produced by using the Briggs weighting. The HI emission is detected in 28 channel maps with a velocity width of 10 $kms^{-1}$ from 1436 $kms^{-1}$ to 1706 $kms^{-1}$. The pixel size is set to CELL = 15" and IMSIZE = 256. Then, the channel map is used as input to generate the integrated intensity HI zeroth moment (mom0) map as shown in Figure 2. The outcome of Figure 1 and Figure 2 is comparable with the previous study (refer to \cite{azeez2016kennicutt} Figure 2 and Figure 3). Next, the integrated-intensity HI map is used to generate PV diagram and obtain the HI velocity of the galaxy as shown in Figure 3.
\newline

\graphicspath{ {./Figure/} }
\begin{figure}
\includegraphics[width=3.9cm]{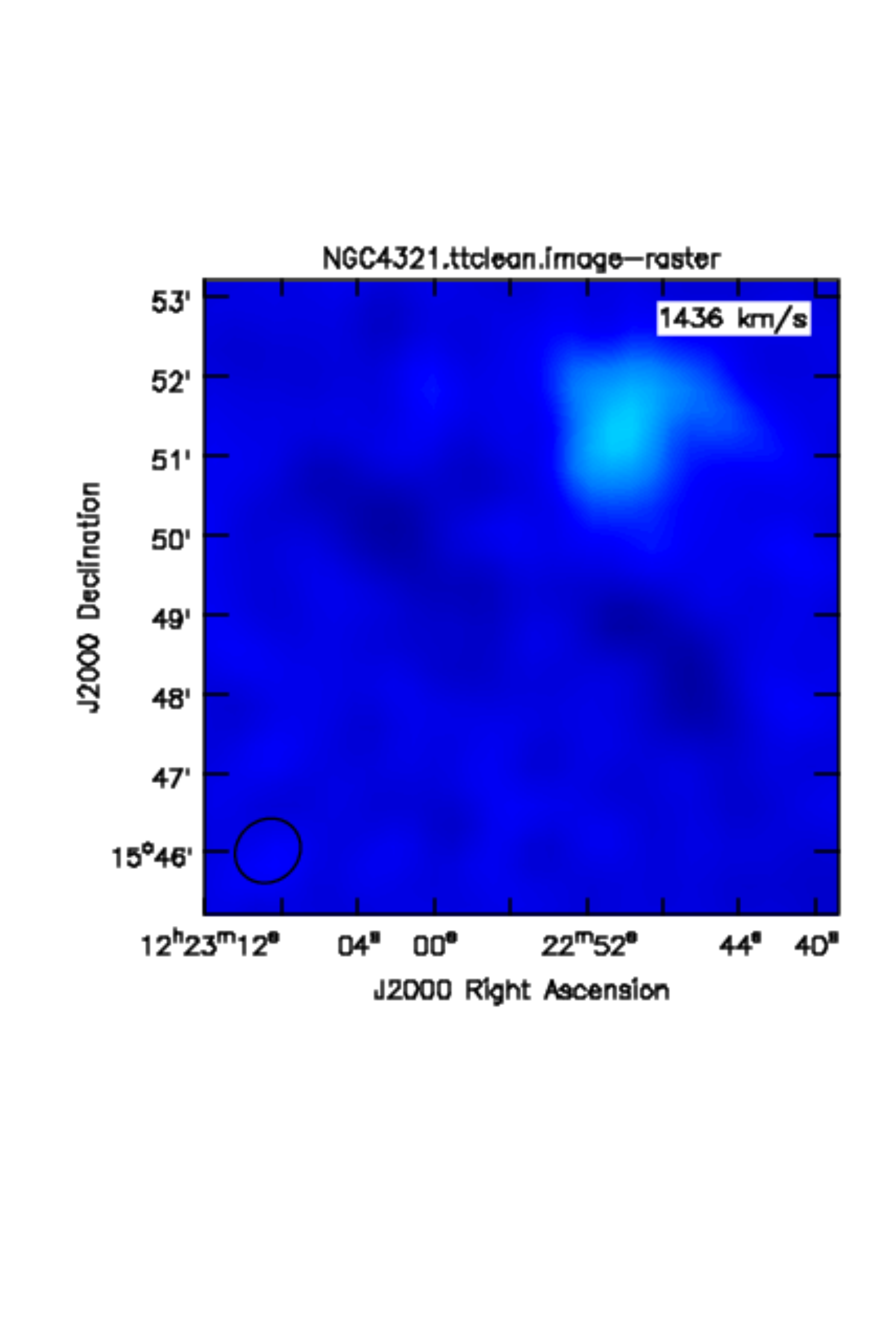}
\includegraphics[width=3.9cm]{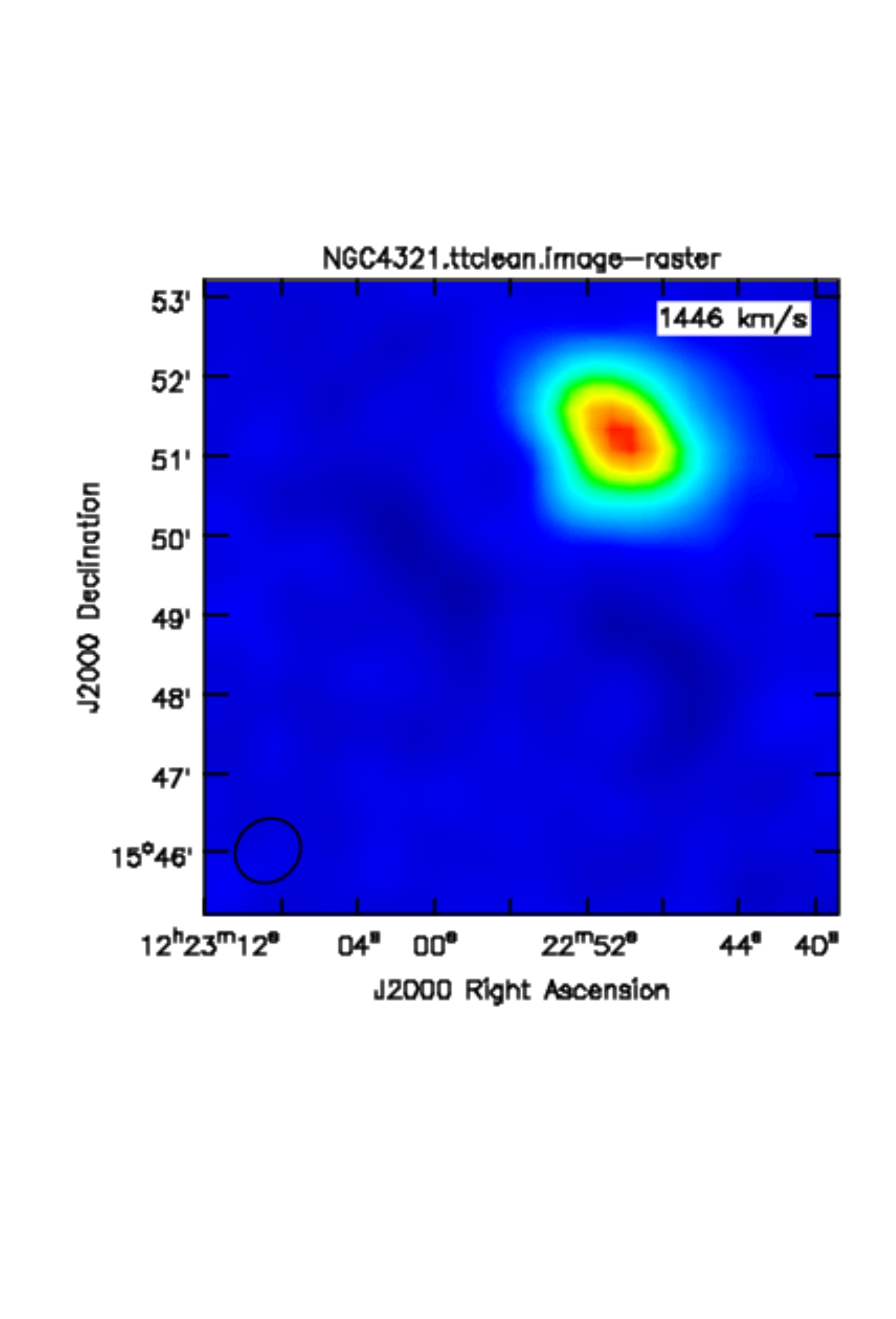}
\includegraphics[width=3.9cm]{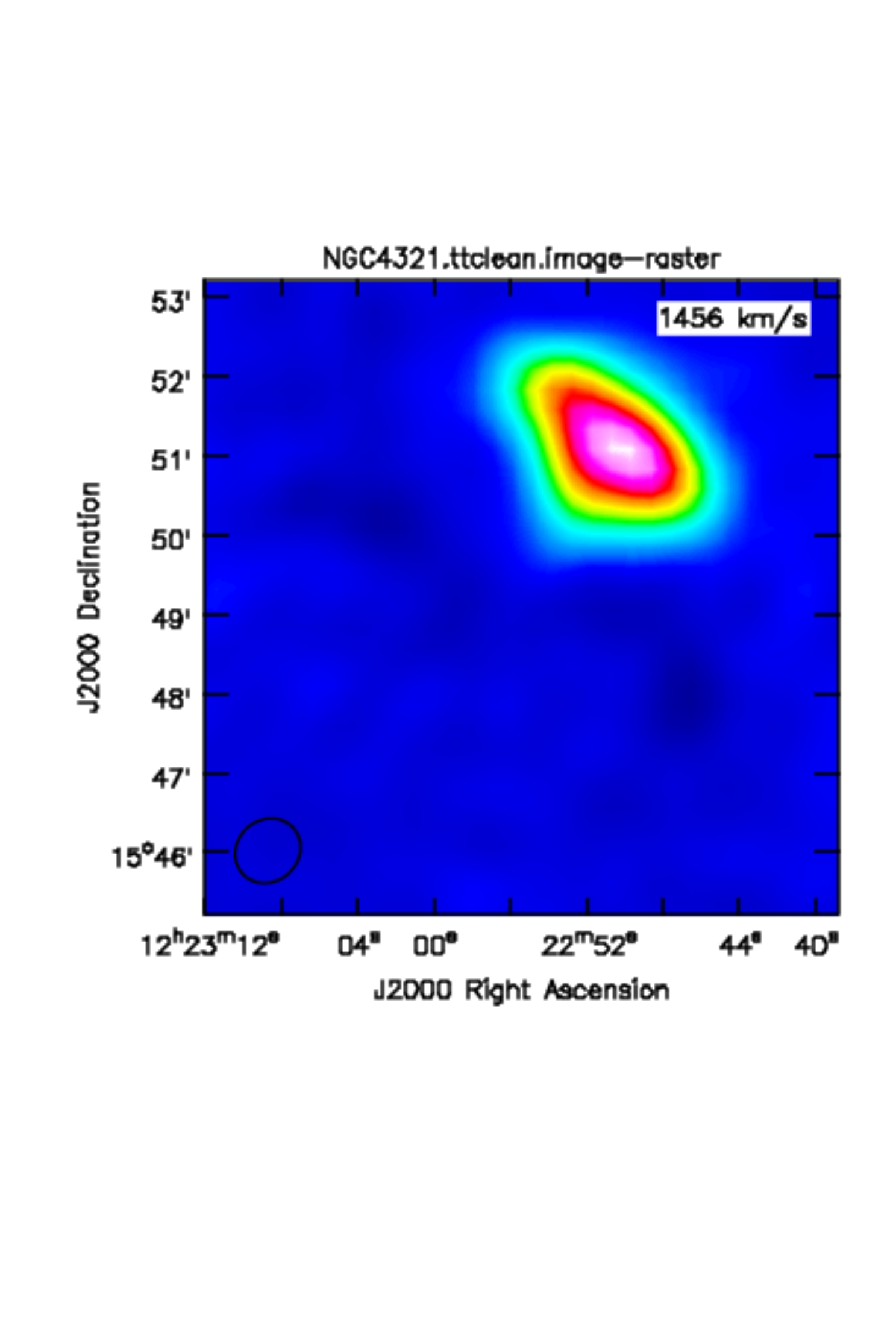}
\includegraphics[width=3.9cm]{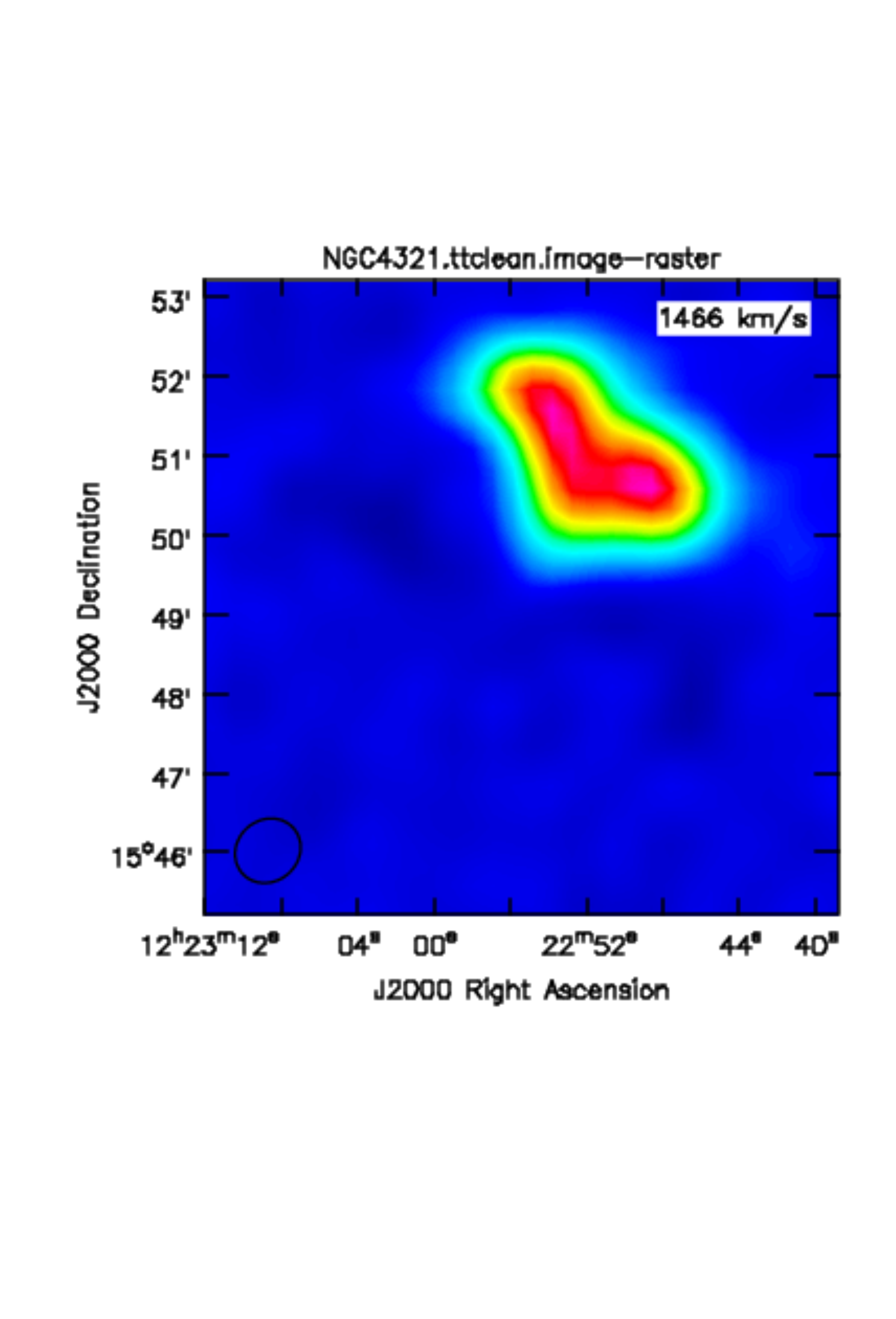}
\includegraphics[width=3.9cm]{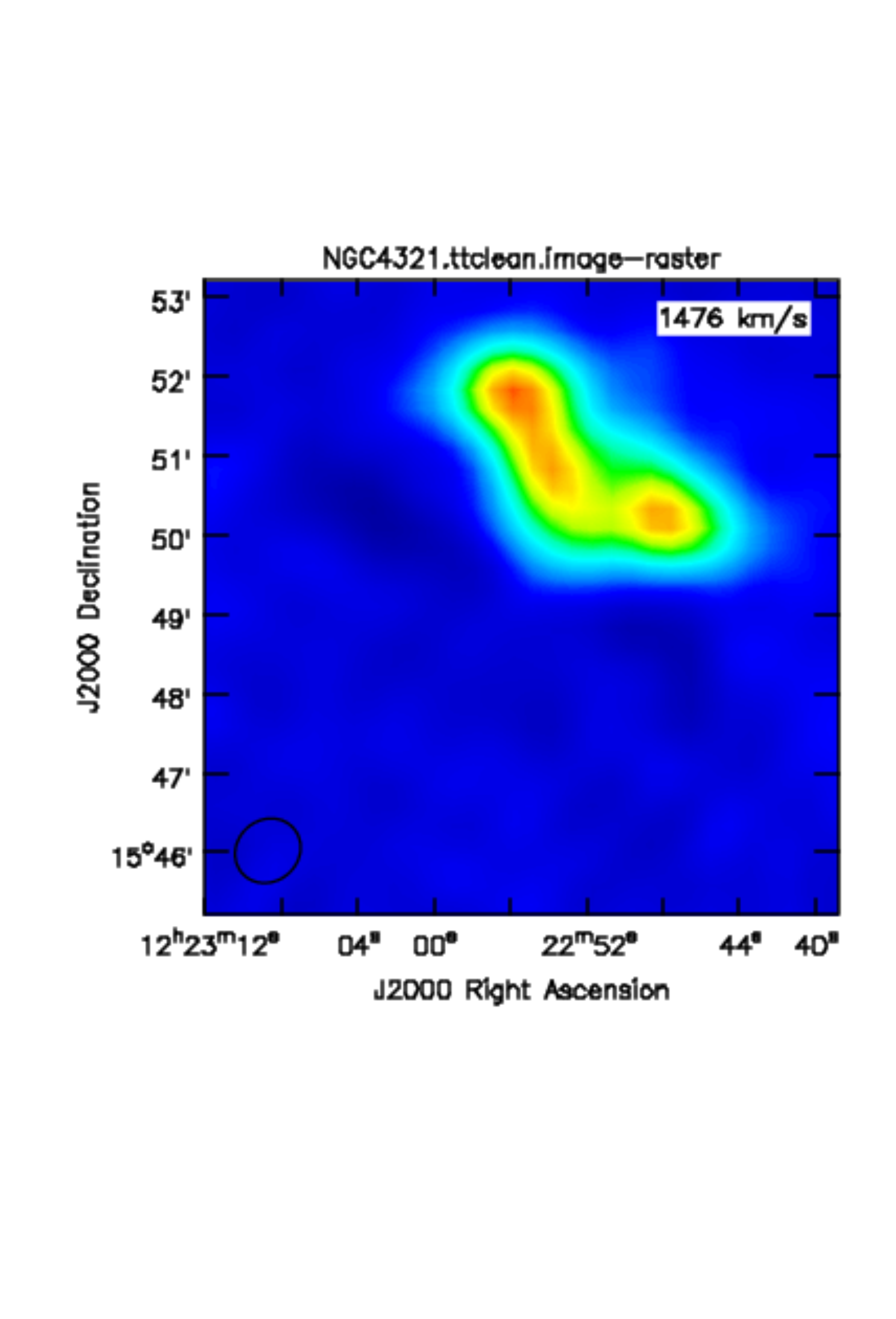}
\includegraphics[width=3.9cm]{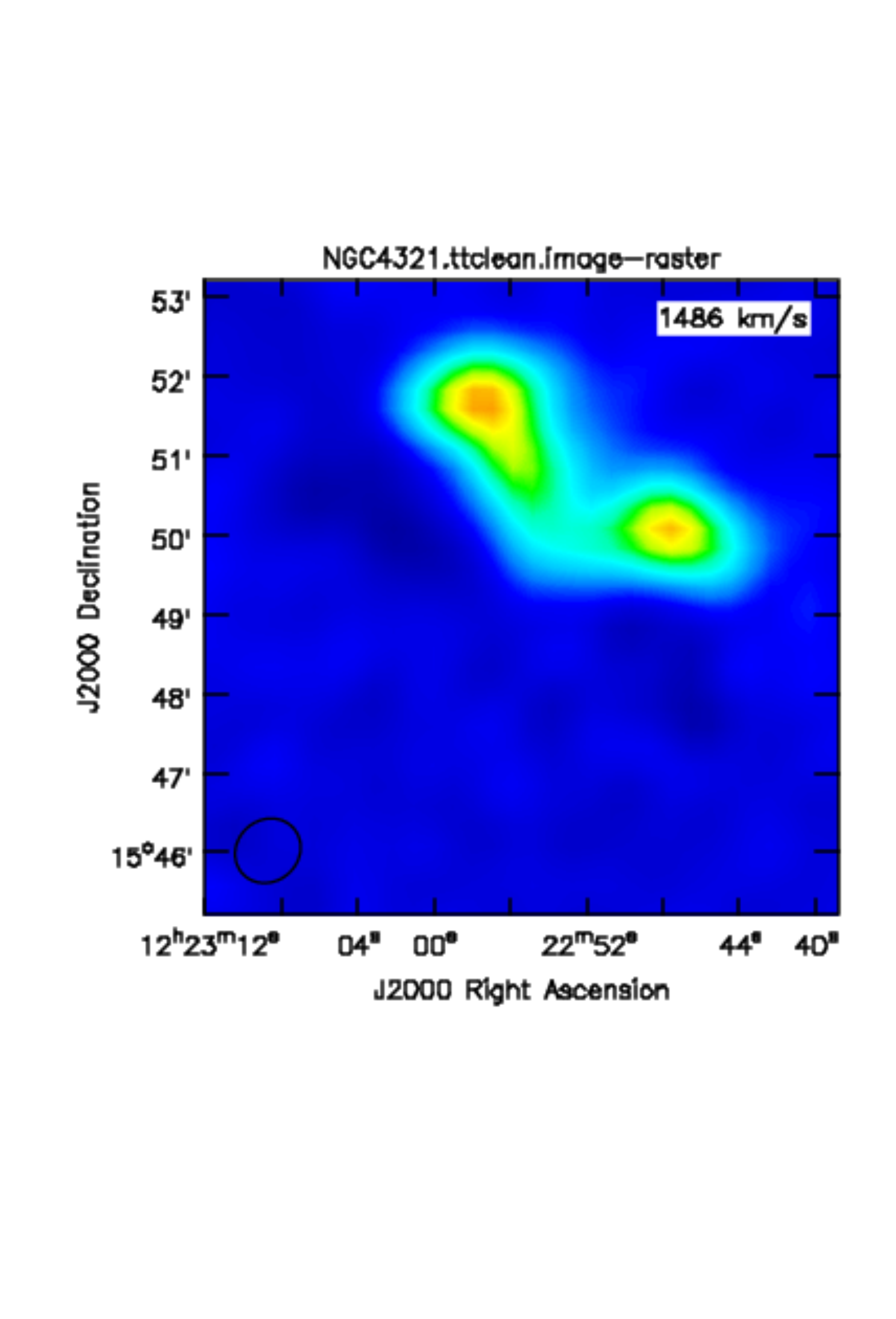}
\includegraphics[width=3.9cm]{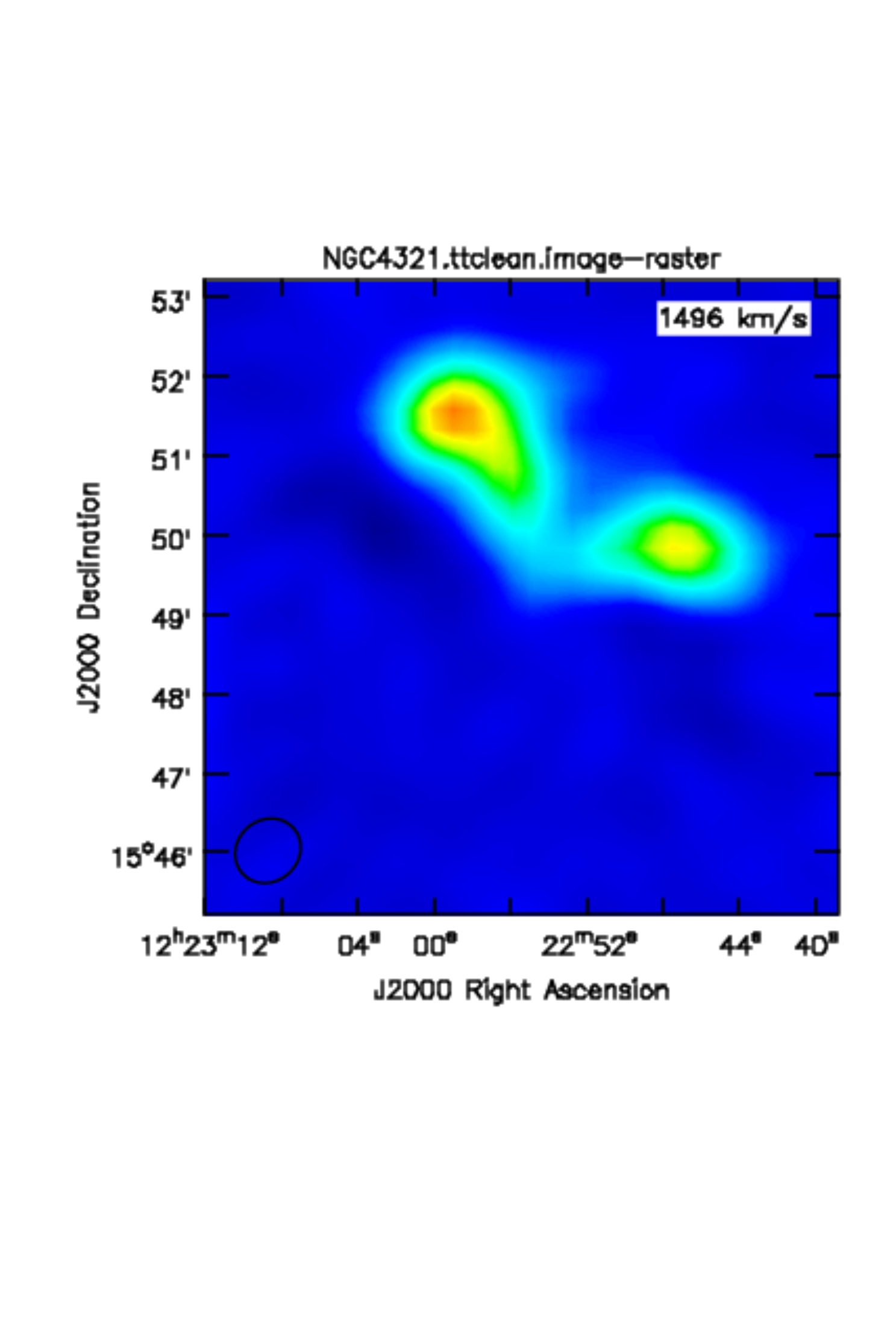}
\includegraphics[width=3.9cm]{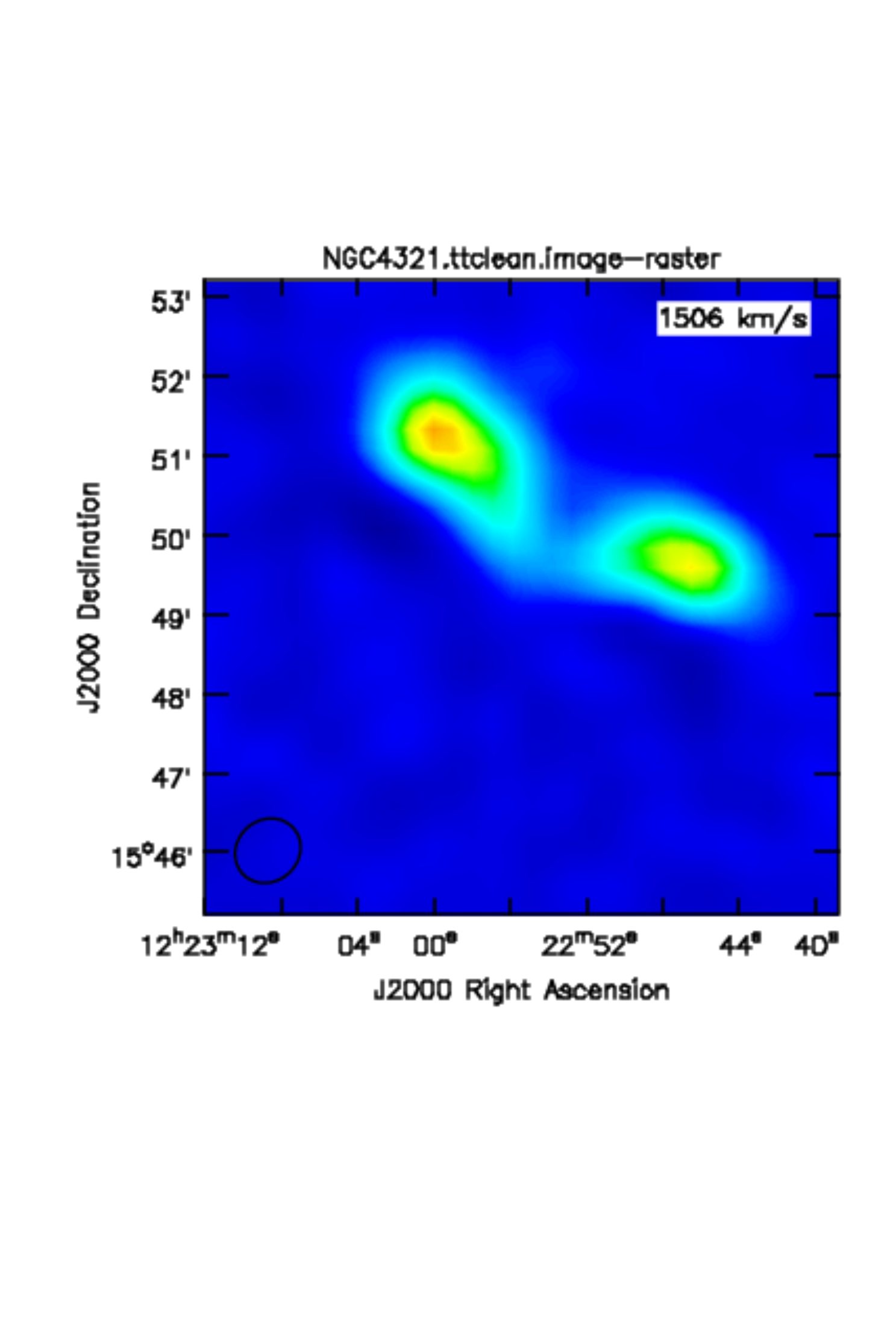}
\includegraphics[width=3.9cm]{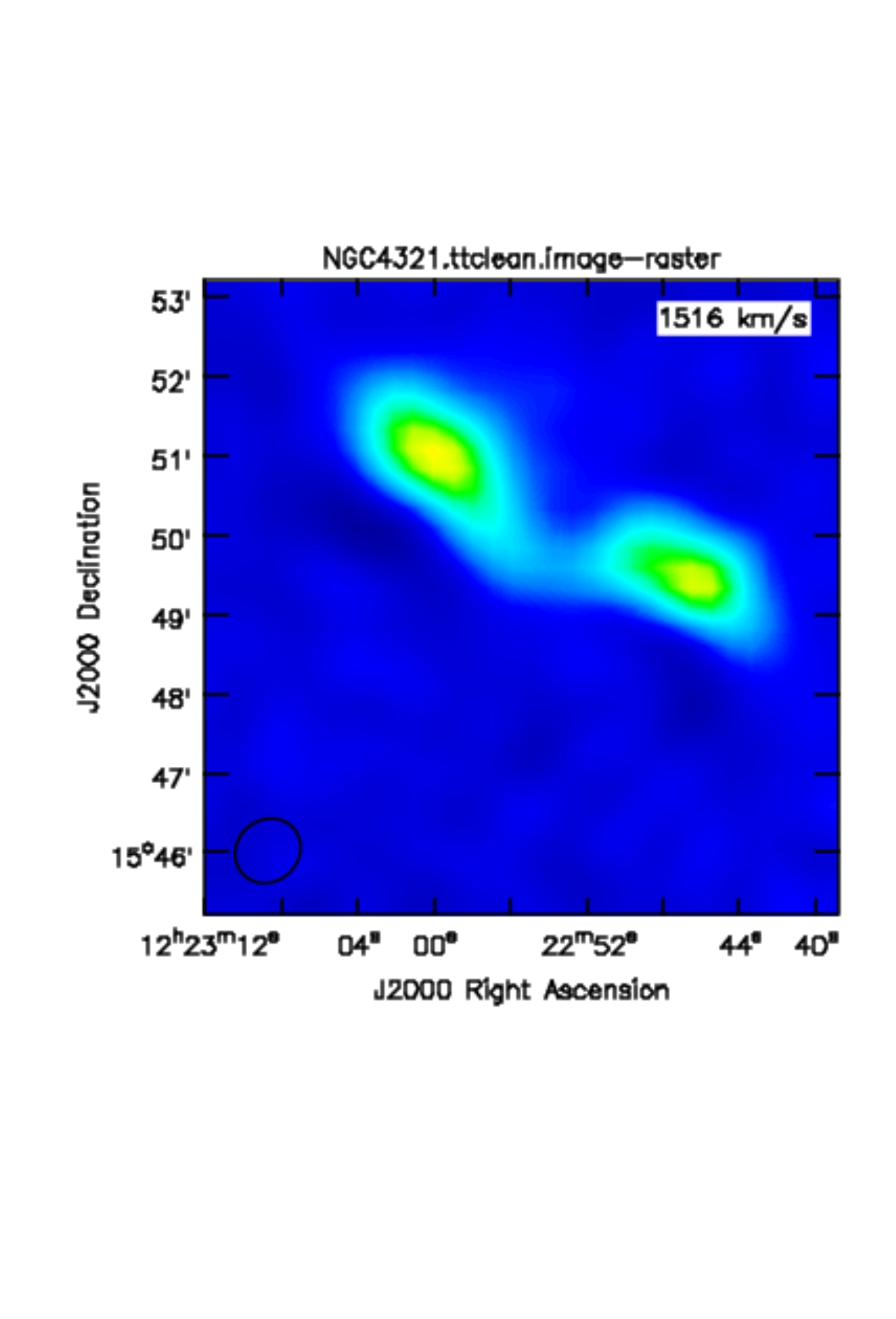}
\includegraphics[width=3.9cm]{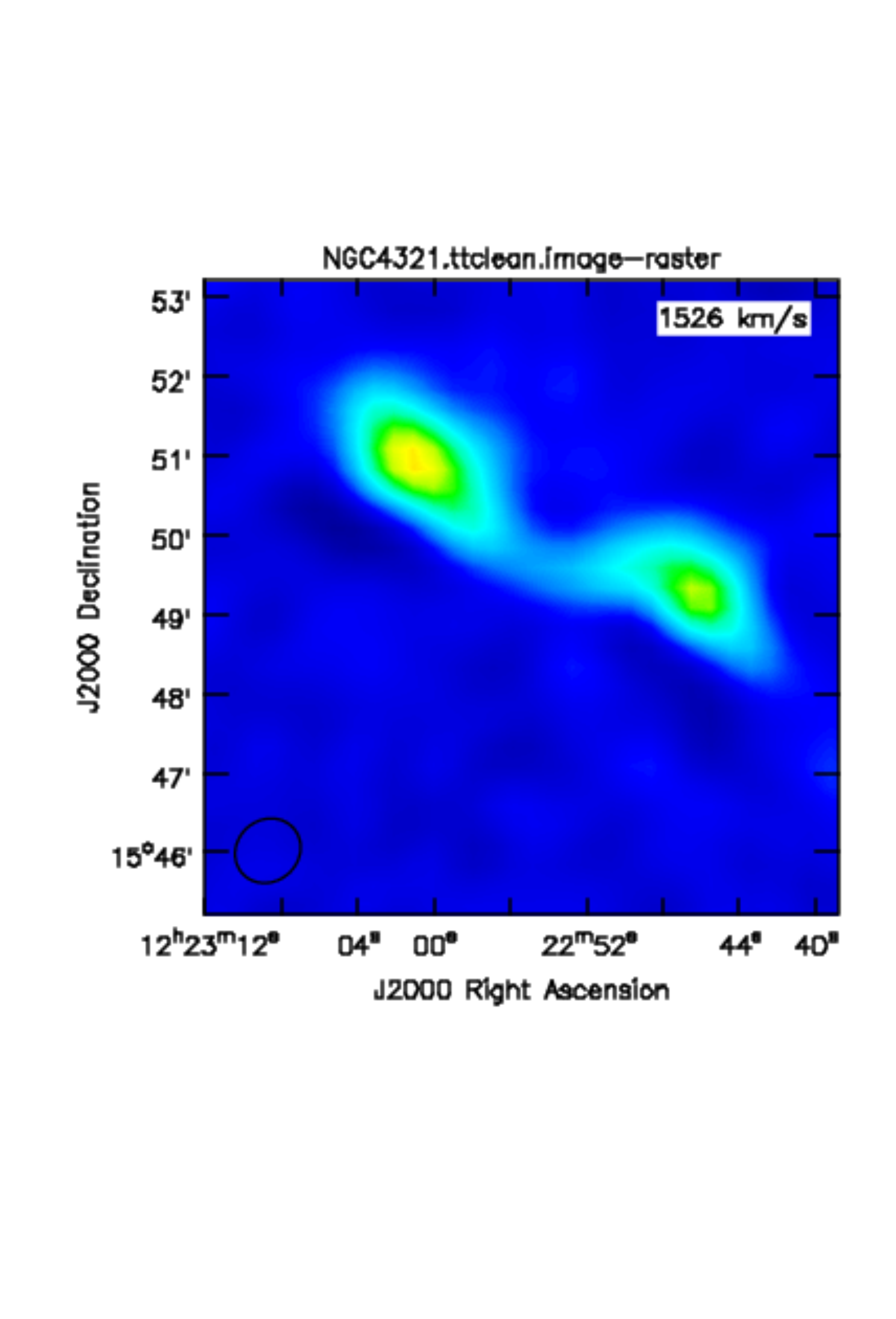}
\includegraphics[width=3.9cm]{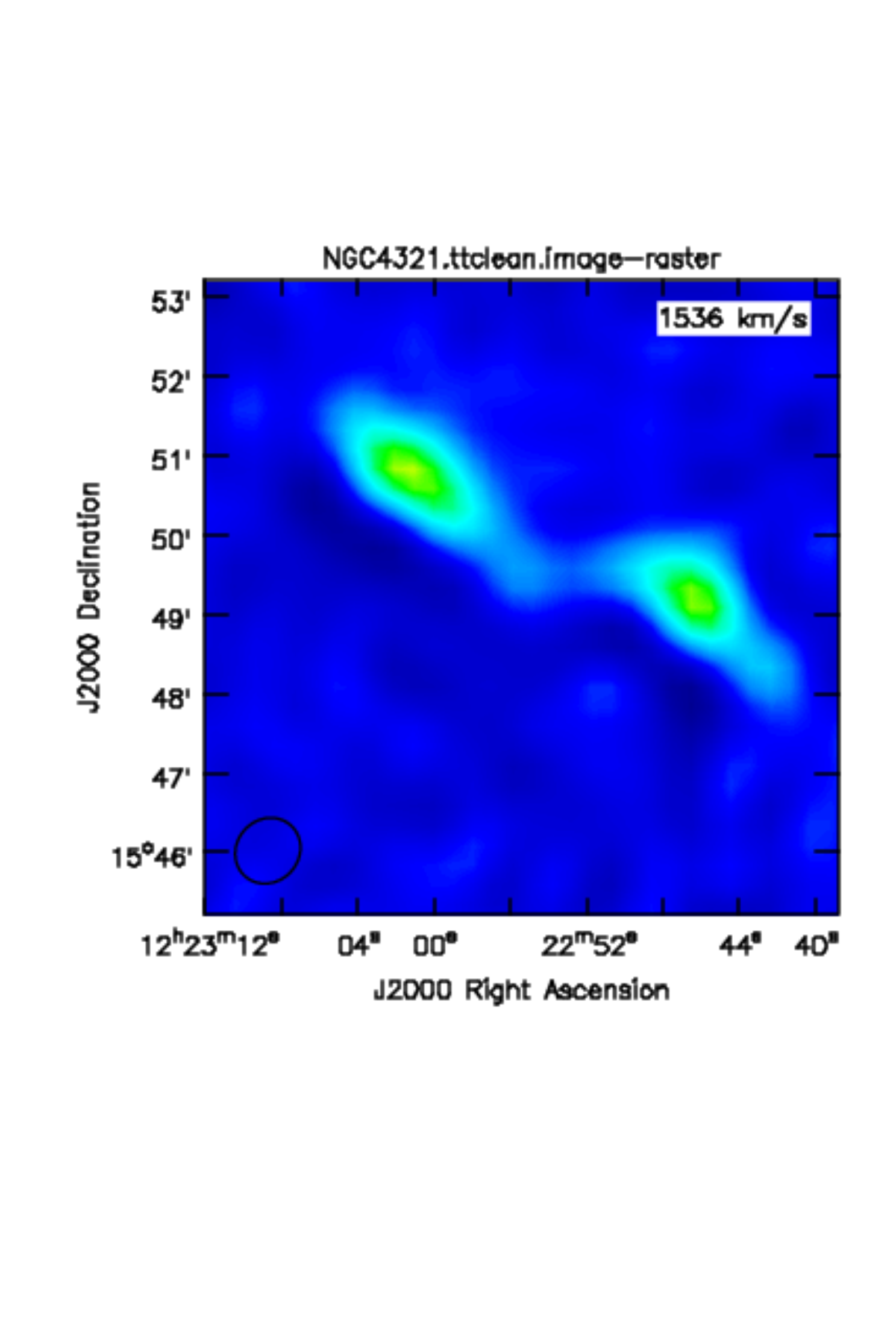}
\includegraphics[width=3.9cm]{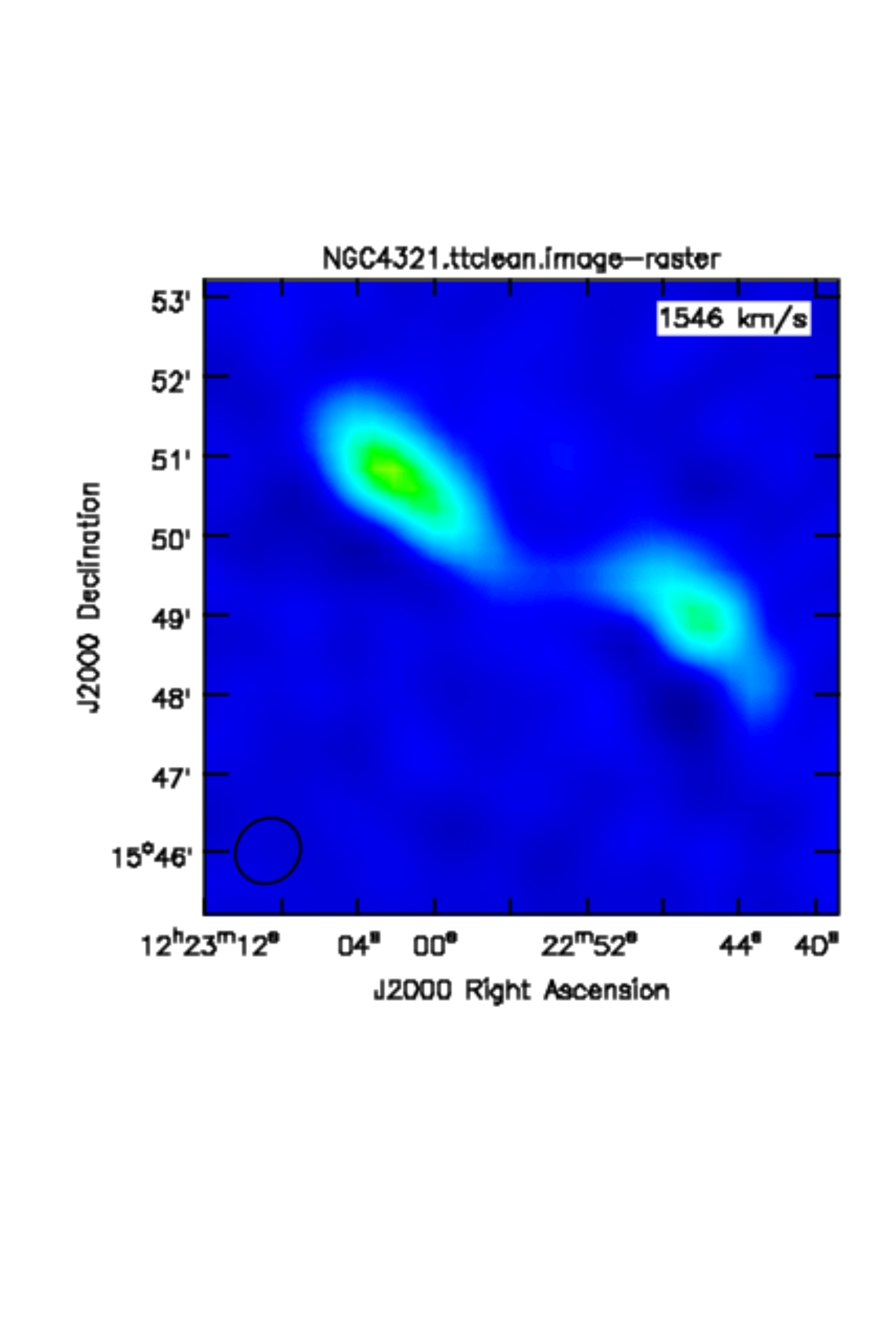}
\end{figure}

\begin{figure}

\includegraphics[width=3.9cm]{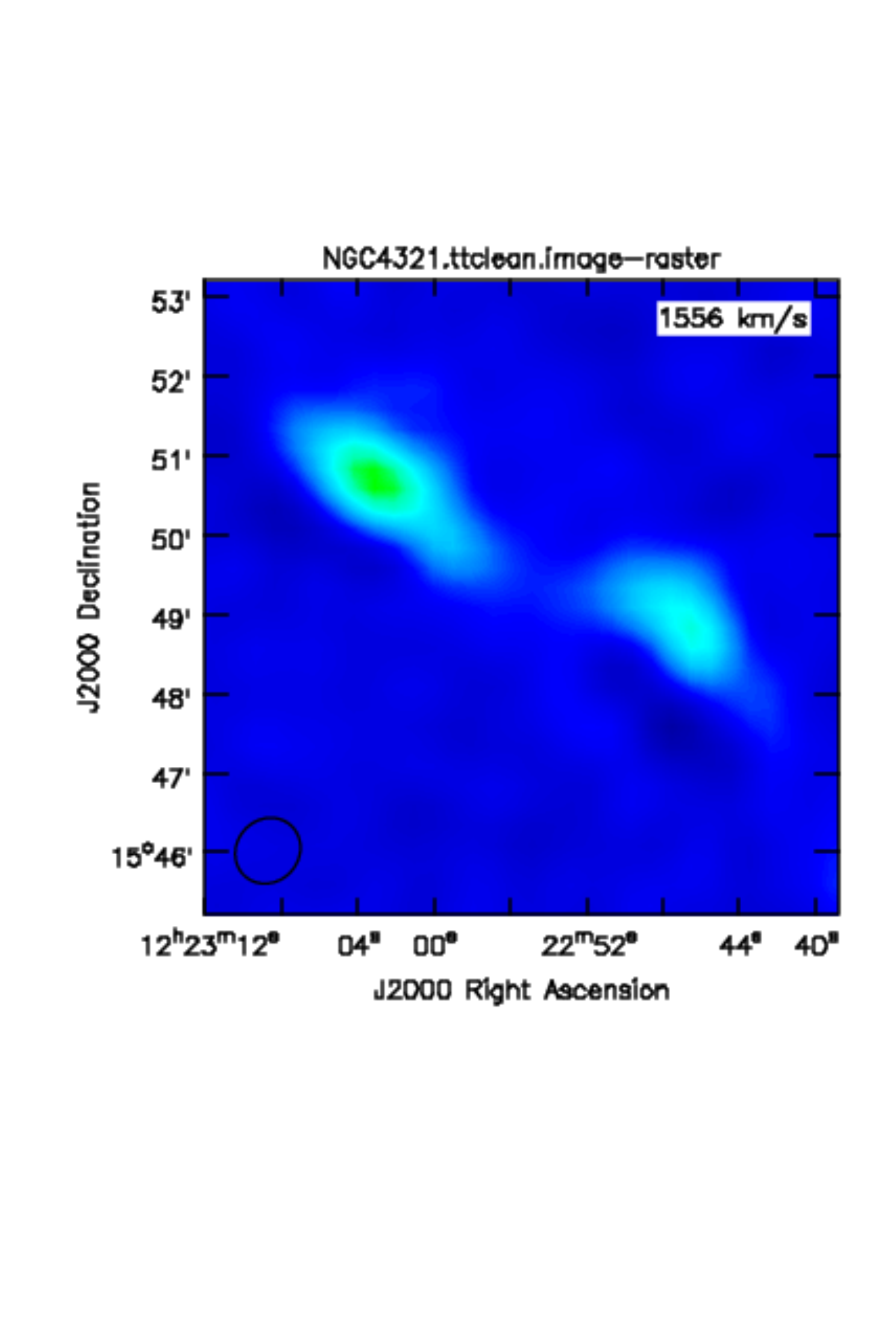}
\includegraphics[width=3.9cm]{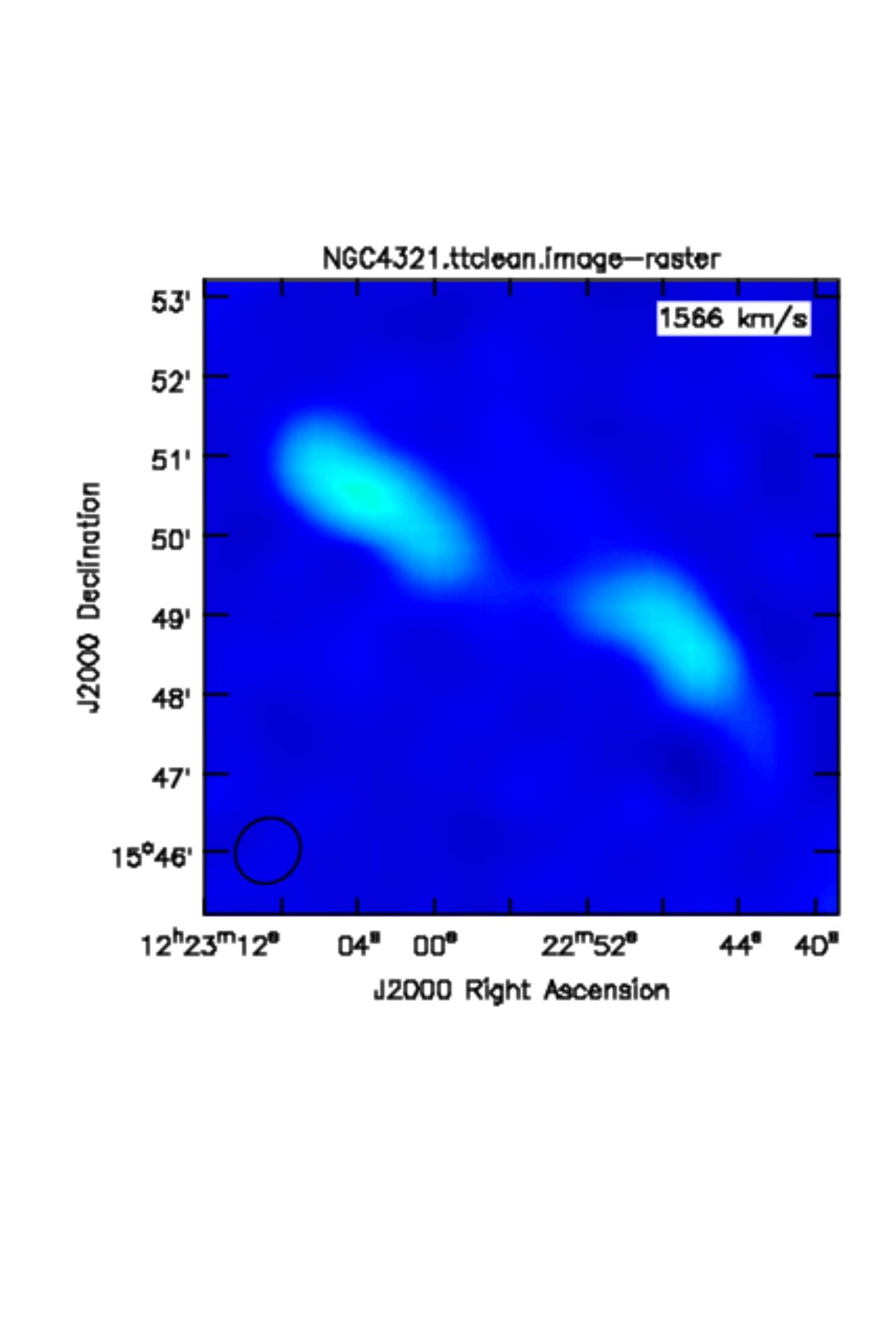}
\includegraphics[width=3.9cm]{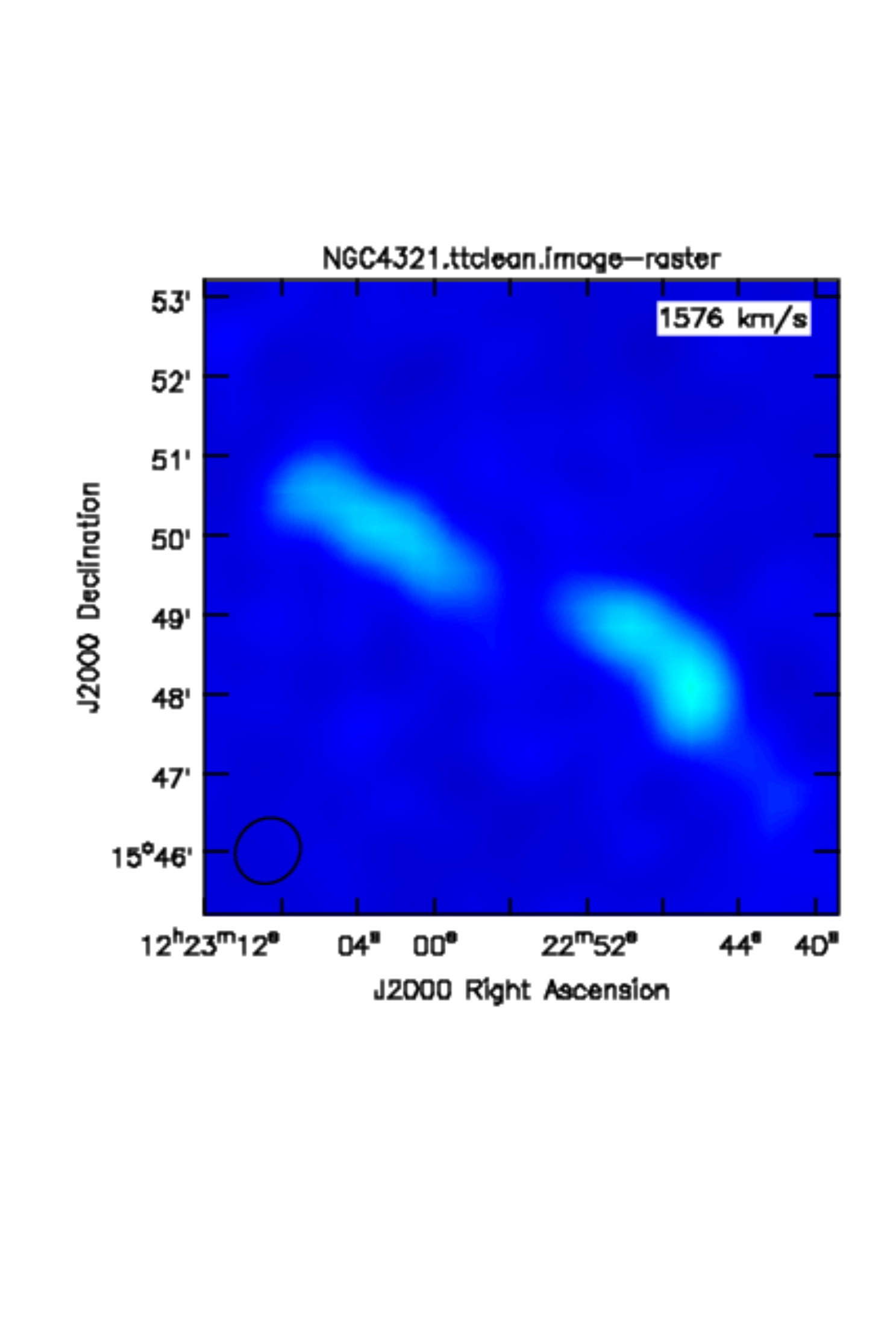}
\includegraphics[width=3.9cm]{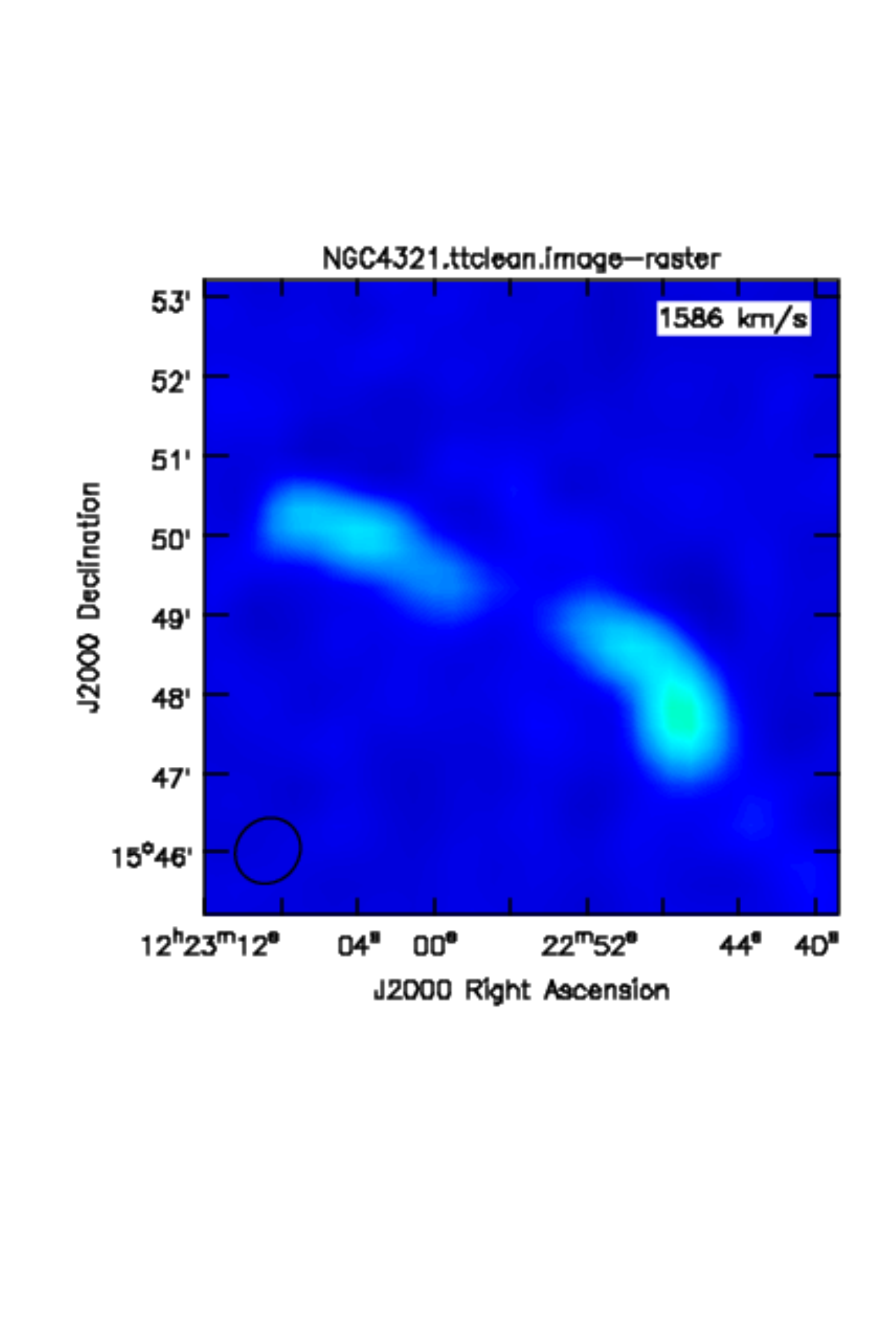}
\includegraphics[width=3.9cm]{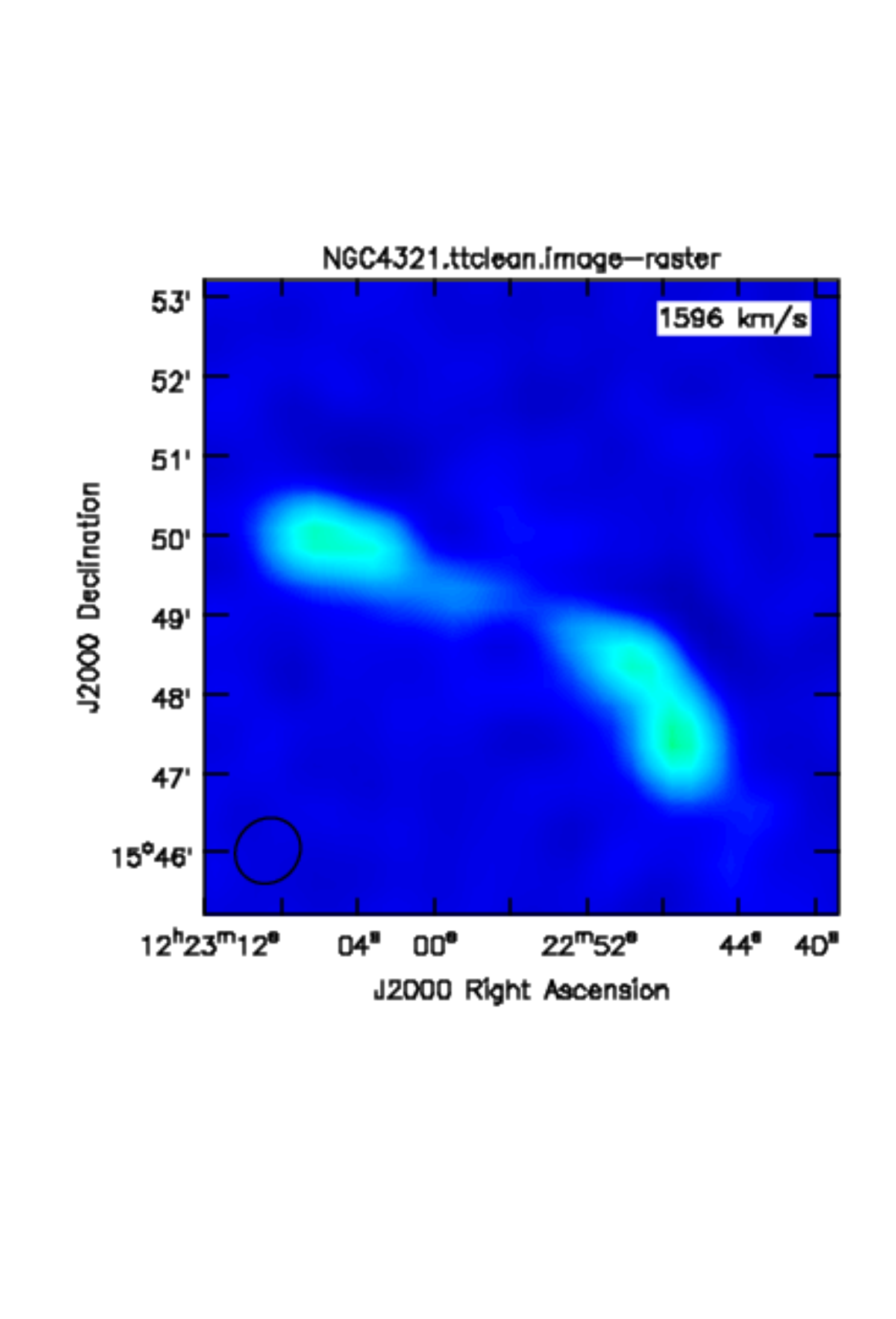}
\includegraphics[width=3.9cm]{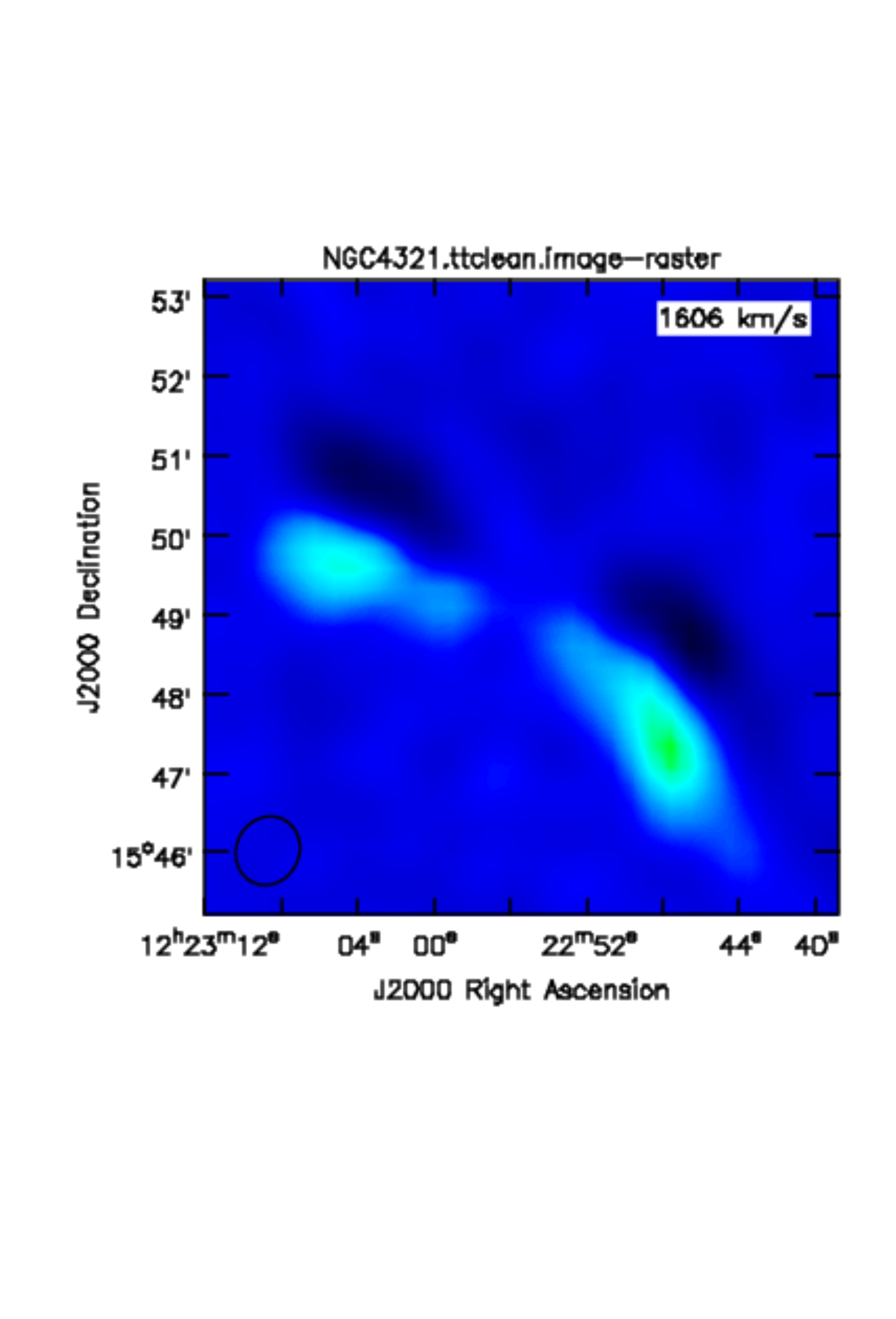}
\includegraphics[width=3.9cm]{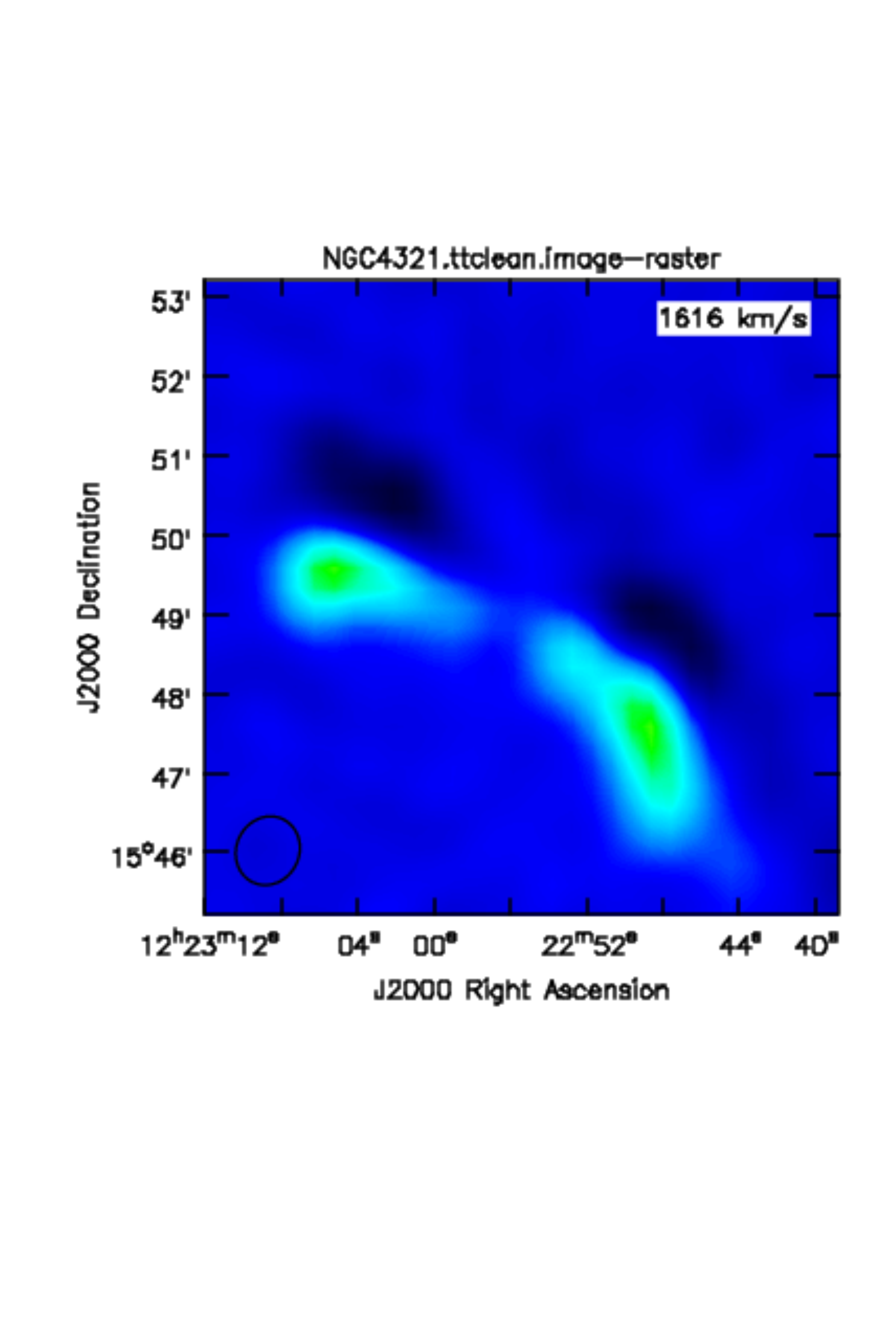}
\includegraphics[width=3.9cm]{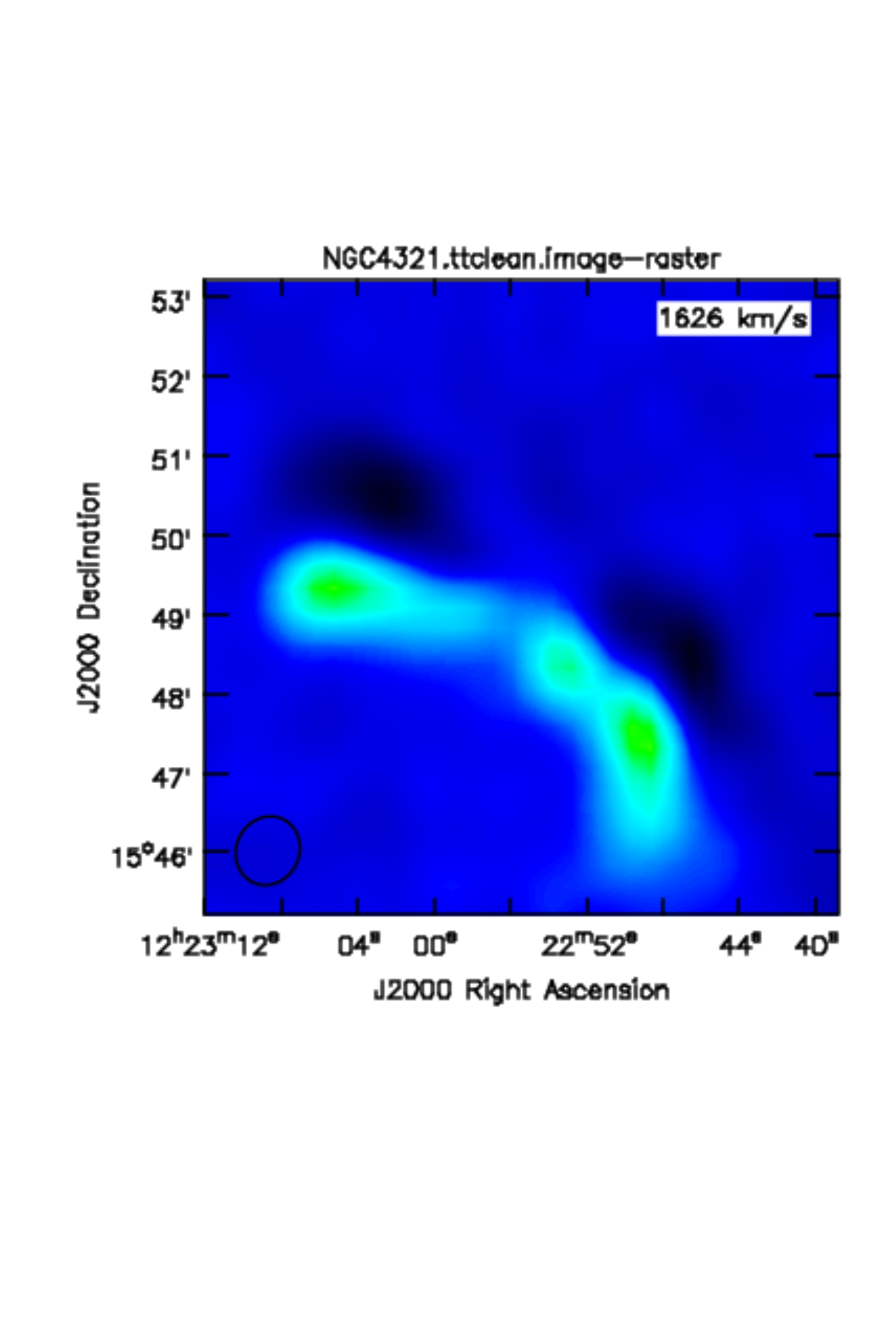}
\includegraphics[width=3.9cm]{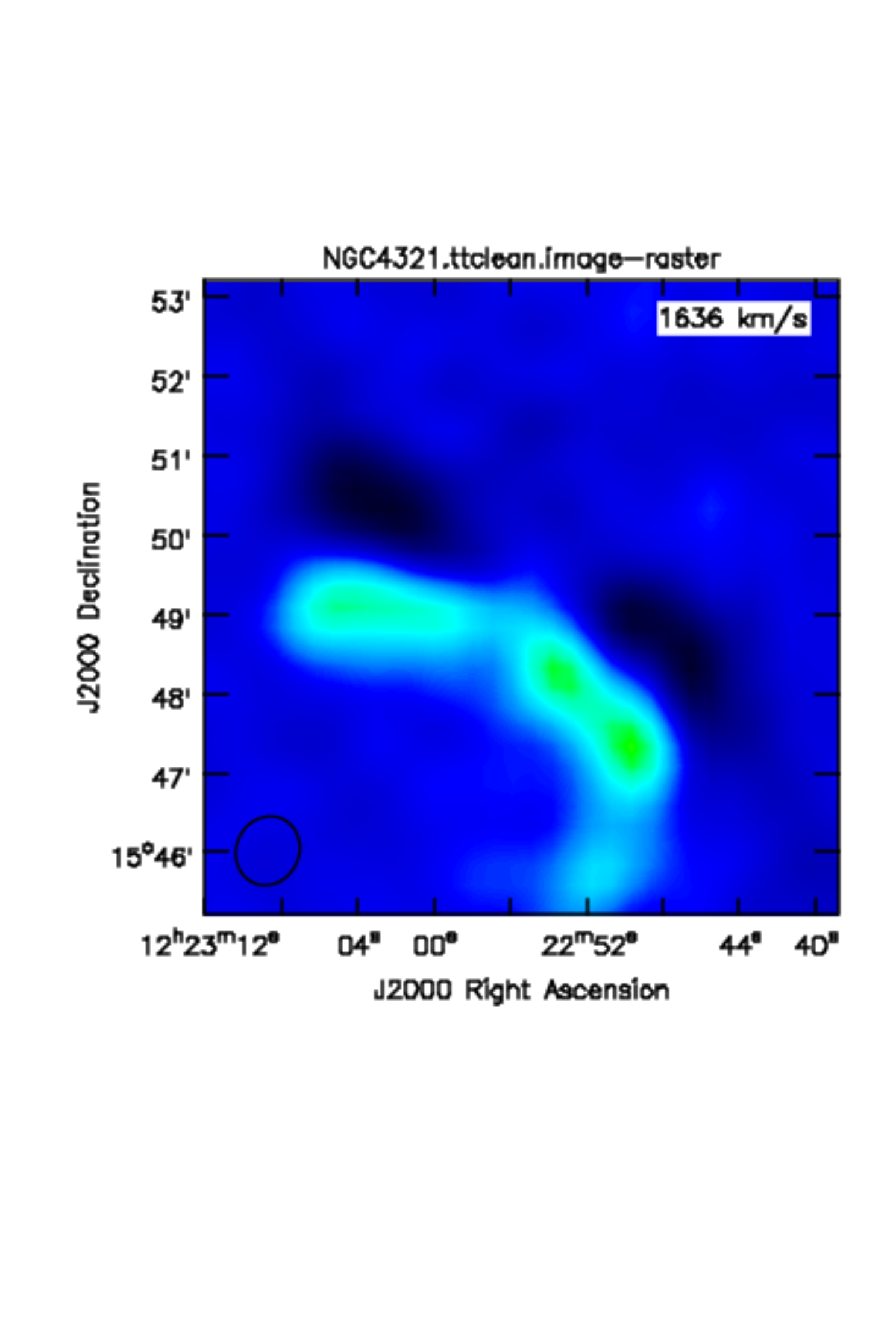}
\includegraphics[width=3.9cm]{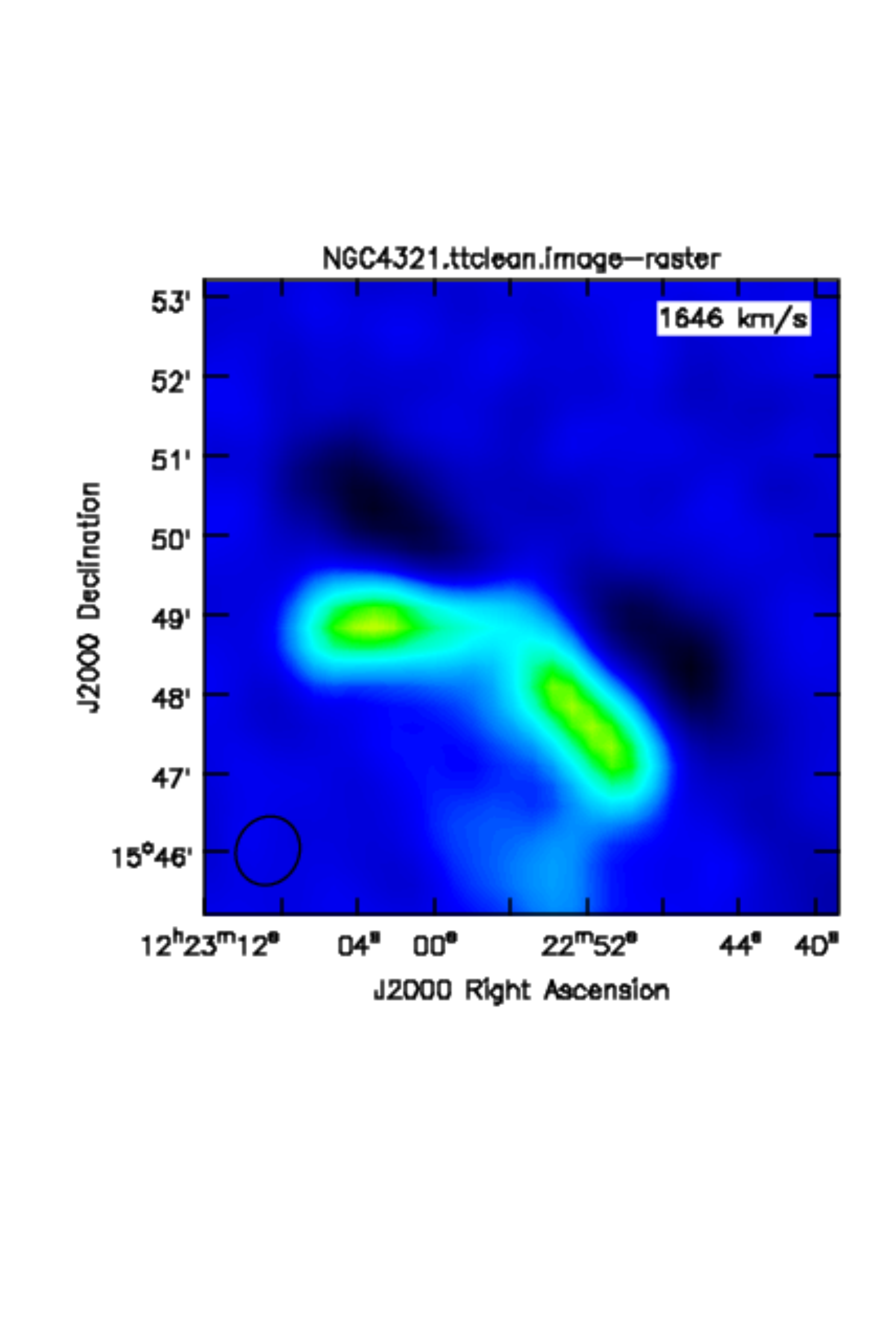}
\includegraphics[width=3.9cm]{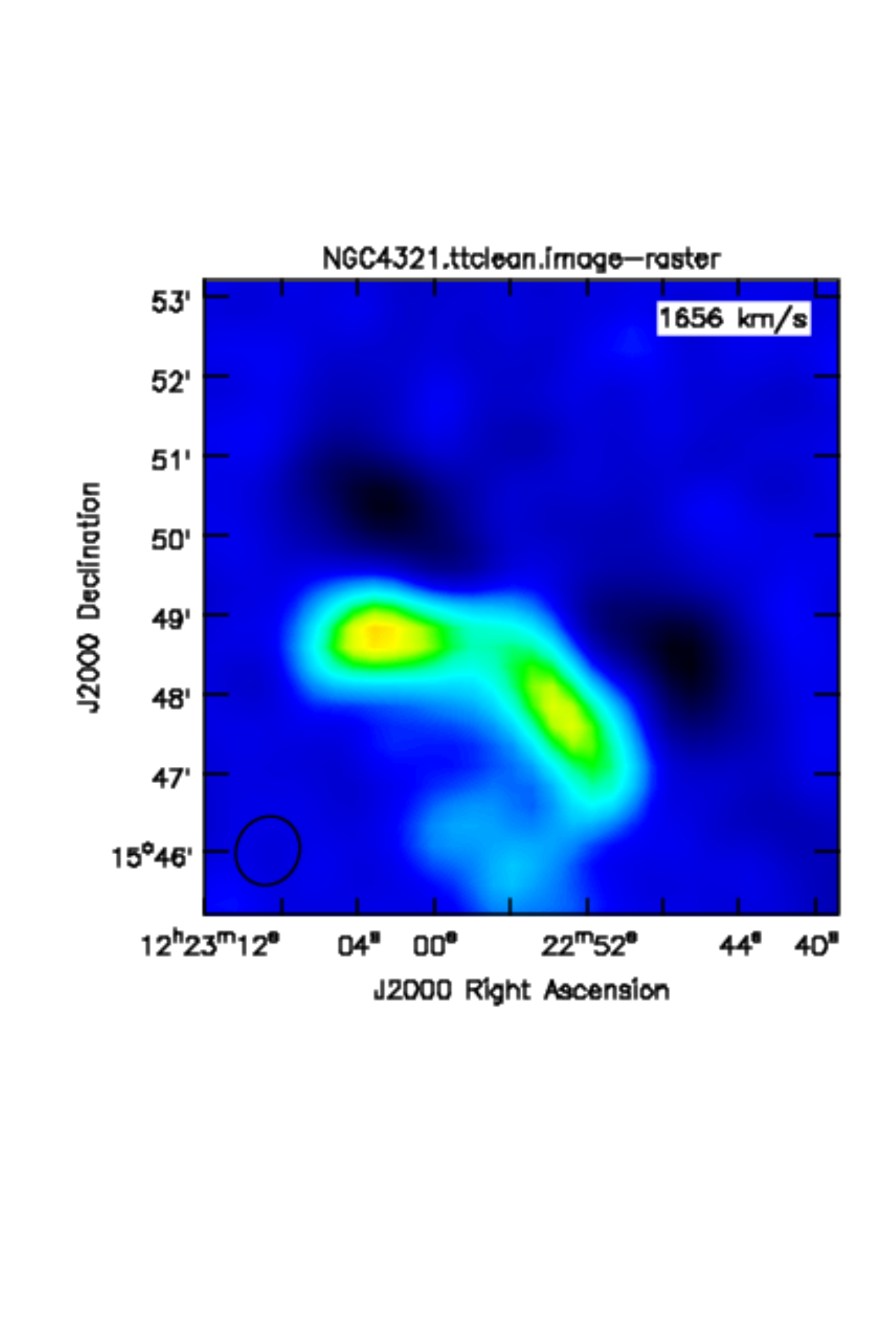}
\includegraphics[width=3.9cm]{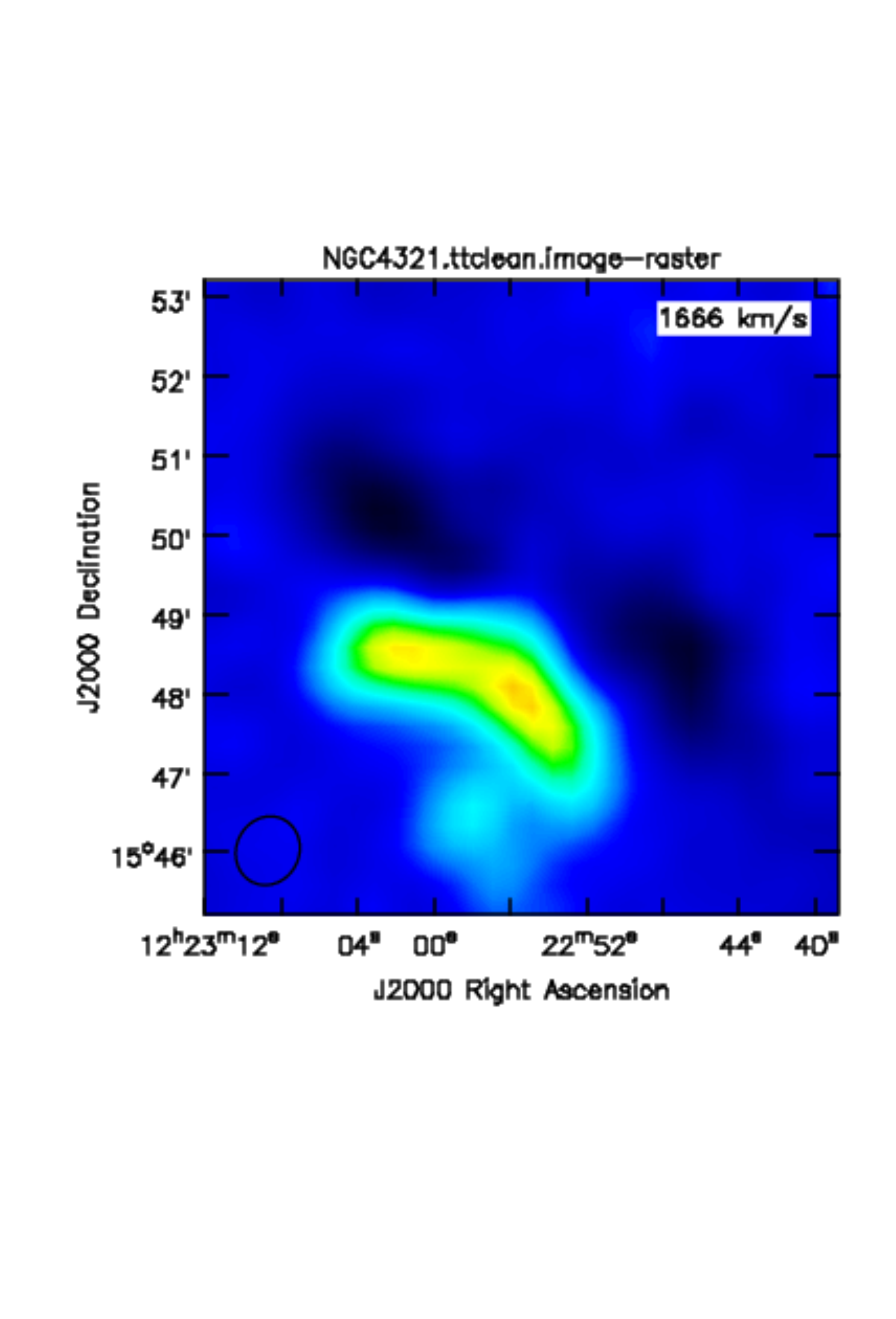}
\end{figure}

\begin{figure}
\includegraphics[width=3.9cm]{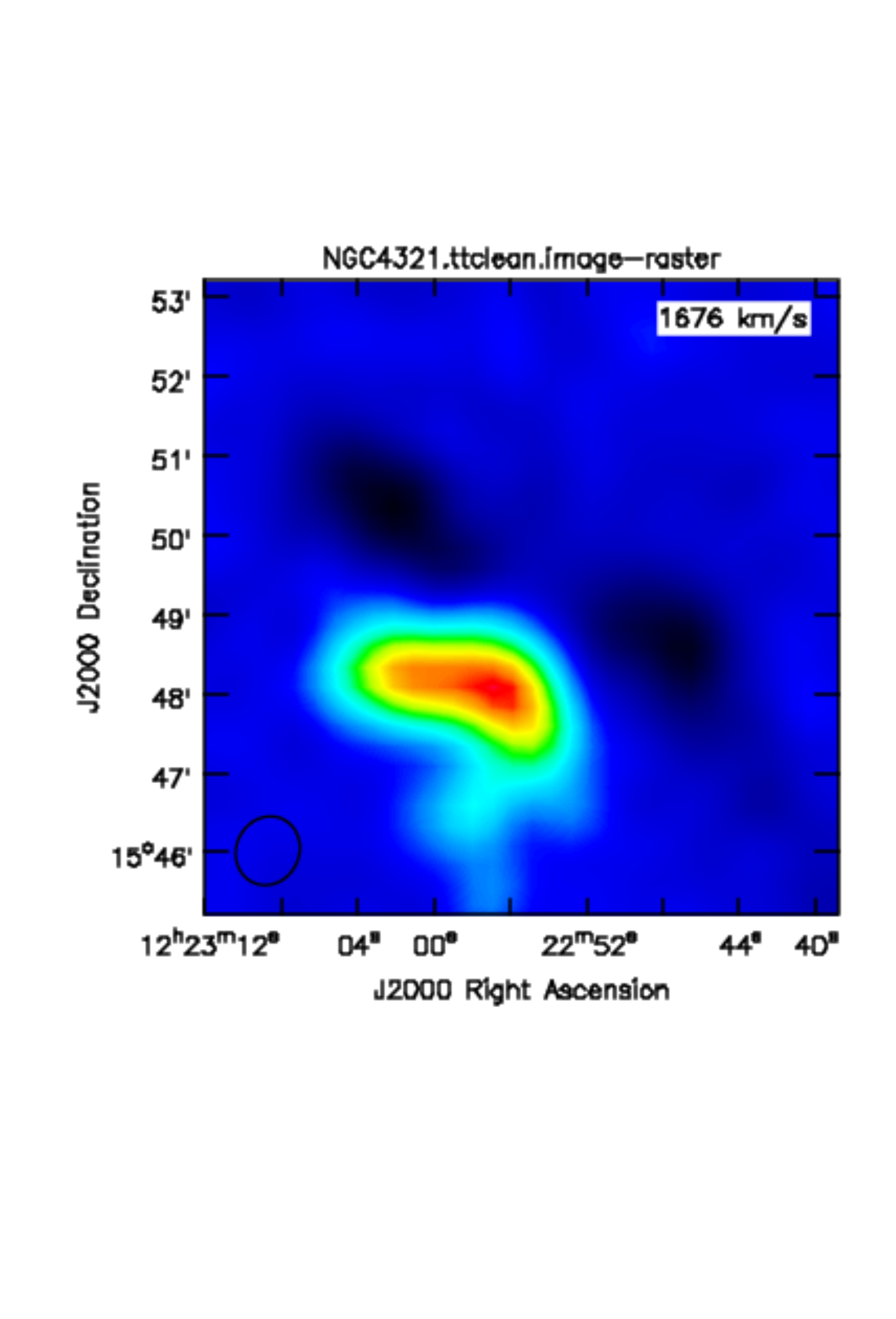}
\includegraphics[width=3.9cm]{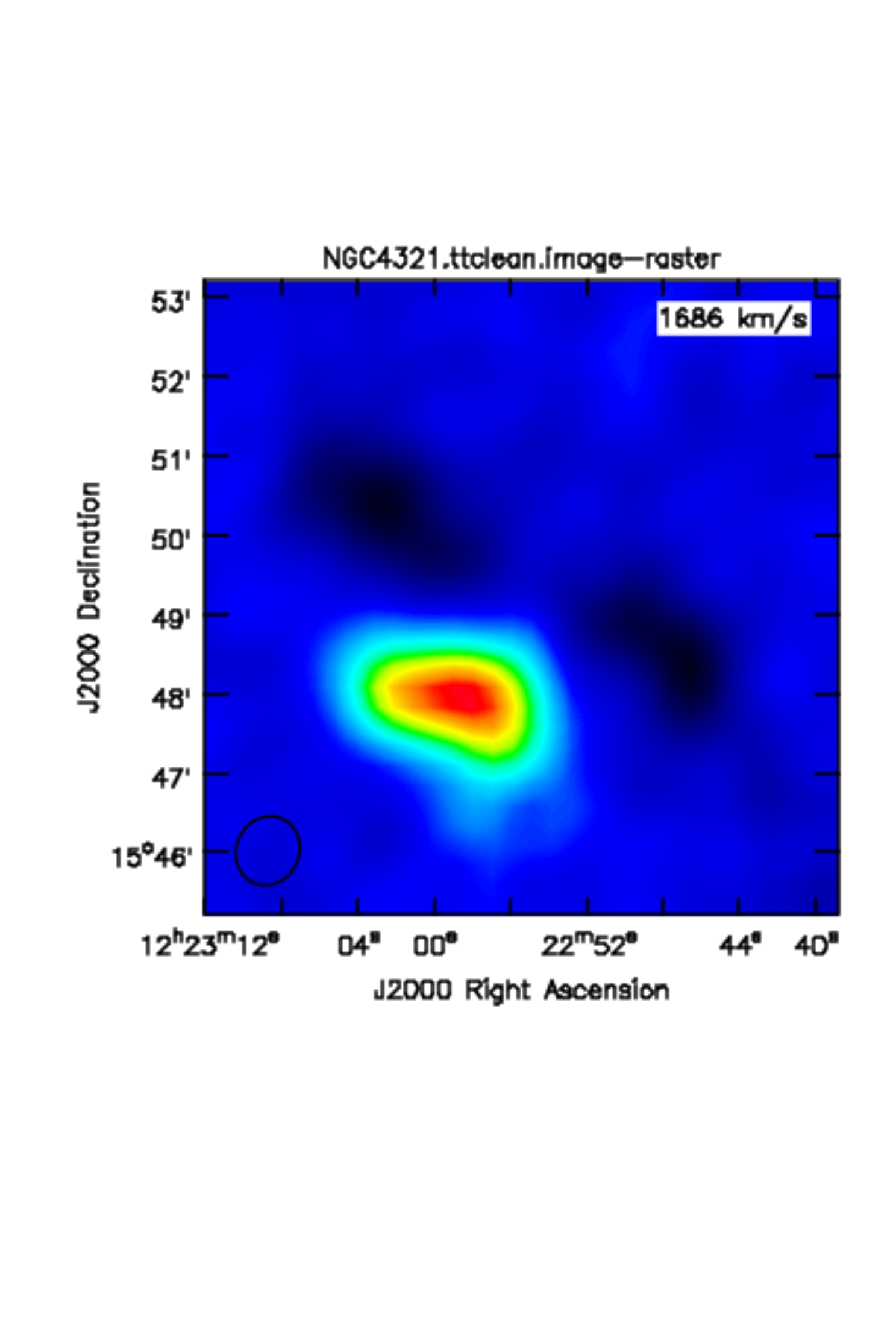}
\includegraphics[width=3.9cm]{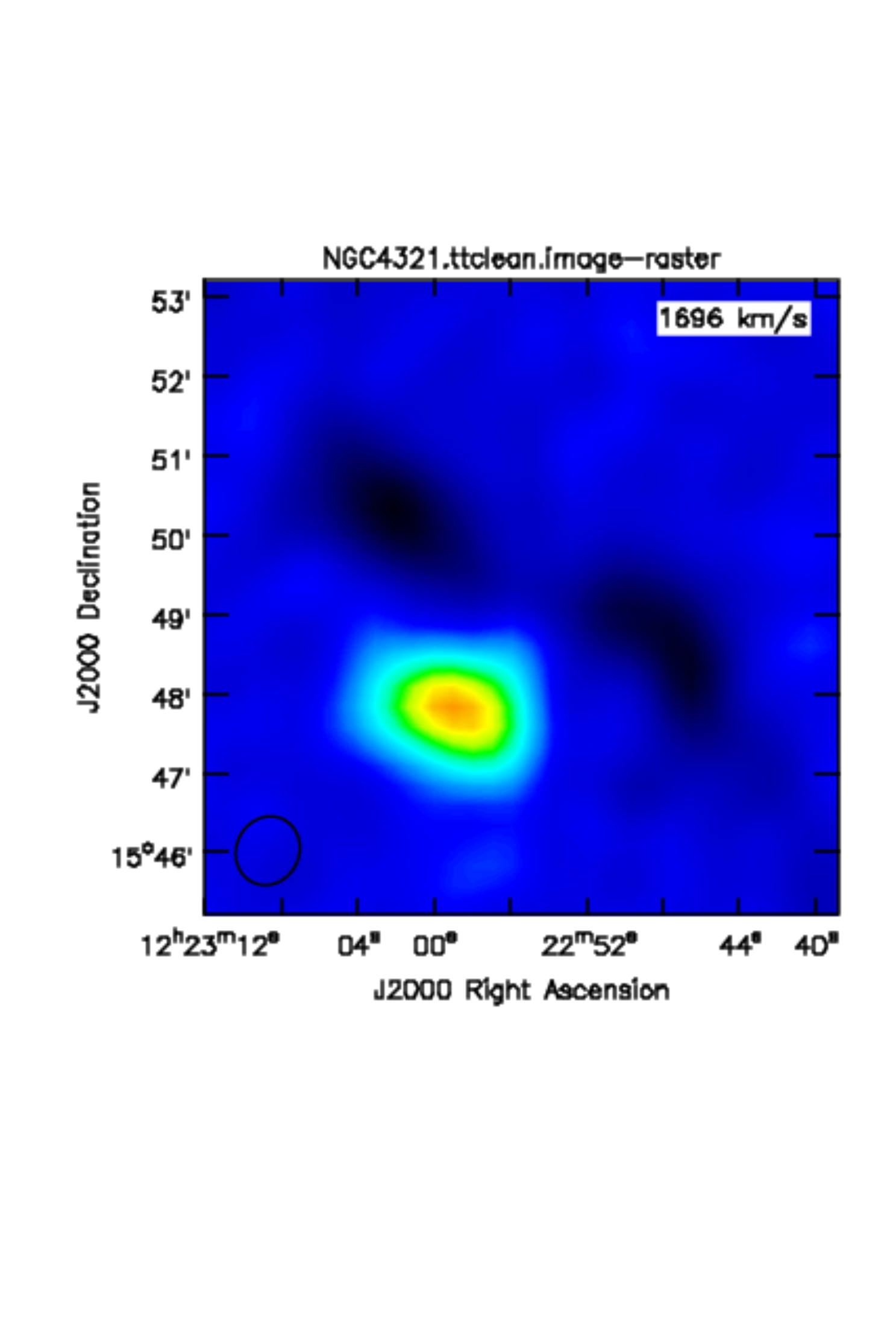}
\includegraphics[width=3.9cm]{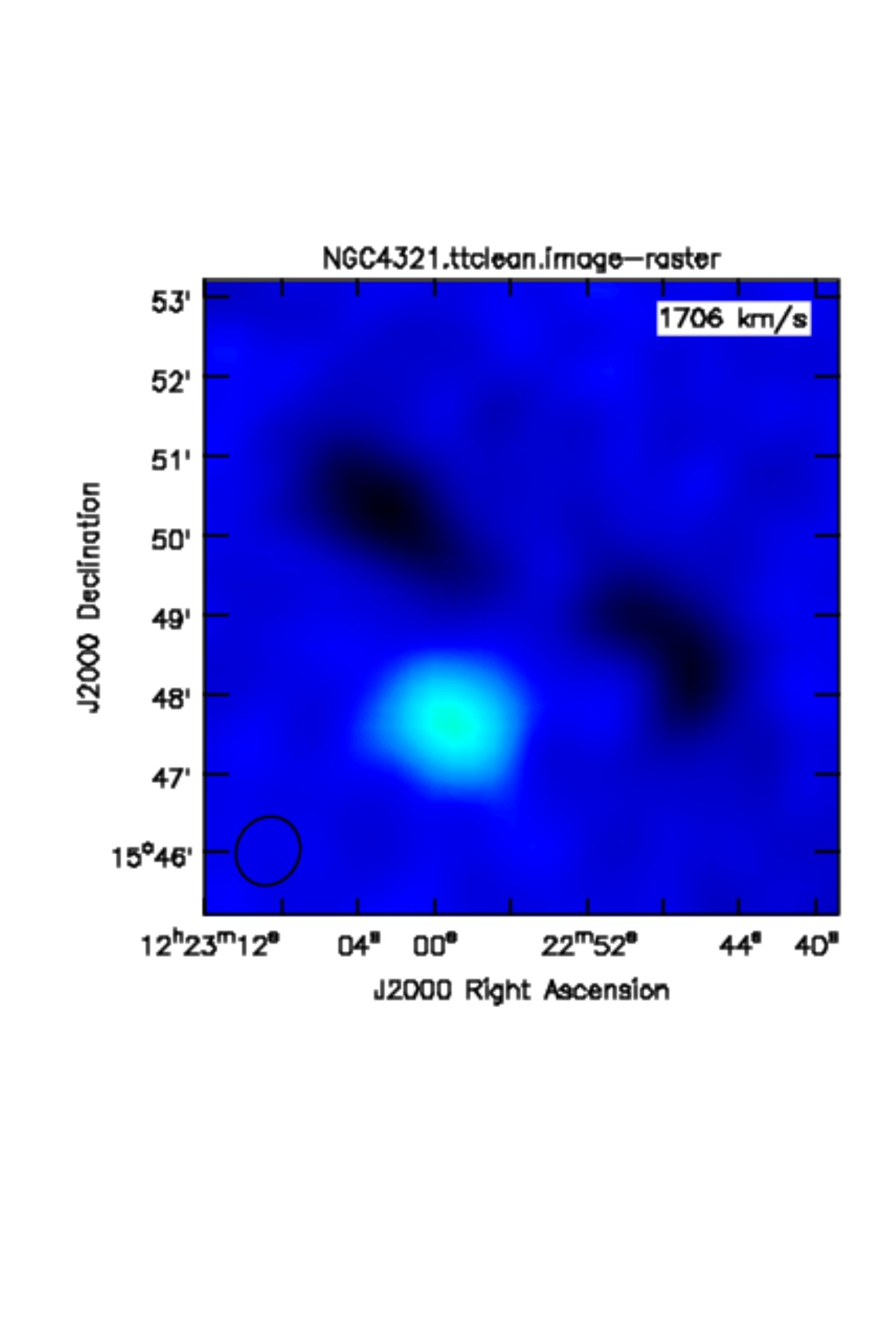}
\includegraphics[width=7.7cm, height=3.5cm]{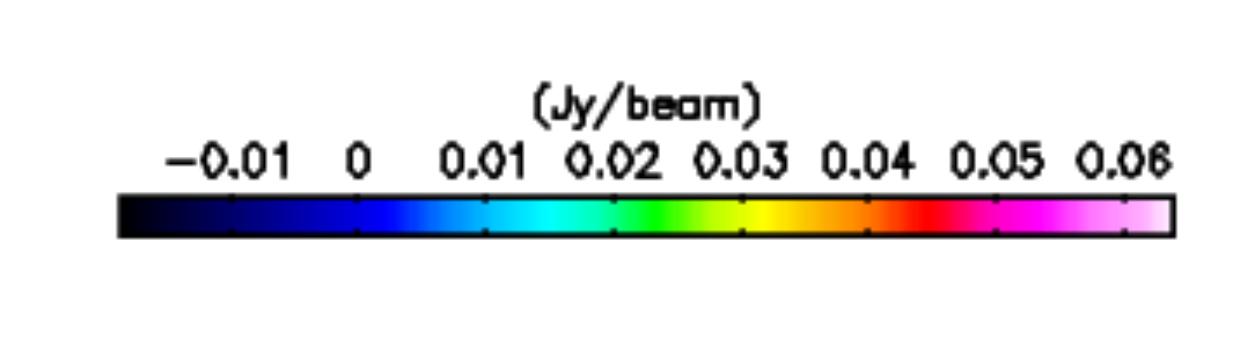}

\caption{\texttt{Velocity channel map of HI in the central region of NGC 4321}}
\end{figure}

\graphicspath{ {./Figure/} }
\begin{figure}
\centering
\includegraphics[width=10cm]{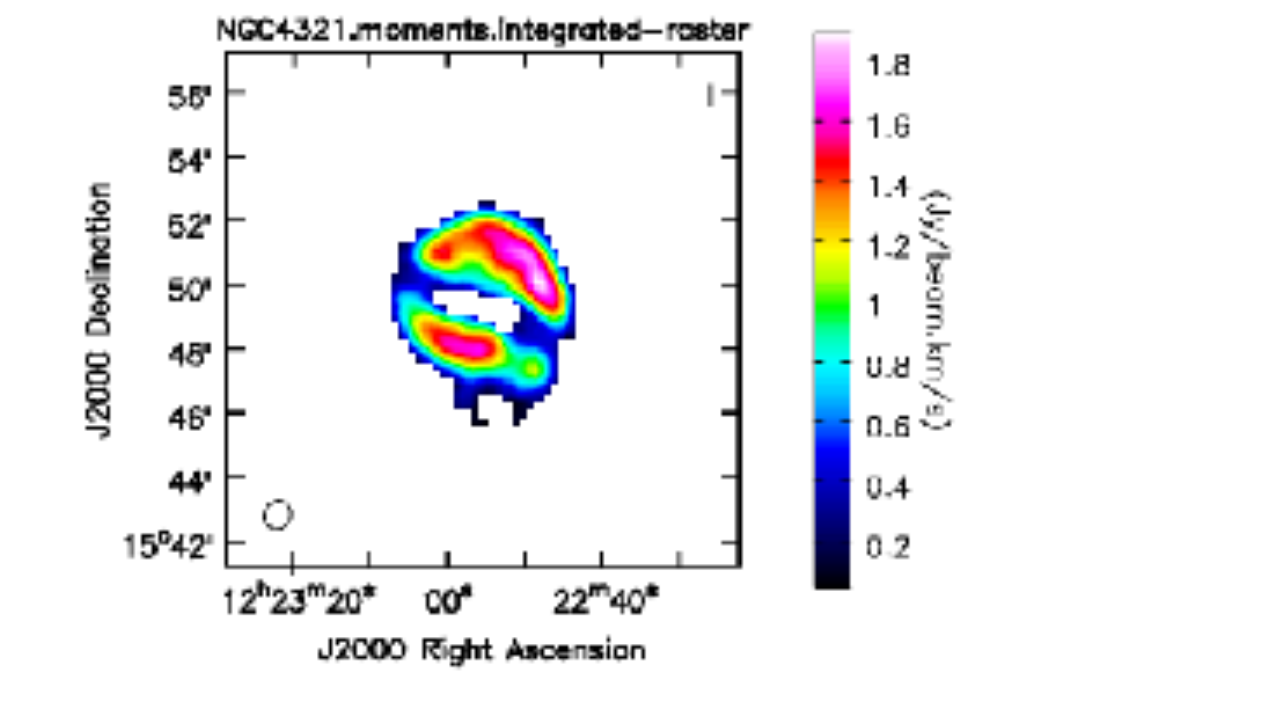}
\caption{\texttt{Integrated intensity (mom0) map of the HI 21cm line emission of NGC 4321}}
\end{figure}

\begin{figure}
\centering
\includegraphics[width=\columnwidth, height=15cm]{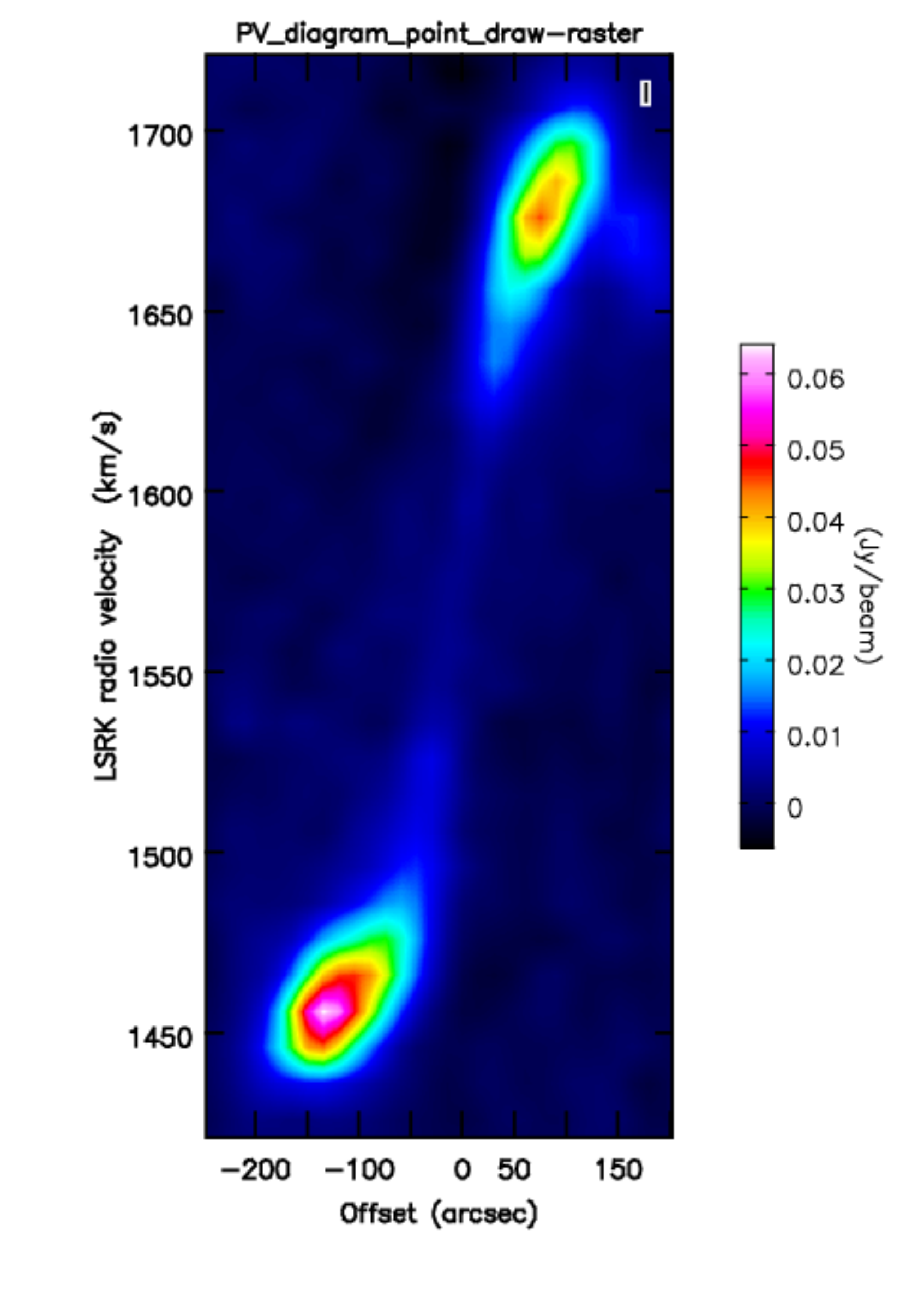}
\caption{\texttt{NGC 4321 PV diagram}}
\end{figure}

\subsection{\textit{Data calculation by using tilted-ring method}}

Next, the tilted-ring method is used in order to obtain the rotation velocity. By using the tilted-ring method, the rotation velocity, radial velocity and inclination angle in a galactic disk are coupled to each other \cite{oikawa2014rotation} as below:

\begin{equation}
V_{r}(r,\theta) = V_{obs}(r,\theta) - V_{sys} = V_{rot}(r)cos\hspace{0.1cm}\theta \hspace{0.1cm} sin \hspace{0.1cm} \emph{i}\hspace{0.5cm}
\end{equation}
where $V_{r}(r,\theta)$ is the radial velocity due to rotational motion, $V_{obs}(r,\theta)$ is the observed radial velocity, \emph{r} is the radius from the center of the galaxy, $\theta$ is the azimuth angle in the disk of a measured point from the major axis, $V_{sys}$ is the systemic velocity of the galaxy, $V_{rot}(r)$ is the rotation velocity and \emph{i} is the inclination angle of the galaxy.

Any coupling of rotation velocity and the inclination can be solved by using the tilted-ring method if a velocity field is observed \cite{rogstad1974aperture, bosma198121, begeman1989hi, jozsa2007kinematic}. This is because of the functional shape of the variation of $V_{r}(r,\theta)/V_{r}(r,0)$ is against the position angle on the sky. In this situation, $V_{r}(r,0)$ is the maximum value of $V_{r}$ along an initially chosen ring \cite{oikawa2014rotation}. For  $\theta = 0 $, \space $V_{rot}(r)$ = $V_{r}(r,0)$/cos 0 sin \emph{i}, where $V_{r}(r,0)$ = $V_{obs}(r,0)$ - $V_{sys}$. The systemic velocity and inclination angle of M100 is 1575 $kms^{-1}$ \cite{knapen2000kinematics} and 27$^{\circ}$ \cite{knapen1993star} respectively. The complete rotation velocity calculation of M100 is shown in Table 2.

\begin{table}
\centering
\begin{tabular}{|c|c|c|c|}
\hline
Radius, r (kpc) & $V_{obs}$ ($kms^{-1}$) & $V_{obs} - V_{sys}$ ($kms^{-1}$) & \begin{tabular}[c]{@{}l@{}} $V_{rot} = (V_{obs} - V_{sys})/sin\hspace{0.1cm} \emph{i}$ \\  ($kms^{-1}$) \end{tabular} \\ \hline 

0.02            & 1586.42      & 11.42               & 25.15                                \\ \hline
0.41            & 1603.5       & 28.50               & 62.78                                \\ \hline
0.83            & 1614.27      & 39.27               & 86.50                                \\ \hline
1.24            & 1622.81      & 47.81               & 105.31                               \\ \hline
1.66            & 1631.35      & 56.35               & 124.12                               \\ \hline
2.07            & 1637.49      & 62.49               & 137.65                               \\ \hline
2.49            & 1643.8       & 68.80               & 151.55                               \\ \hline
2.90            & 1649         & 74.00               & 163.00                               \\ \hline
3.31            & 1652.71      & 77.71               & 171.17                               \\ \hline
3.73            & 1656.42      & 81.42               & 179.34                               \\ \hline
4.14            & 1660.5       & 85.50               & 188.33                               \\ \hline
4.56            & 1664.96      & 89.96               & 198.15                               \\ \hline
4.97            & 1668.3       & 93.30               & 205.51                               \\ \hline
5.39            & 1671.27      & 96.27               & 212.05                               \\ \hline
5.80            & 1674.24      & 99.24               & 218.59                               \\ \hline
6.21            & 1675.73      & 100.73              & 221.88                               \\ \hline
6.63            & 1678.33      & 103.33              & 227.60                               \\ \hline
7.04            & 1679.81      & 104.81              & 230.86                               \\ \hline
7.46            & 1681.66      & 106.66              & 234.94                               \\ \hline
7.87            & 1684.71      & 109.71              & 241.66                               \\ \hline
8.29            & 1686.55      & 111.55              & 245.71                               \\ \hline
8.70            & 1688.38      & 113.38              & 249.74                               \\ \hline
9.11            & 1690.21      & 115.21              & 253.77                               \\ \hline
9.53            & 1691.44      & 116.44              & 256.48                               \\ \hline
9.94            & 1693.27      & 118.27              & 260.51                               \\ \hline
\end{tabular}
\newline
\newline
Table 2: The radius, observed velocity, radial velocity and rotation velocity of galaxy M100
\end{table}

In Table 2, column 1 illustrates the radius, \emph{r} of M100, column 2 reveals the observed velocity, $V_{obs}$ of M100, obtained from Figure 3 VLA PV diagram. Column 3 demonstrates the radial velocity, $V_{r}$ of M100 by using observed velocity, $V_{obs}$ minus systemic velocity, $V_{sys}$ (1575 $kms^{-1}$). While column 4 shows the rotation velocity, $V_{rot}$ of M100 by using radial velocity, $V_{r}$ divides sin \emph{i} (27$^{\circ}$). Then, a graph of the total rotation velocity against the radius of M100 is drawn as shown in Figure 4. In Figure 4, the error bars of the observed data are obtained from the rms values of the VLA data, which the root mean square is often use as a synonym for standard deviation of a signal from a given baseline.

\graphicspath{ {./Figure/} }
\begin{figure}
\centering
\includegraphics[width=\columnwidth]{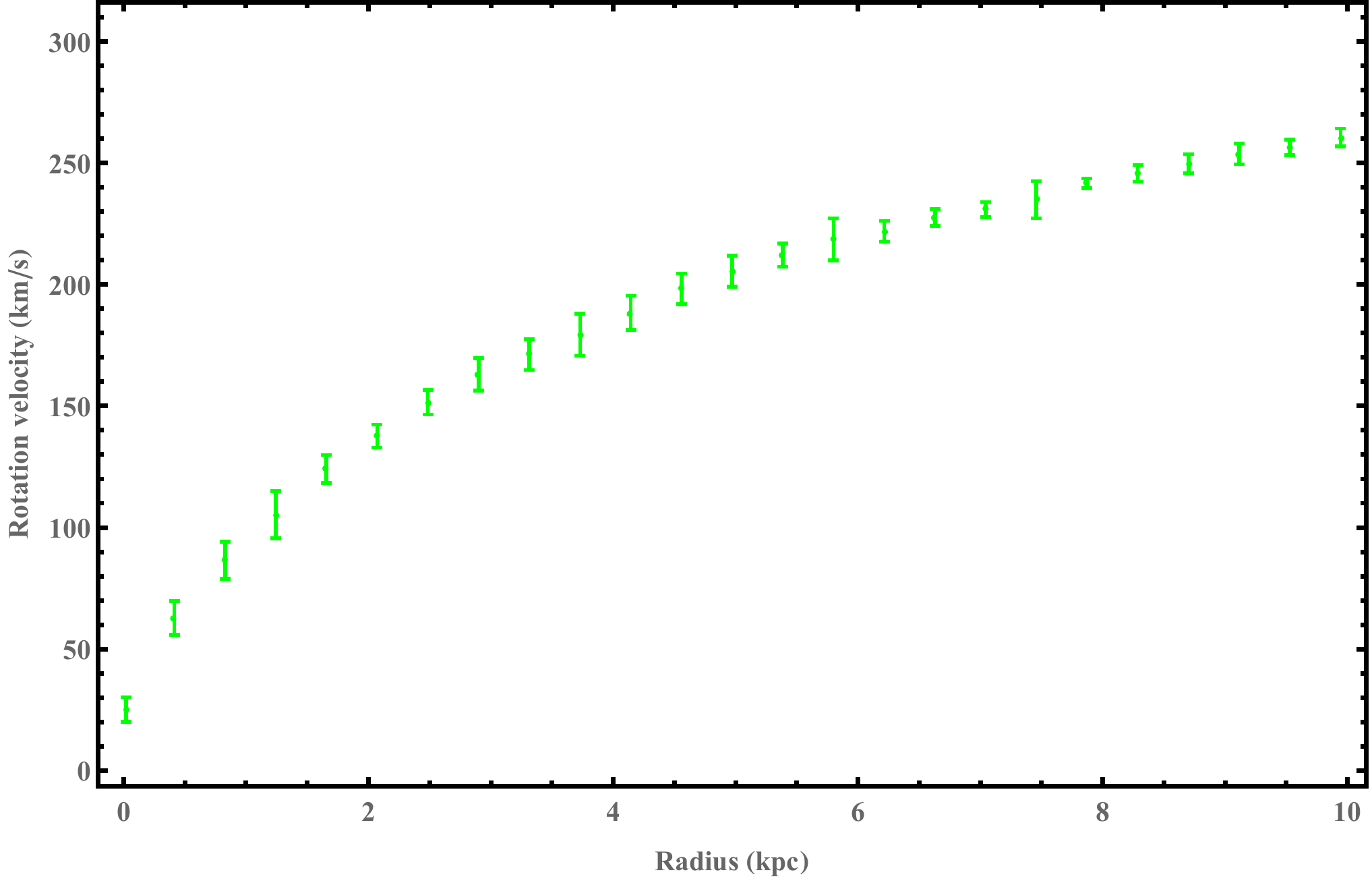}
\caption{\texttt{The total rotation velocity against the radius of M100}}
\end{figure}

\subsection{\textit{Rotation curve comparison}}

Our rotation curve is then compared with Rubin et al. \cite{rubin1980rotational} and Knapen et al. \cite{knapen1993star} rotation curves. Rubin and Knapen only showed rotation velocity and did not decompose it into separate contribution of gas, star and dark matter. Hence, we were only able to compare the total rotation velocities between Rubin, Knapen and our rotation curves.

For Rubin rotation curve, the values of the rotation velocities are available in their paper, we have included their rotation velocities in the Table 3. For Knapen rotation curve, they did not include the values of rotation velocities from their paper, so we will trace the rotation velocity value from the figure in their paper. We standardized all the relevant units from the two said papers and our manuscript. The total rotation velocity against the radius of Rubin, Knapen and our rotation curve is hence shown in Table 3. The three rotation curves are plotted together in Figure 5.

\begin{table}
\centering
\begin{tabular}{|c|c|c|c|}
\hline
Radius (kpc) & \multicolumn{3}{c|}{Total rotation velocity ($kms^{-1}$)}    \\ \hline
             & Rubin et al. 1980 \cite{rubin1980rotational} & Knapen et al. 1993 \cite{knapen1993star} & Our rotation curve \\ \hline
0            & 0                      & 0                       & 0                    \\ \hline
1            & 133                    & 159                     & 96                   \\ \hline
2            & 124                    & 205                     & 137                  \\ \hline
3            & 158                    & 211                     & 168                  \\ \hline
4            & 182                    & 216                     & 185                  \\ \hline
5            & 188                    & 229                     & 206                  \\ \hline
6            & 190                    & 231                     & 220                  \\ \hline
7            & 193                    & 233                     & 230                  \\ \hline
8            & 197                    & 247                     & 243                  \\ \hline
9            & 199                    & 259                     & 252                  \\ \hline
10           & 201                    & 265                     & 262                  \\ \hline
\end{tabular}
\newline
\newline
Table 3: The comparison of the rotation velocity obtained by Rubin et al., Knapen et al. and our rotation curve
\end{table}

\graphicspath{ {./Figure/} }
\pdfsuppresswarningpagegroup=1
\begin{figure}
\centering
\includegraphics[width=\columnwidth]{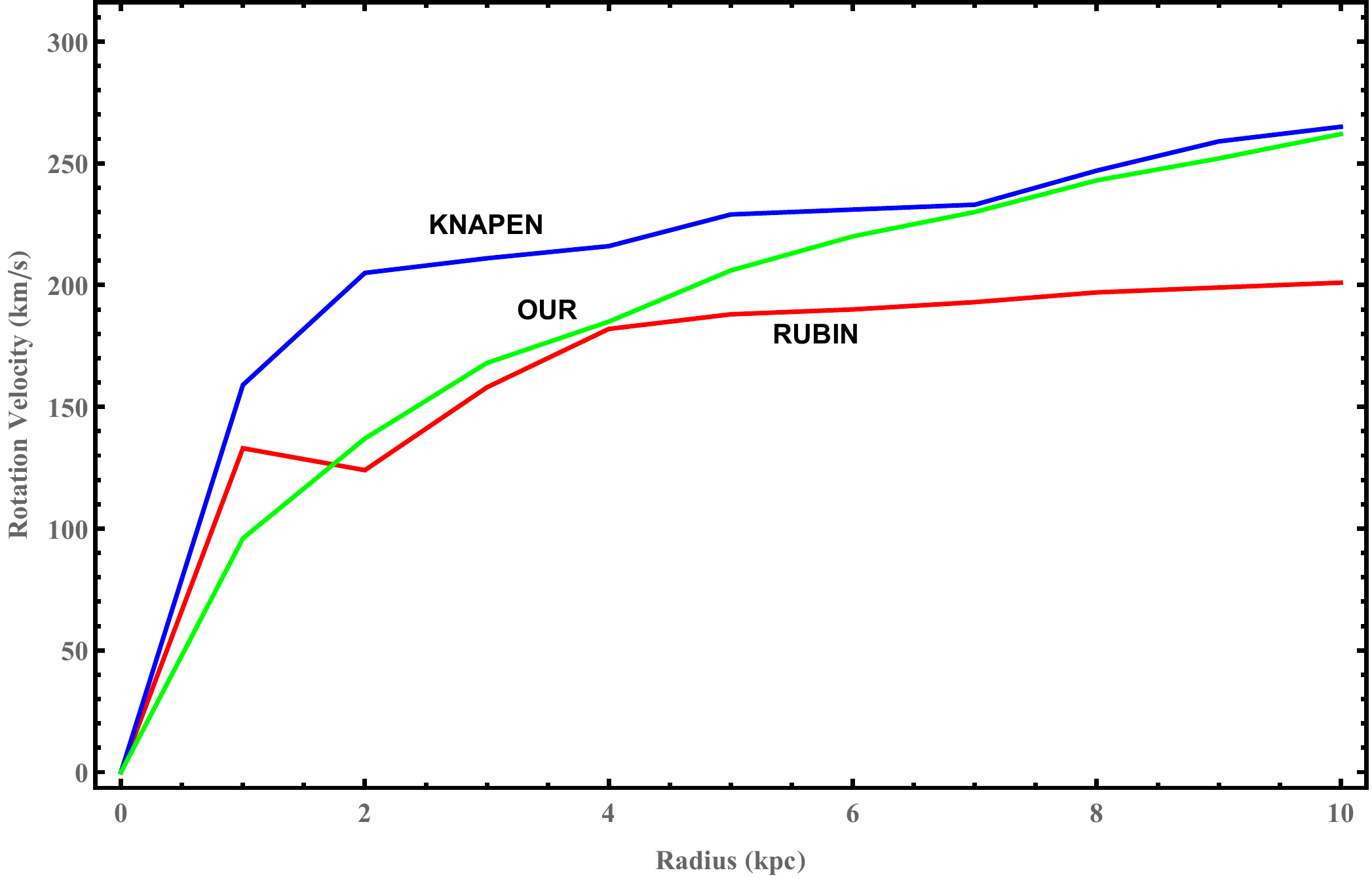}
\caption{\texttt{The total rotation velocity comparison between Rubin, Knapen and our rotation curve. The red line represents Rubin rotation curve, the blue line represents Knapen rotation curve and the green line represents our rotation curve.}}
\end{figure}

By referring to Figure 5, the reasons of Rubin rotation curve to be different from Knapen and our rotation curves are due to the differences of systemic velocity, position angle and inclination angle that were applied. The comparison is shown in Table 4.

\begin{table}
\centering
\begin{tabular}{|l|l|l|l|}
\hline
Parameter & Rubin et al. 1980 \cite{rubin1980rotational} & Knapen et al. 1993 \cite{knapen1993star} & This paper  \\ \hline
\begin{tabular}[c]{@{}l@{}} Systemic Velocity \\ ($kms^{-1}$) \end{tabular} & $1545\pm25$ & $1570.8\pm0.8$ & 1575                \\ \hline
Position angle   ($^{\circ}$)      & 140 & $153\pm 1$      & \begin{tabular}[c]{@{}l@{}} -26 \\ (same as 154) \end{tabular} \\ \hline
\begin{tabular}[c]{@{}l@{}} Inclination \\ angle ($^{\circ}$) \end{tabular}   & 35          & 27             & 27                  \\ \hline
\end{tabular}
\newline
\newline
Table 4: The comparison of parameter adopted between Rubin et al, Knapen et al. and in this paper
\end{table}

The rotation curve of Vera Rubin was observed in the HII regions emission from the Kitt Peak 4 m RC spectrograph in 1978 \cite{rubin1980rotational}, while the rotation curve of Knapen and ours were observed in the 21 cm line of neutral hydrogen with Very Large Array (VLA) in 1990 and 2003, respectively. In Table 4, the difference in position angle will affect the Position-Velocity diagram obtained during data reduction. It is worth noting that the differences in systemic velocity and inclination angle will affect the calculation for rotation velocity. This is because the tilted ring method formulation involves both systemic velocity and inclination angle. The uncertainty of Rubin is far larger than the uncertainty of Knapen. The reason is that different instrument is used and normally instruments improved in their uncertainty calculations over the years.

Furthermore, the measurement methods vary between Knapen and ours. They produced maps with a resolution of $12.7" \times 13.3"$ uniform weighting and $31.75" \times 28.75"$ natural weighting by using AIPS software \cite{knapen1993star}. However, we produced images with a resolution of $52.95" \times 47.66"$ Briggs weighting by using CASA software. Natural weighting is applied to all visibilities equally and results in maximum point-source sensitivity in images. However, this weighting produces poor synthesized beam-shape and side-lobe levels. The uniform weighting gives the visibilities a weight that is inversely proportional to the sampling density function. This causes a reduction in the side-lobes of the PSF. The uniform weighting scheme provides better resolution but lowers the sensitivity of the image \cite{vafaei2019deepsource}. Both natural and uniform weightings have some weak points but the use of Briggs weighting presents a way to overcome these weaknesses. Briggs weighting creates a PSF that smoothly varies between natural and uniform weightings based on the signal-to-noise ratio of the measurements and a tunable parameter that defines a noise threshold \cite{briggs1995high}, which produced better image. Furthermore, for Knapen rotation curve central region, they did mention that their curve rises more steeply near the center. This is because their profile was measured by a high full width at half maximum within a beam of half-power width, which corresponds to a rapidly rising rotation curve \cite{knapen1993star}. Beyond 2 kpc radius (0.4 arcmin), the rapidly rising rotation velocity has slowed down and, eventually, starting from 6 kpc radius, the rotation velocity of Knapen rotation curve is comparable with the rotation velocity of our rotation curve.

\subsection{\textit{Dark Matter Profile}}

Researchers expand the dark matter study with a larger theoretical framework to provide better ways of directly investigating it, since then a lot of dark matter profiles have been derived. In our research of rotation curve-fitting on the galaxy NGC 4321, we will consider the nine dark matter profiles including seven cored and two cuspy profiles to analyze the distribution and mass of dark matter halo. Cored profiles that we used are Pseudoisothermal, Burkert, Einasto, core-modified, DC14, coreNFW and Lucky13 profiles. While for cuspy profiles, we used Navarro, Frenk and White (NFW) and Moore profiles.

\subsubsection{\textit{Pseudoisothermal profile}}
Pseudoisothermal profile is a singular density profile that approaches a power law at the centre \cite{martel2002gravitational} and the mass distribution for larger radii would diverge proportional to the radius. For Pseudoisothermal profile, the density, mass and velocity equations of dark matter \cite{jimenez2003dark} are as follows:
\begin{equation}
\rho_{Iso}( r )= \frac{\rho_{0}}{1+(r/r_{s})^2}
\end{equation}
\begin{equation}
M_{Iso}( r ) = 4\pi \rho_{0} r_{s}^{2}( r- {r_{s}} \tan^{-1} (\frac{r}{r_{s}}))
\end{equation}
\begin{equation}
V_{Iso}^{2}( r ) = 4\pi G\rho_{0} r_{s}^{2}( 1- \frac{r_{s}}{r} \tan^{-1} (\frac{r}{r_{s}}))
\end{equation}
where parameter \textit{G} is the universal gravitational constant, $\rho_{0}$ is the scale density, $r_{s}$ is the scale radius and \textit{r} is the radius from the centre of the galaxy. These parameter definitions are the same for all others profiles except Einasto and DC14 profiles. Hence, these parameter definitions will not be mentioned again in the following profiles.

\subsubsection{\textit{Burkert profile}} 
Burkert profile revises density law of Pseudoisothermal profile in the inner region. This represents the mass profile for larger radii that diverge logarithmically with increasing radius. This is in agreement with the predictions of cosmological CDM calculations \cite{burkert1995structure}. For Burkert profile, the density \cite{burkert1995structure}, mass \cite{salucci2000dark} and velocity equations \cite{hashim2015nonlinear} of dark matter are as follows:
\begin{equation}
\rho_{Bur}( r )=\frac{ {\rho_{0}r_{s}}^{3}}{{(r+ r_{s}) (r^2+r_{s}^2)}}
\end{equation}
\begin{equation}
M_{Bur}( r ) = 6.4{\rho_{0}}r_{s}^{3} [\ln(1+ \frac{r}{r_{s}}) - \tan^{-1}(\frac{r}{r_{s}}))+ 0.5 \ln(1+(\frac{r}{r_{s}})^2)]
\end{equation}
\begin{equation}
V_{Bur}^{2}( r ) = {\frac{6.4G\rho_{0} r_{s}^{3}}{r}}( \ln[(1+ \frac{r}{r_{s}})(1+(\frac{r}{r_{s}})^2)^{0.5}] -\tan^{-1}(\frac{r}{r_{s}}))
\end{equation}

\subsubsection{\textit{Navarro, Frenk and White (NFW) profile}}
NFW profile is a traditional benchmark profile motivated by N-body simulations, which is shallower than Pseudoisothermal near the centre, and steeper than Pseudoisothermal in the outer regions \cite{navarro1996structure}. For NFW profile, the density \cite{navarro1996structure}, mass \cite{coe2010dark} and velocity equation \cite{hashim2015nonlinear} of dark matter are as follows:
\begin{equation}
\rho_{NFW}( r )= {\rho_{0}}\frac{r_{s}}{r}{(1+\frac{r}{r_{s}})^{-2}} 
\end{equation}
\begin{equation}
M_{NFW}( r ) = 4\pi \rho_{0} r_{s}^{3}( \ln(1+\frac{r}{r_{s}})  - \frac{r}{r_{s}+r})
\end{equation}
\begin{equation}
V_{NFW}^{2}( r ) =\frac{12.6G \rho_{0} r_{s}^{3}}{r}( \ln(1+\frac{r}{r_{s}})  - \frac{r}{r_{s}+r})
\end{equation}

\subsubsection{\textit{Moore profile}}
Moore profile behaves similarly to the NFW profile at large radii but is steeper than NFW profile at smaller radii \cite{diemand2004convergence}. For Moore profile, the density \cite{moore1999cold}, mass \cite{klypin2001resolving} and velocity equations of dark matter are as follows:
\begin{equation}
\rho_{Moo}( r )= \frac{\rho_{0}}{{(\frac{r}{r_{s}})^{1.5}} {(1+(\frac{r}{r_{s}})^{1.5}})} 
\end{equation}
\begin{equation}
M_{Moo}( r ) = \frac{8}{3}\pi \rho_{0} r_{s}^{3}(\ln(1+(\frac{r}{r_{s}})^{1.5}))
\end{equation}
\begin{equation}
V_{Moo}^{2}( r ) = \frac{8.38G\rho_{0} r_{s}^3}{r}(\ln(1+(\frac{r}{r_{s}})^{1.5}))
\end{equation}

\subsubsection{\textit{Einasto profile}}
Einasto profile is emerging as a better fit for more recent numerical simulations and provides the most accurate description of dark matter haloes \cite{sereno2016comparison}. Compared to other dark matter profiles use two free parameters, Einasto profile uses three free parameters to describe the halo mass profile instead and hence significantly improves the accuracy of the fitting to the inner density profiles of simulated haloes \cite{navarro2010diversity}. For Einasto profile, the density, mass \cite{chemin2011improved} and velocity equations of dark matter are as follows:
\begin{equation}
\rho_{Ein}( r )= {\rho_{-2}} exp[-2n((\frac{r}{r_{-2}})^ {1/n}  -1)]
\end{equation}
\begin{equation}
M_{Ein}( r )= 4\pi nr_{-2}^{3}{\rho_{-2}}e^{2n}(2n)^{-3n}\gamma( 3n,\frac{r}{r_{-2}})
\end{equation}
\begin{equation}
V_{Ein}^{2}( r ) = {\frac{4\pi Gnr_{-2}^{3}{\rho_{-2}}}{r}}e^{2n}(2n)^{-3n}\gamma( 3n,\frac{r}{r_{-2}})
\end{equation}
where $\gamma(3n,x)= \int_{0}^{x}e^{-t}t^{3n-1}dt$, parameter \textit{G} is the universal gravitational constant, $r_{-2}$ is the radius where the density profile has a slope of -2, $\rho_{-2}$ is the local density at that radius and \textit{r} is the radius from centre of the galaxy. While other dark matter models are described by two free parameters, $r_{s}$ and $\rho_{0}$, a characteristic scale and a characteristic density at that radius, Einasto model involves a third parameter, \textit{n}, the Einasto index which describes the shape of the density profile \cite{chemin2011improved}.

\subsubsection{\textit{Core-modified profile}}
The NFW profile is singular at the galactic center. To avoid the singularity, Brownstein proposed the core-modified profile. The core-modified profile is a profile with constant density in the central core \cite{brownstein2009modified}. The density, mass and velocity \cite{li2017comparing} equations of core-modified profile are as follows:

\begin{equation}
{\rho_{com}  (r) = \frac{\rho_0 r_s^3}{r^3+r_s^3} }
\end{equation}

\begin{equation}
{M_{com}  ( r ) = \frac{4}{3} \pi \rho_0 r_s^3 [\ln (r^3+r_s^3) - \ln (r_s^3)]}
\end{equation}

\begin{equation}
{V_{com}^2  ( r  ) = \frac{4}{3} \pi G\rho_0 \frac{r_s^3}{r} [\ln (r^3+r_s^3) - \ln (r_s^3)]}
\end{equation}

\subsubsection{\textit{DC14 profile}}

DC14 profile considers the baryonic feedback on the halo due to the supernovae, and hence modifies the halo profiles \cite{cintio2014mass}. Cintio et al. established the DC14 model, whose profile is defined in terms of the model class ($\alpha$, $\beta$, $\gamma$) \cite{hernquist1990analytical}, \cite{zhao1996analytical}:

\begin{equation}
\rho_{ \alpha \beta \gamma }  ( r  ) = \frac{\rho_0}{( \frac{r}{r_s})^\gamma [1+( \frac{r}{r_s})^\alpha]^{(\beta-\gamma)/\alpha}}
\end{equation}
where $\beta$ and $\gamma$ are the inner and outer slopes, respectively, and $\alpha$ describes the transition between the inner and outer regions. These parameters are represented by the equations below:

\begin{align}
\label{eqn:eqlabel}
\begin{split}
&\alpha = 2.94 - \log (10^{(X+2.33)-1.08}+10^{(X+2.33)2.29}),\\ \\
&\beta = 4.23 + 1.34X + 0.26X^{2},\\ \\
&\gamma = -0.06 + \log 10^{(X+2.56)-0.68}+10^{X+2.56}\\ \\
\end{split}
\end{align}
where X = $\log (M_{star}/M_{halo})$ , is the stellar-to-halo mass (SHM) ratio in logarithm space. The mass \cite{li2020comprehensive} and velocity of DC14 profile are as follows:

\begin{equation}
{M_{DC14}  ( r  ) =4 \pi \rho_0 r_s^3 \frac{1}{\alpha} (B[a,b+1, \epsilon]+ B[a+1,b,\epsilon])}
\end{equation}

\begin{equation}
{V_{DC14}^2  ( r  ) =4 \pi G \rho_0 \frac{r_s^3}{r} \frac{1}{\alpha} (B[a,b+1, \epsilon]+ B[a+1,b,\epsilon])}
\end{equation}
\newline
where $B(a,b,x) = \int_{0}^{x} t^{ \alpha - 1 } (1-t)^{ b-1 } dt$ is the incomplete Beta function and $a = ( 3- \gamma  )/\alpha$, $b = ( \beta -3  )/\alpha$ and $\epsilon = \frac{( r/r_{s} )^ \alpha}{(1+(r/r_{s})^ \alpha )}$. This equation only works for the SHM ratio within -4.1 $<$ X $<$ -1.3, due to the fact that this is the range where the supernovae feedback is significant and dominant \cite{cintio2014mass}. At X $<$ -4.1, the energy released by supernova is insufficient to modify the initial cuspy profile, while at X $>$ -1.3, the feedback due to active galactic nuclei might start to dominate \cite{li2020comprehensive}.

\subsubsection{\textit{CoreNFW}}

A coreNFW halo \cite{read2016dark} is essentially a NFW halo which transforms an inner cusp into a finite central core by a spherically symmetric function $f^{n}$ that models the effects of supernova feedback \cite{allaert2017testing}. The mass of coreNFW is defined as:

\begin{equation}
{M_{cN F W}  ( <r  )} = M_{N F W} ( <r  ) f^{n} ( r )
\end{equation}

with

\begin{equation}
f( r )= [\tanh (\frac{r}{r_{s}})  ]
\end{equation}
The strength of the core is determined by the parameter \emph{n}, which ranges between 0 $<$ \emph{n} $\leq$ 1. The equation of n is as follows:

\begin{equation}
n = \tanh( \kappa \frac{t_{SF}}{t_{dyn}}) 
\end{equation}
where $\kappa$ is a tunning parameter and $t_{SF}$ is the star formation time of the galaxy. We set $\kappa$ = 0.04 and $t_{SF}$ = 14 Gyrs as suggested by the simulations of Read et al. \cite{read2016dark}. The dynamical time, $t_{dyn}$ is the duration of 1 circular orbit at the scale radius in the unmodified NFW halo:

\begin{equation}
t_{dyn} = 2 \pi \sqrt \frac{r_{s}^3}{(G M_{NFW} ( r_{s}))} 
\end{equation}

Hence, the mass and velocity of coreNFW profile are defined as:

\begin{equation}
{M_{cN F W}  ( <r  )} = 4 \pi \rho_0 r_s^3  ( \ln  ( 1 + \frac{r}{r_{s}}) - \frac{r}{r_{s}+r}) f^{n} ( r )
\end{equation}

\begin{equation}
{V_{cN F W}^2  ( <r  )} = 4 \pi G \rho_0 \frac{r_s^3}{r}  ( \ln  ( 1 + \frac{r}{r_{s}}) - \frac{r}{r_{s}+r}) f^{n} ( r )
\end{equation}

\subsubsection{\textit{Lucky13}}

Lucky13 is a new semi-empirical profile constructed from Equation (21), the ($\alpha$, $\beta$, $\gamma$) models by Li et al. \cite{li2020comprehensive}. They  considered the transition parameter $\alpha$ = 1, $\gamma$ =  0 to reach a finite core and $\beta$ = 3 to get the same decreasing rate as the NFW profile at large radii. The density, mass and velocity of Lucky13 \cite{li2020comprehensive} are as follows:

\begin{equation}
\rho_{130}= \frac{\rho_{0}}{[ 1+ (\frac{r}{r_{s}}) ]^3}
\end{equation}

\begin{equation}
{M_{130}  ( r  )} = 4 \pi \rho_0 r_s^3  [ \ln( 1+ \frac{r}{r_s} ) + \frac{2}{(1+ \frac{r}{r_s} ) } - \frac{2}{2(1+ \frac{r}{r_s})^2  }  - \frac{3}{2}]
\end{equation}

\begin{equation}
{V_{130}^2  ( r  )} = 4 \pi G \rho_0 \frac{r_s^3}{r}  [ \ln( 1+ \frac{r}{r_s} ) + \frac{2}{(1+ \frac{r}{r_s} ) } - \frac{2}{2(1+ \frac{r}{r_s})^2  }  - \frac{3}{2}]
\end{equation}

\subsection{\textit{Star Velocity}}
Most of the luminous mass is occupied by stars, and the rest (i.e. less than 10 \%) is filled with interstellar gases. Hence, the star disc luminosity distribution roughly represents the luminous mass distribution \cite{sofue2013mass}. To estimate the stellar star luminosity, we use the velocity of star disc derived by Freeman \cite{freeman1970disks,di2019universal}, which is described by the equation below:
\begin{equation}
V^{2}_{star}=\frac{G M_{D}x^{2}} {2R_{D}}(I_{o}K_{o}-I_{1}K_{1})
\end{equation}
where $\textit{G}$ is the universal gravitational constant, $I_{n}$ and $K_{n}$ are the modified Bessel functions of the first and second kinds computed at $\textit{x}/2$ and $x = \frac{r}{R_{D}}$ , $M_{D}$ is the star mass and $R_{D}$ is the star scale length with value in kpc. The star scale length for NGC 4321 that we adopted is 75 arcsec \cite{koopmann2001atlas}, which is equivalent to 6.22 kpc.

The bulge in M100 is small compared to the disks hence no bulge-disk decomposition was adopted \cite{beckman1996scale}. The bulge of M100 does not contribute much to the surface brightness of the disk and the surface brightness have a linear correlation with stellar mass \cite{ichikawa2010moircs}. This indicates that the bulge of M100 does not contribute much to the stellar mass and can be omitted.

\subsection{\textit{Gas velocity}}
Taking account of the gas component and if its mass is significant, it is possible to use photometry to calculate the rotation curve of the baryonic disc component by matching it with observations to estimate the effect of dark matter on the disc dynamics \cite{zasov2017dark}. We compute the mass of gas from the equation \cite{hashim2014rotation} that is given below:
\begin{equation}
M_{gas}(r)=2\pi \int_{0}^{r} r \sigma (r) dr
\end{equation}
where \textit{r} is the radius from centre of the galaxy and $\sigma(r)$ is the surface density. From Equation (35), we applied the non-linear fitting method to the curve of total gas surface density. The steps of obtaining the gas velocity by using software Mathematica are as below:
\begin{enumerate}[label=\roman*]
\item Adopt the surface density profile total gas data from T. Wong and L. Blitz \cite{wong2002relationship} as shown in Figure 6. 
\item Plot the best fitting curve that computed from several trial values of the parameter and equation as shown in Figure 7.
\item Calculate the goodness of fit, $\chi^{2}$ to indicate the best-fitting curves and it will show the best model of $\Sigma$gas surface density, $\sigma(r)$.
\item Apply the best model of $\Sigma$gas surface density, $\sigma(r)$ to Equation (35) to calculate the mass of gas for each radius.
\item From the mass of gas for each radius, then derive the velocity for each radius by considering the equation \begin{equation} v=\sqrt{\frac{GM{(<r)}}{r}} \end{equation} and the result is as shown in Figure 8.
\end{enumerate}

\graphicspath{ {./Figure/} }
\pdfsuppresswarningpagegroup=1
\begin{figure}
\centering
\includegraphics[width=\columnwidth]{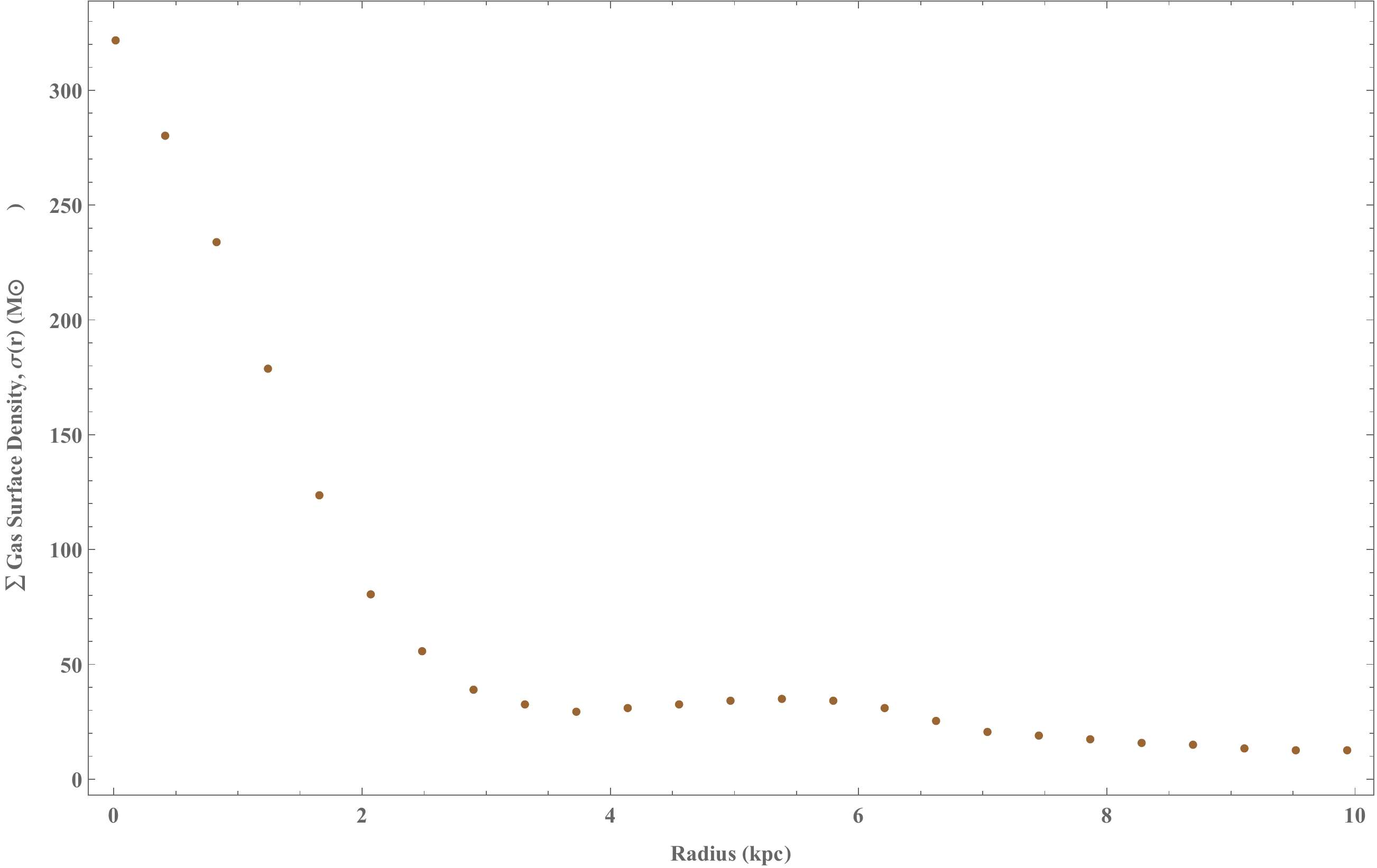}
\caption{\texttt{$\sum Gas$ surface density graph}}
\end{figure}

\begin{figure}
\centering
\includegraphics[width=\columnwidth]{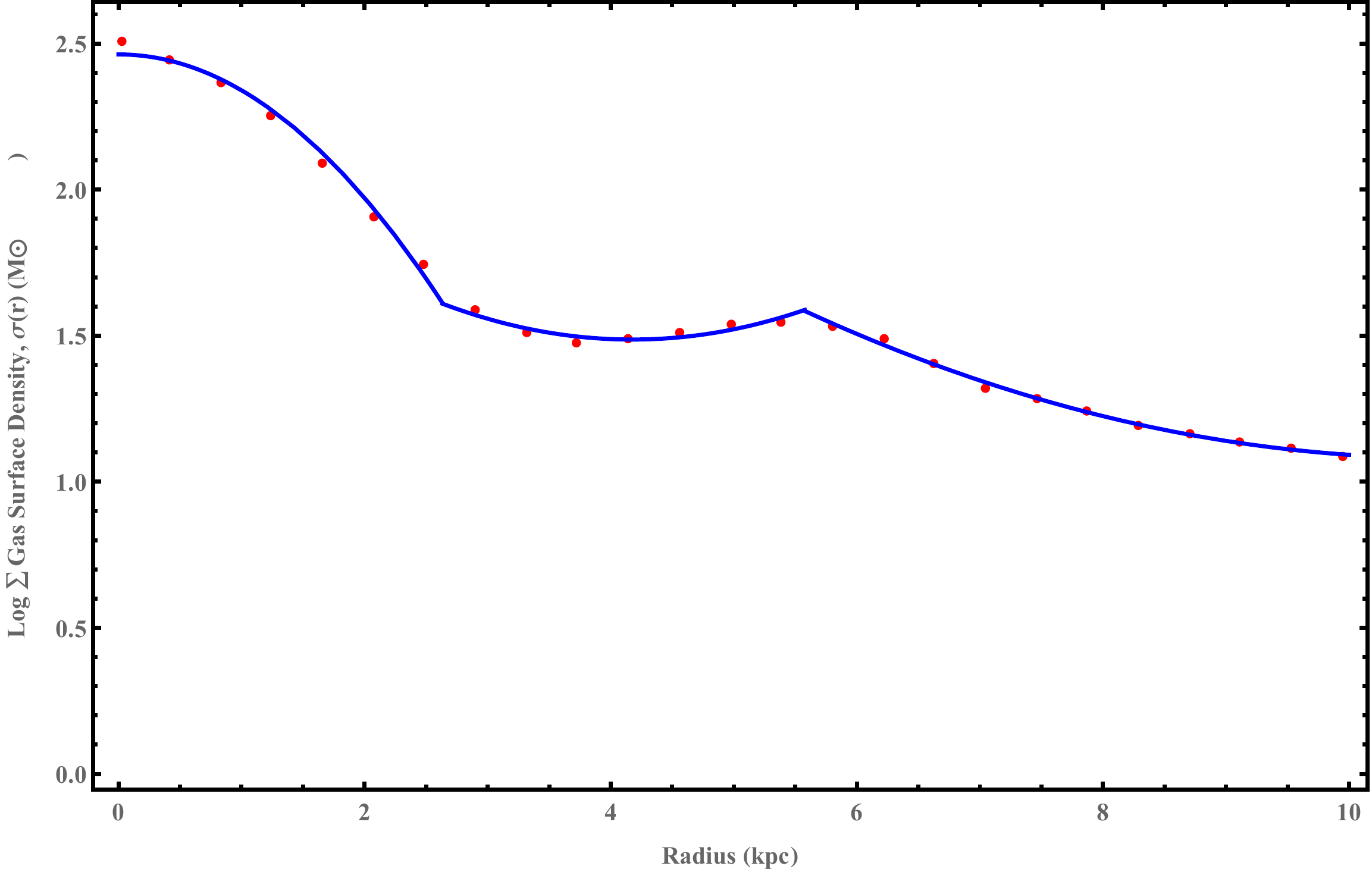}
\caption{\texttt{The best-fitting graph with $\chi^2 = 0.94$. The red dot point represents gas surface density and the blue line represents the best fitting line.}}
\end{figure}

\begin{figure}
\centering
\includegraphics[width=\columnwidth]{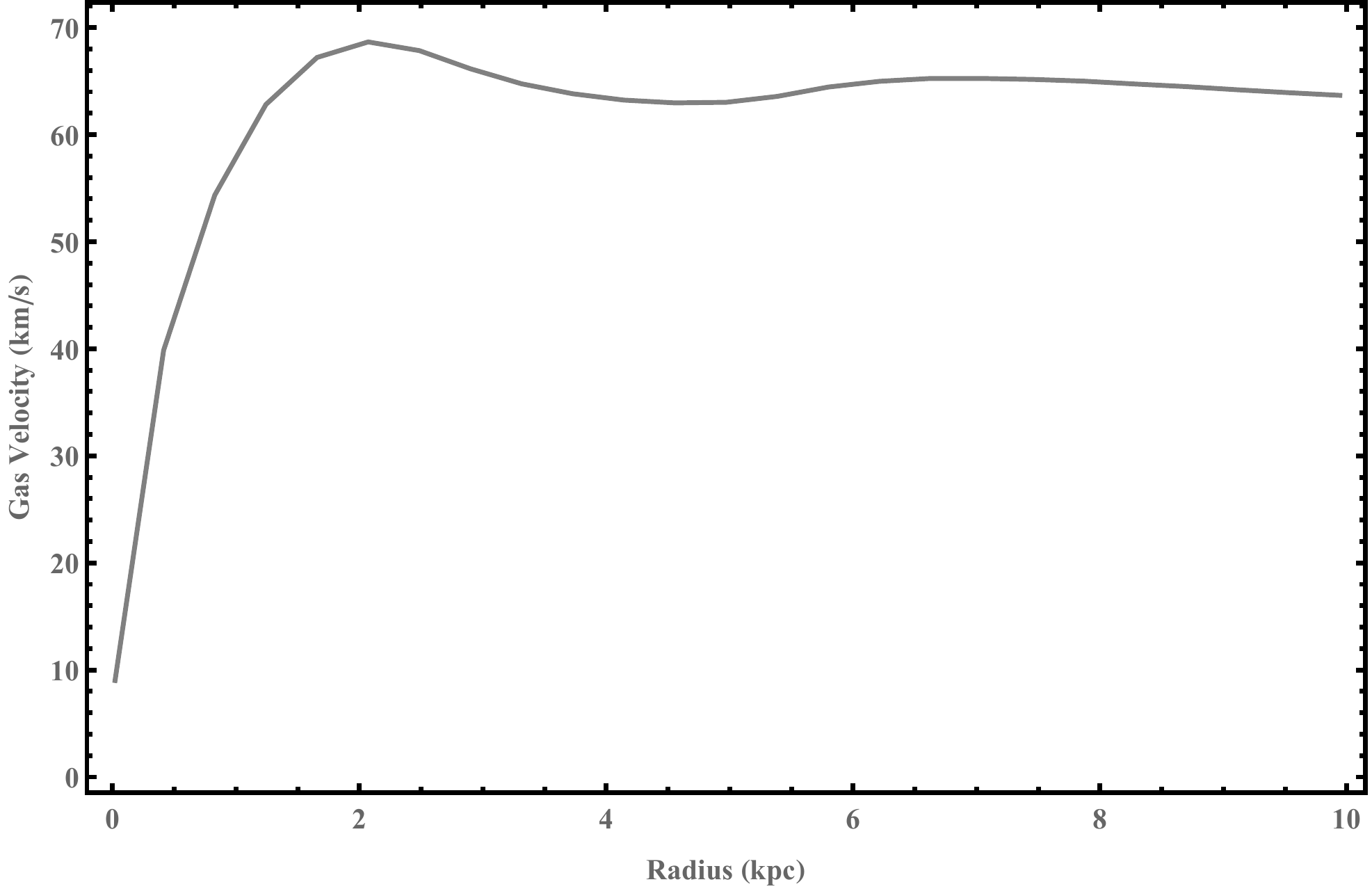}
\caption{\texttt{Gas velocity graph}}
\end{figure}

\section{Result \& Discussion}

This section is organized as follows: In subsection 3.1 we present the free parameter obtained and nonlinear fitting rotation curve. Then we illustrate the chi-square test and the mass of dark matter in subsection 3.2. The nine dark matter profiles analysis will be discussed one by one in subsections 3.1 and 3.2.

\subsection{\textit{Free parameter and nonlinear fitting rotation curve}}

By referring to Equation (1), where $V_{rot}^2=V_{HI}^2$ from VLA data, $V_{gas}^2$ derived from total gas surface density data, while for star and dark matter velocity we will use nonlinear fitting rotation curve method to find the most accurate free parameter. For star velocity $V_{star}^2$, we will use Equation (34) while for dark matter velocity $V_{DM}^2$, we will use Equations (5), (8), (11), (14), (17), (20), (24), (30), (33) for Pseudoisothermal, Burkert, NFW, Moore, Einasto, core-modified, DC14, coreNFW and Lucky13 profiles respectively.

We calculated the free parameters from the best fit that we have achieved. The free parameters of star mass $M_{D}$, scale density $\rho_{0}$ and scale radius $r_{s}$ are obtained using the Pseudoisothermal, Burkert, NFW, Moore, core-modified, DC14, coreNFW and Lucky13 profiles. Meanwhile the radius where density profile has a slope of -2, $r_{-2}$ and the local density at that radius, $\rho_{-2}$ is obtained using the Einasto profile. The third free parameter of Einasto profile is obtained using Einasto index that is fitted to n = $0.77\pm0.027$. The other free parameters fitting for the nine dark matter profiles are listed in Table 5. Next, by using the free parameter obtained in Table 5, we plot the rotation curves of the total rotational velocity with gas, star and dark matter halo velocities for the nine dark matter profiles, are given in Figure 9. 

\begin{table}
\centering
\begin{tabular}{|c|c|c|c|c|}
\hline
Dark Matter Profile & $M_{D}$ ($M_\odot$) & \begin{tabular}[c]{@{}l@{}}$\rho_{0}$ ($\rho_{-2}$ for Einasto) \\ ($M_\odot$ $kpc^{-3}$)\end{tabular} & \begin{tabular}[c]{@{}l@{}}$r_{s}$ ($r_{-2}$ for \\ Einasto) (kpc)\end{tabular}    \\ \hline
\begin{tabular}[c]{@{}l@{}} Pseudoisothermal \\ profile \end{tabular}  & $(5.46\pm 12.97) \times 10^{10}$    & $(1.98\pm 1.23) \times10^{8}$ & $2.55\pm 0.24$ \\ \hline
Burkert profile     & $(5.56\pm 4.53) \times10^{10}$    & $(1.62\pm 7.93) \times10^{7}$ & $5.72\pm 14.61$ \\ \hline
NFW profile     & $(5.43\pm 2.82) \times10^{10}$   & $(1.48\pm 0.53) \times10^{6}$  & $66.77\pm 21.88$ \\ \hline
Moore profile       & $(5.56\pm 7.94) \times10^{10}$  & $(1.20\pm 10.11) \times10^{6}$    & $43.88\pm 251.20$ \\ \hline
Einasto profile     & $(5.50\pm 6.21) \times10^{10}$    & $(3.87\pm 1.38) \times10^{7}$ & $8.44\pm 1.84$ \\ \hline
core-modified profile         & $(5.26\pm 6.88) \times10^{10}$   & $(5.57\pm 2.00) \times10^{7}$  & $4.10\pm 0.25$   \\ \hline
DC14 profile         & $(2.43\pm 0.024) \times10^{11}$   & $(7.47\pm 0.0061) \times10^{8}$  & $0.037\pm 0.0049$   \\ \hline
coreNFW profile         & $(5.57\pm 12.89) \times10^{10}$   & $(6.43\pm 4.17) \times10^{7}$  & $5.23\pm 0.36$   \\ \hline
Lucky13 profile         & $(2.43\pm 0.044) \times10^{11}$   & $(0.42\pm 5.48) \times10^{7}$  & $5.08\pm 32.96$   \\ \hline
\end{tabular}
\newline
\newline
Table 5: Free parameter obtained for the nine dark matter profiles
\end{table}

\graphicspath{ {./Figure/} }
\pdfsuppresswarningpagegroup=1
\begin{figure}
\centering
\includegraphics[width=5.8cm]{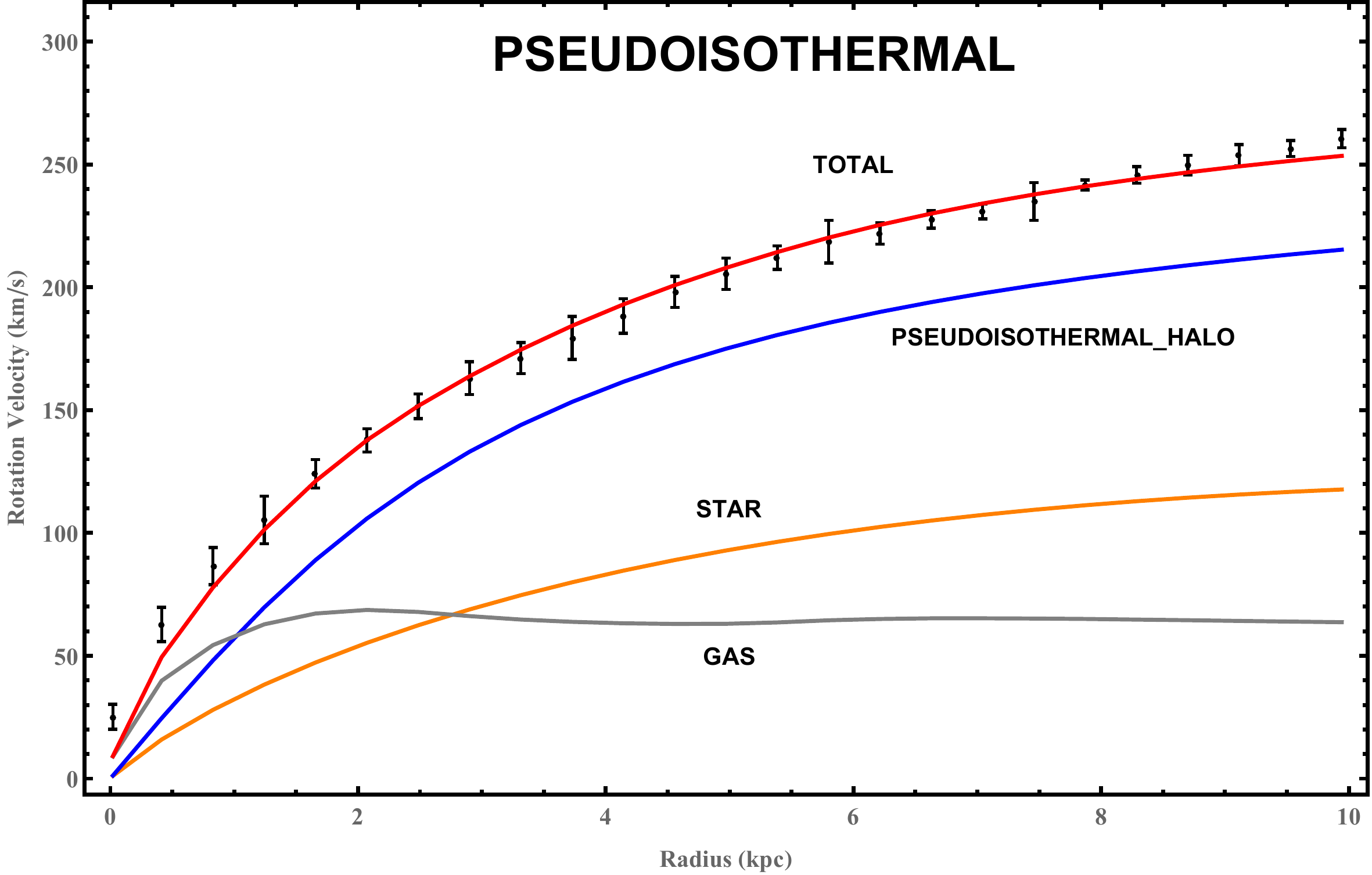}
\includegraphics[width=5.8cm]{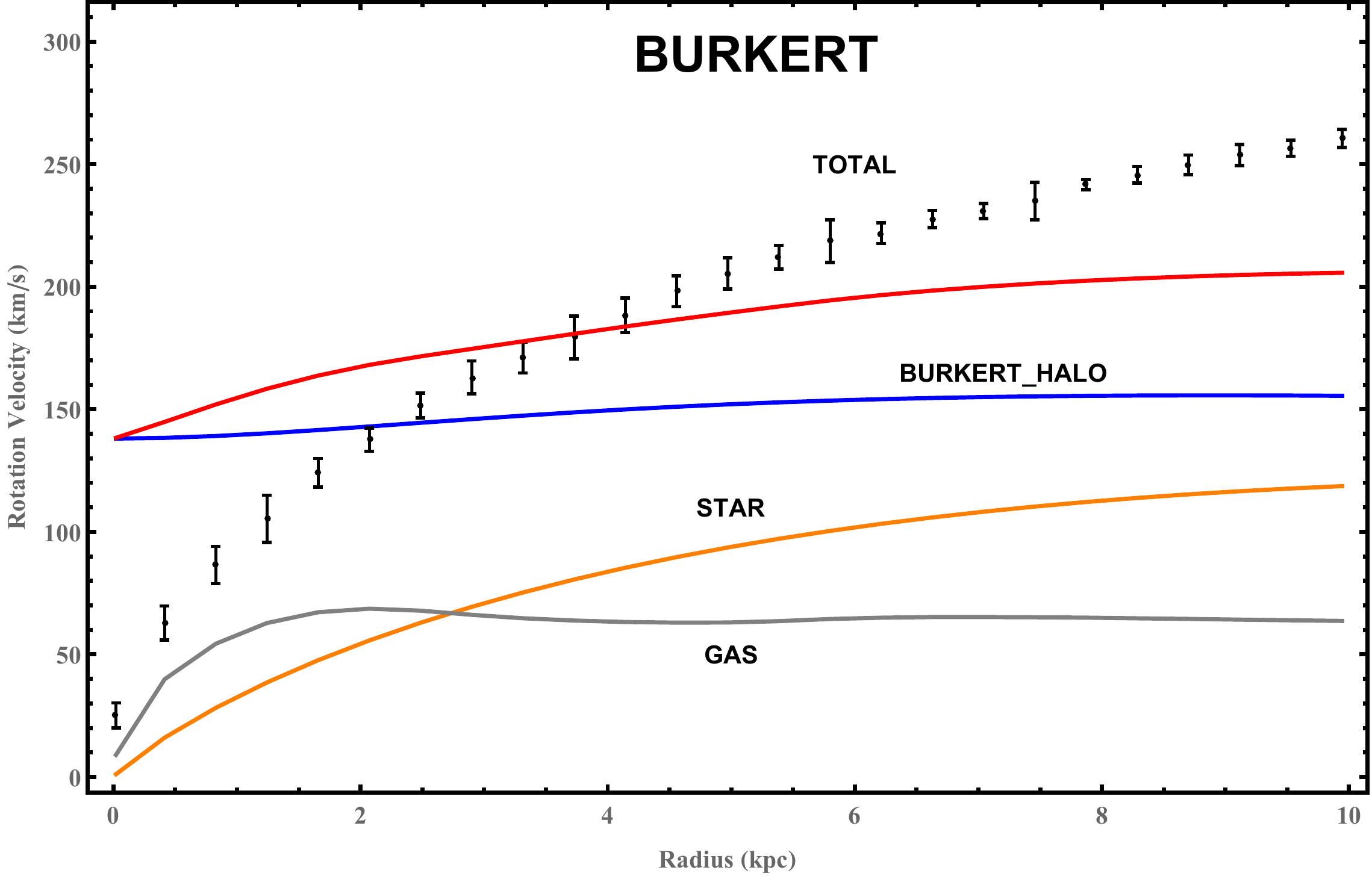}
\includegraphics[width=5.8cm]{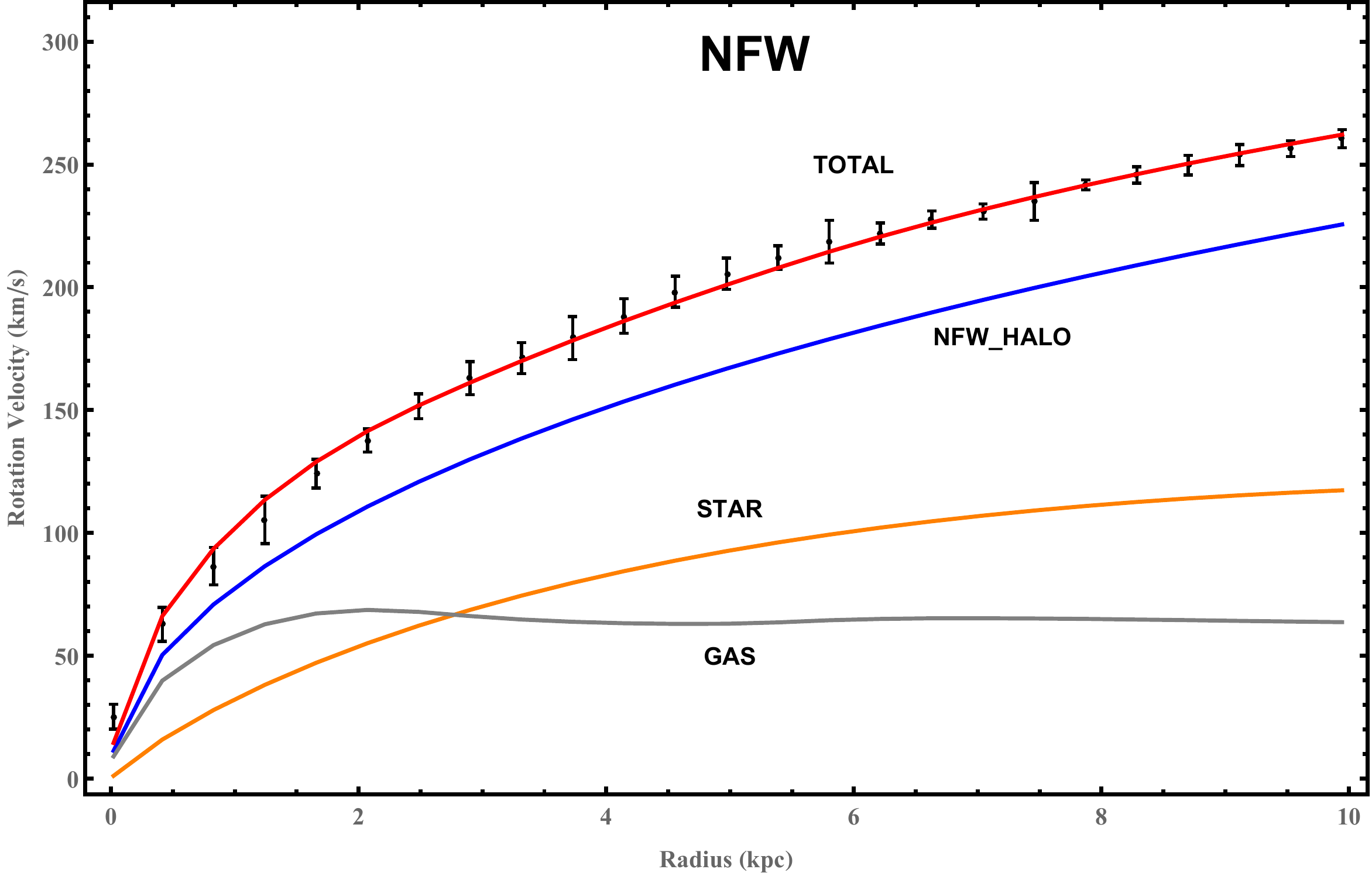}
\includegraphics[width=5.8cm]{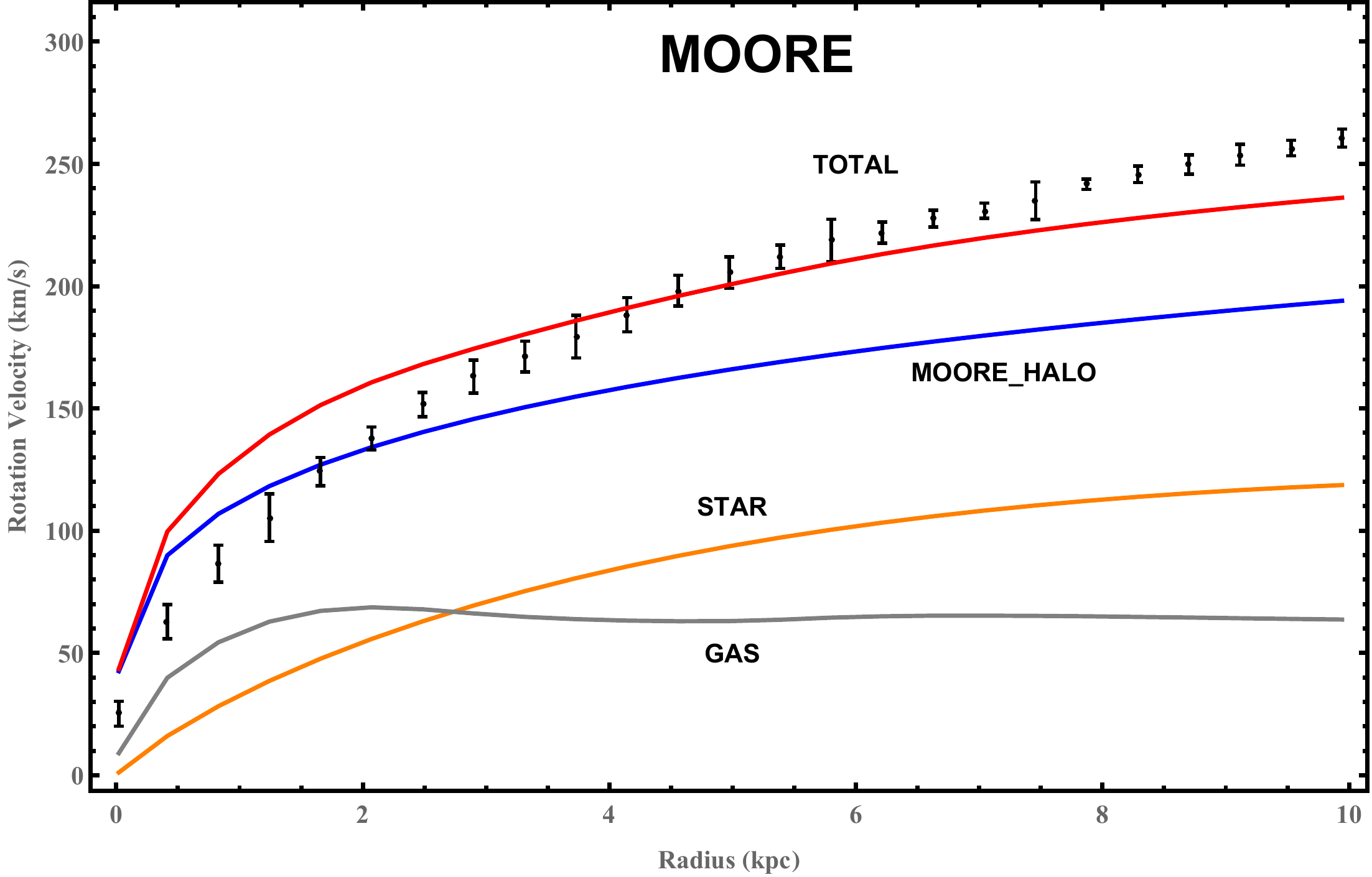}
\includegraphics[width=5.8cm]{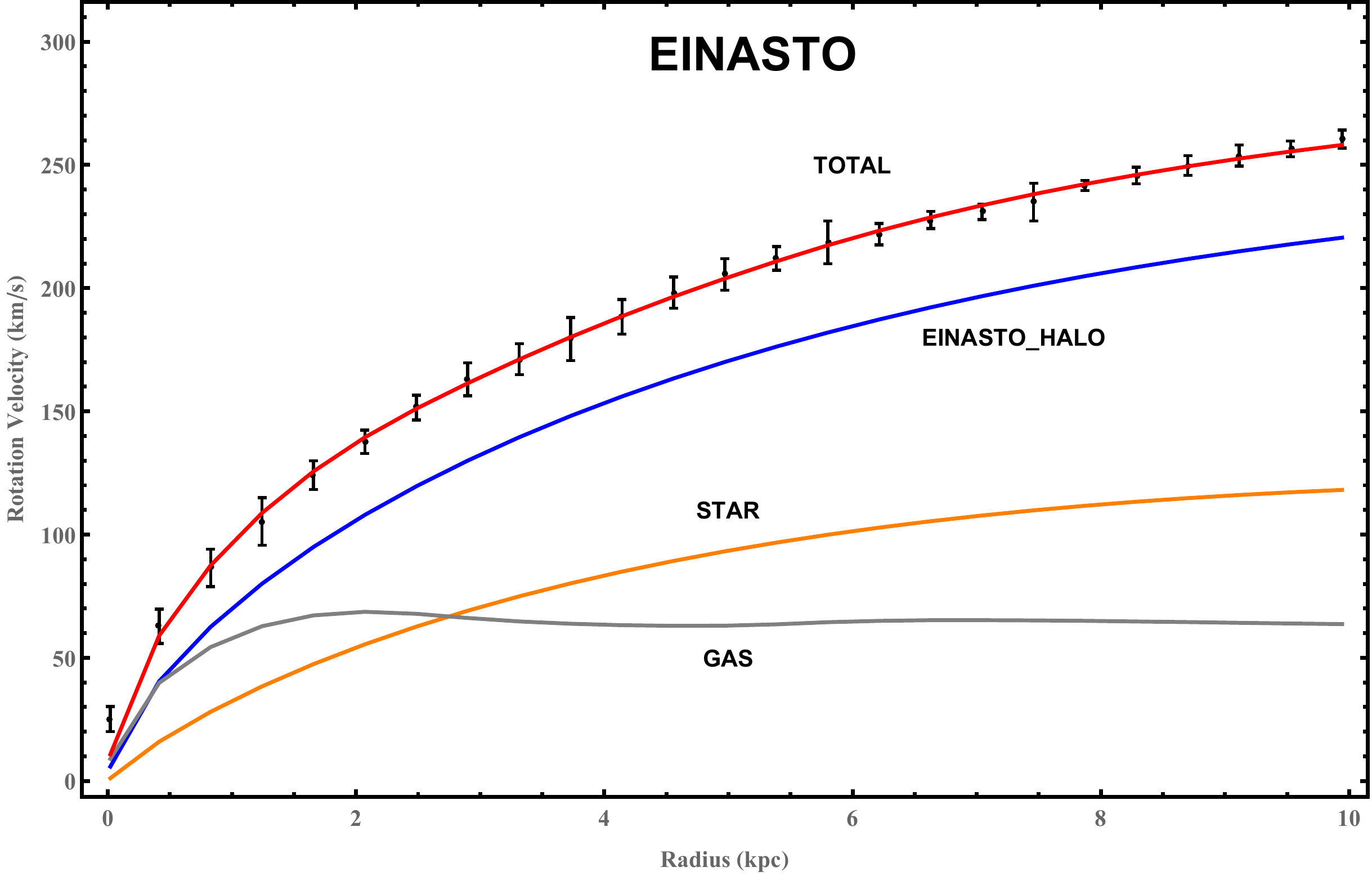}
\includegraphics[width=5.8cm]{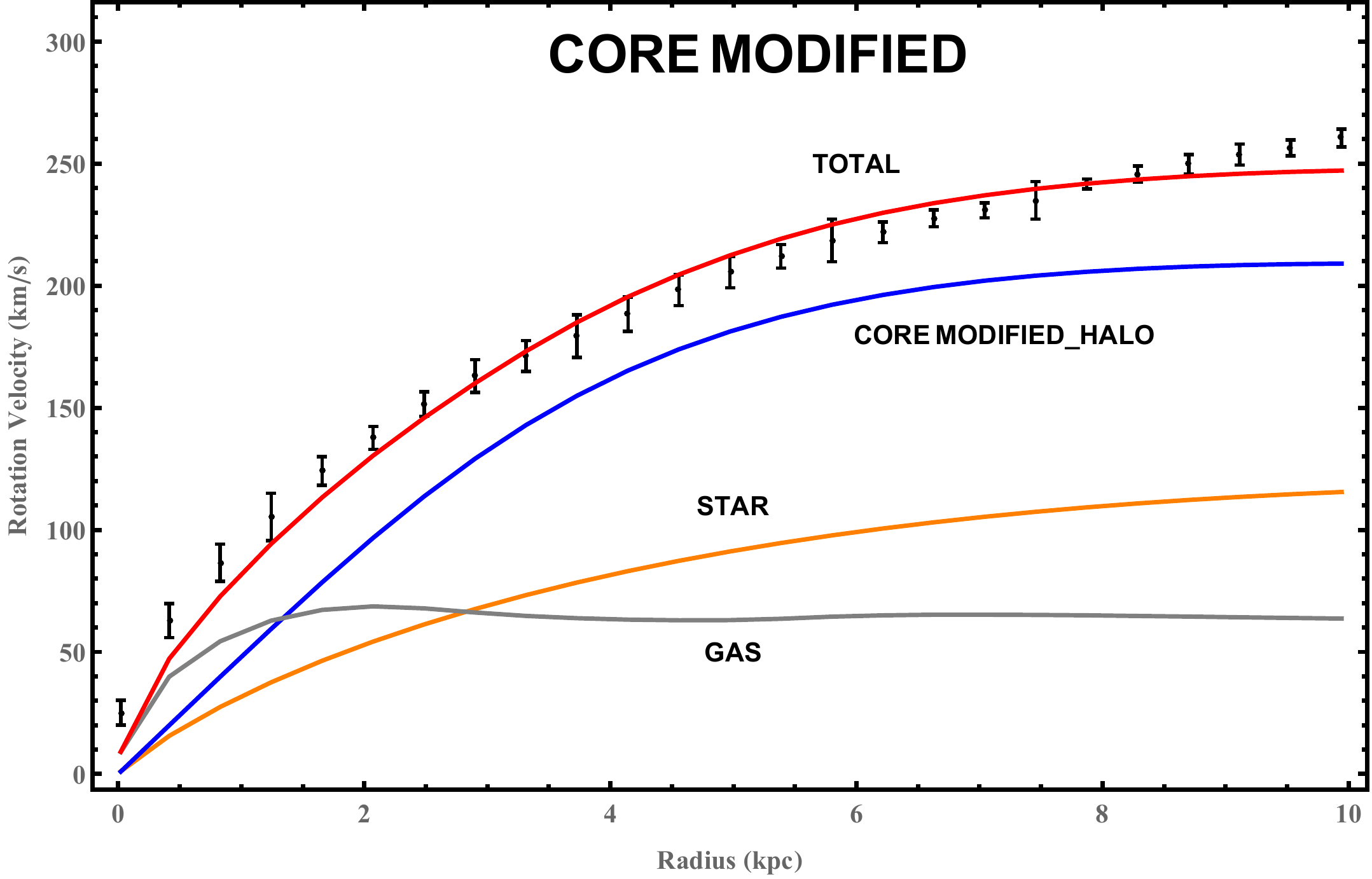}
\includegraphics[width=5.8cm]{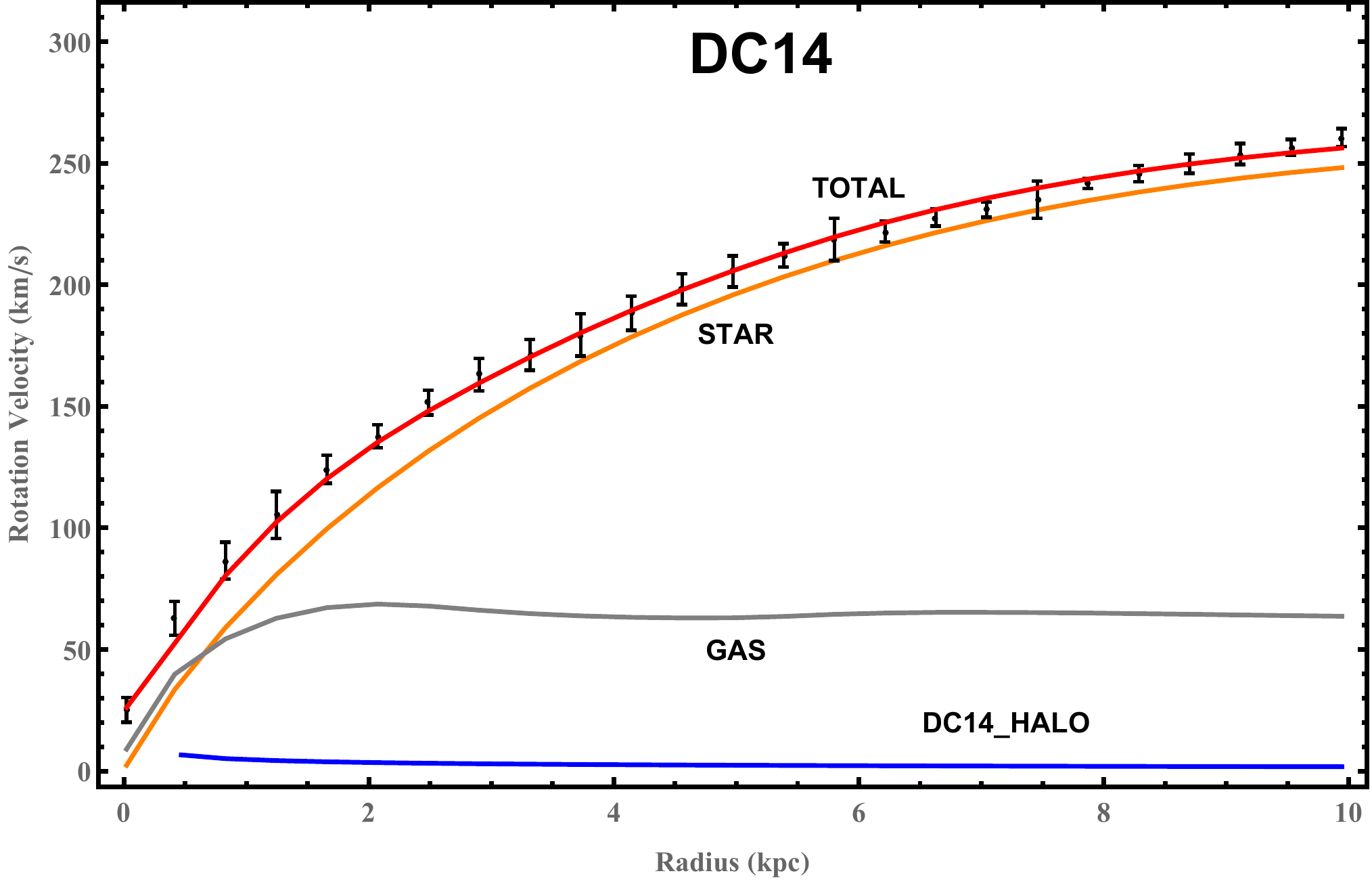}
\includegraphics[width=5.8cm]{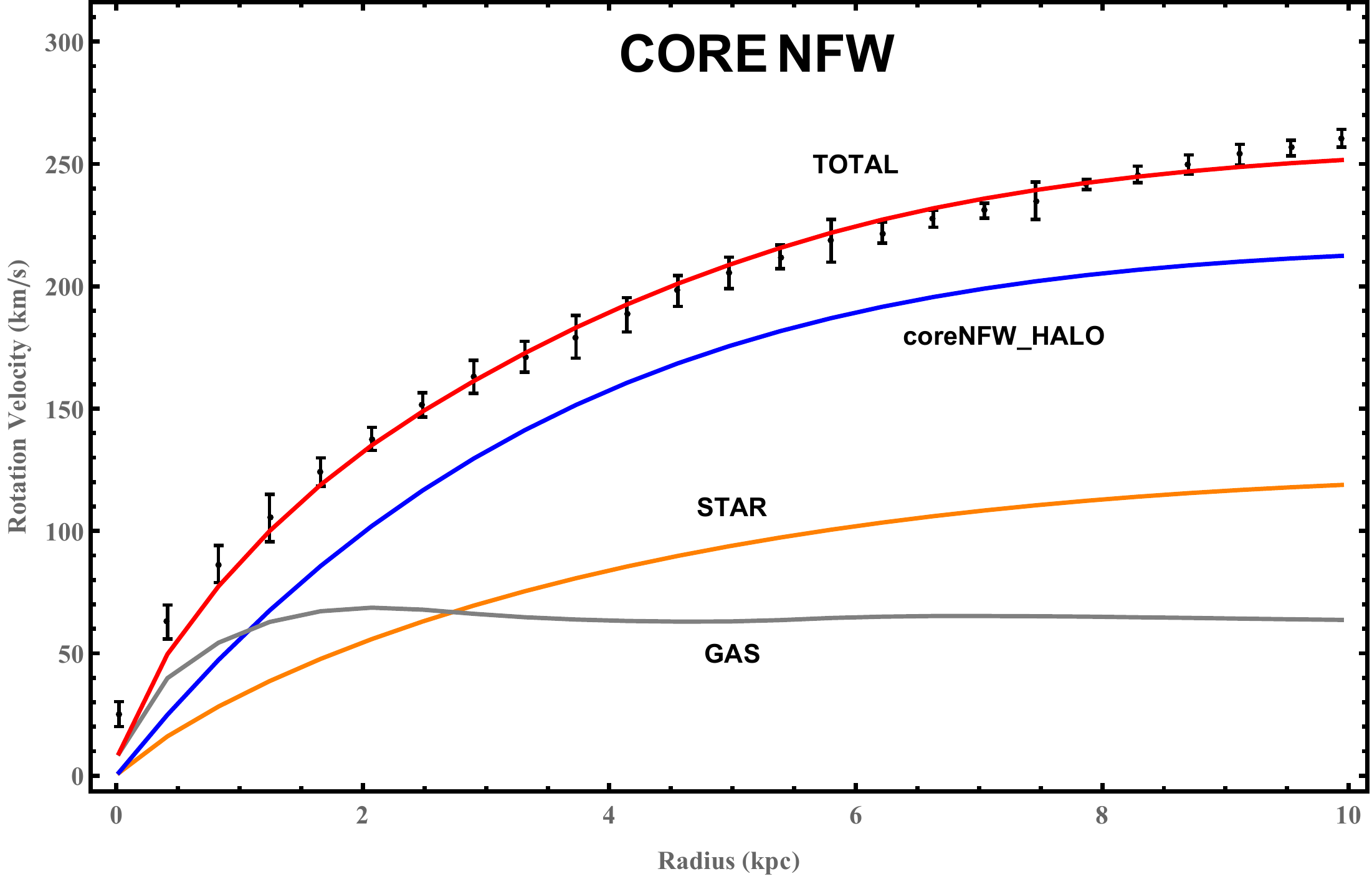}
\includegraphics[width=5.8cm]{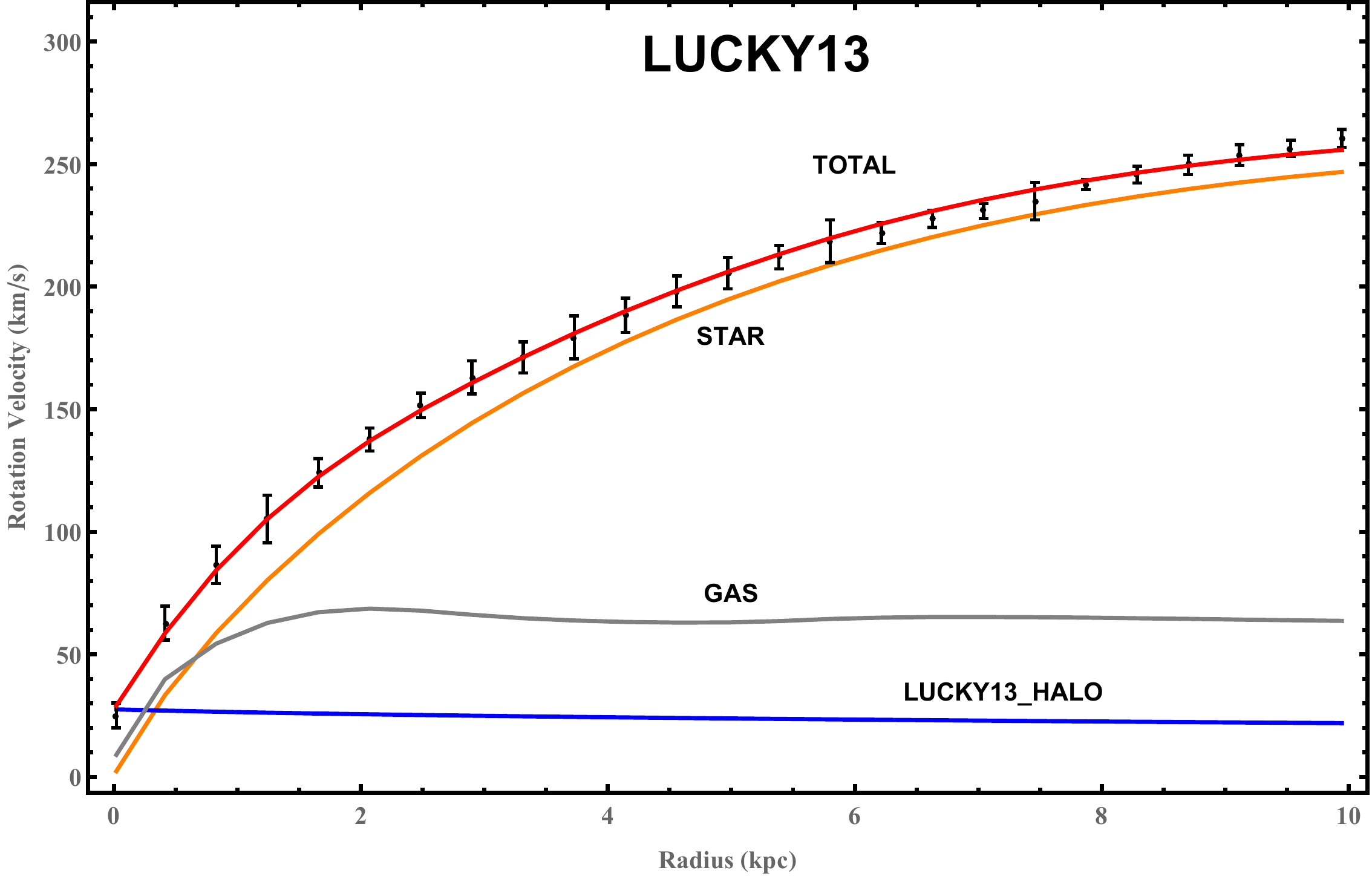}

\caption{\texttt{The nonlinear rotation curve of NGC 4321 by
using the nine dark matter profiles. The black and red color
represents the total rotation velocity and best fitting line,
while the grey, orange and blue color represent gas, star and
dark matter halo velocities respectively.}}
\end{figure}

By referring to Figure 9, the rotation curve fitting of DC14 and Lucky13 is rejected, due to the unrealistic low of dark matter velocity for the overall 10 kpc radius. The reason behind the unsuitability of DC14 for this galaxy is because the DC14 equation only works for the SHM ratio within -4.1 $<$ X $<$ -1.3 \cite{cintio2014mass}. During the fitting process, we used X as a free parameter and performed the fitting within the range of -4.1 and -1.3 but none of the values were able to fit the rotation curve well. We then calculated the SHM ratio by applying the relation X = log (Mstar/Mhalo) and the masses of the star and dark matter from the five accepted dark matter profiles (Table 7). The calculation for all five accepted profiles showed that the limit of X $>$ -1.3 for this galaxy, which was found to be beyond the limit of the SHM ratio range. Meanwhile, for the Lucky13 profile, this new semi-empirical profile changes the value of alpha, beta and gamma of the ($\alpha, \beta. \gamma$) model.  The gamma set to 0 to reach a finite core, beta set to 3 to get the same decreasing rate as the NFW profile at large radii and transition parameter alpha is set to 1 \cite{li2020comprehensive}. These changes within the new semi-empirical profile are not suitable for this galaxy.

According to the previous researches, the Burkert profile is statistically more suitable for the dark matter-dominated dwarf galaxies \cite{burkert1995structure}. Burkert profile revises the density law of Pseudoisothermal profile in the inner region and diverge logarithmically for larger radii. By referring to Figure 9, the Burkert profile reveals the existence of high dark matter velocity 138 kms$^{-1}$ within the inner region close to 0 kpc and increases gradually until the large radii of this galaxy. However, the HI observation data shows the total rotational velocity is only 25 kms$^{-1}$ within the inner region close to 0 kpc, hence the mismatch between observation data and Burkert profile suggests that Burkert profile is not suitable for this spiral galaxy.

Moore profile behaves similarly to NFW profile at large radii but it is steeper than NFW profile at smaller radii \cite{diemand2004convergence}. By referring to Figure 9, the dark matter velocity of the Moore profile has the steepest rise in the inner region compare to other profiles (except DC14 and Lucky13 profiles). The steepness in the Moore profile leads to its dark matter velocity to be found as higher than the total rotational velocity in the inner region. Even as its steepness reduces after radius 0.5 kpc, the dark matter velocity is still higher than the total rotational velocity below the radius 2.0 kpc. After radius 2.0 kpc, the dark matter velocity of Moore profile is continuously lower than the total rotational velocity until the large radii of the galaxy. However, our result for the dark matter velocity of Moore profile that is found to be higher than the total rotational velocity before radius 2.0 kpc suggests that Moore profile is not suitable for this galaxy too.

For the remaining five accepted dark matter profiles, it is difficult to determine which dark matter model is better by looking at the figure only. Hence, we will do a chi-square test calculation first before making dark matter model analysis.

\subsection{\textit{Chi-square test and the mass of dark matter}}

We implemented the chi-square test to test the goodness-of-fit of the rotation curve fitting. The goodness-of-fit parameter, $\chi^2$ to be minimized is \cite{bevington1969data}:

\begin{equation}
{\chi^2 = \sum_{i=1}^{N} { \frac{1}{\sigma_{i}^2} { [  y_{i} - y  (  x_i , a , b , c  )   ]}^2}}
\end{equation}
where N are observed data points, $\sigma_{i}$ is the uncertainty in $y_{i}$, $y_{i}$ is the observed rotation curve, $y(x_{i},a,b,c)$
are the values of the model function calculated at $x_{i}$, $x_{i}$ is the radius from the galactic centre and a, b, c are the fit parameters.

Then we performed the goodness-of-fit test by implementing the reduced chi-square:

\begin{equation}
{\chi_{red}^2 =\frac{\chi^2}{v}}
\end{equation}
where $v=N-N_{c}$, N is the number of data points and $N_{c}$ is the number of fit parameters. In principle, a value of $\chi_{red}^2$ = 1 indicates that the extent of the match between the observations data and the value of the estimates is in accord with the error variance \cite{bevington1969data}.

All reduced chi-square, $\chi_{red}^2$ for the five accepted dark matter profiles are shown in Table 6. Generally, we achieved 0.40 $<$ $\chi_{red}^2$ $<$ 1.70 for four out of five dark matter profiles. We found one profile that achieved the $\chi_{red}^2$ closest to 1, which is Pseudoisothermal profile. For cored profiles, the core-modified profile achieved the highest $\chi_{red}^2$ with 3.29 and followed by coreNFW profiles with $\chi_{red}^2$ of 1.64. While the Einasto and cuspy NFW profile achieved similar lowest $\chi_{red}^2$ with 0.52 and 0.49, respectively.

\begin{table}
\centering
\begin{tabular}{|l|l|l|}
\hline
Dark Matter Profile    & Model      & $\chi_{red}^2$    \\ \hline
\multirow{4}{*}{Cored} & Pseudoisothermal profile & 1.25 \\ \cline{2-3} 
& Einasto profile   & 0.52 \\ \cline{2-3} 
& core-modified profile & 3.29 \\ \cline{2-3} 
& coreNFW profile   & 1.64  \\ \hline
\multirow{1}{*}{Cuspy} & NFW profile & 0.49 \\ \hline
\end{tabular}
\newline
\newline
Table 6: $\chi_{red}^2$ for five dark matter profiles
\end{table}

Next, we will continue our five accepted dark matter model analysis by referring to the results from the rotation curve in Figure 9 rotation curve and the reduced chi-square test in Table 6.

Core-modified is a profile with constant density in the central core to avoid singularity in the galactic center \cite{brownstein2009modified}. However, the constant density in the central core presented very little flexibility for this profile during the fitting process in the rotation curve modeling. During rotation curve fitting, we found that this profile has certain limitations where less fitting changes that can be made especially in the central core of the galaxy. For the large radii of the galaxy, the increasing rate of dark matter velocity is declining and becomes almost constant after 7.5 kpc radius. The dark matter velocity is normally supposed to continue increasing as the total rotational velocity increases throughout. The increment of dark matter velocity in the large radii of the galaxy can be seen in NFW, coreNFW, Einasto and Pseudoisothermal profiles. This dark matter profile is found to be able to fit the galaxy and the calculated dark matter velocity is reasonably better when compare to the other four rejected dark matter profiles. However, the limited flexibility in the core and the constant dark matter velocity in the large radii make the core-modified profile is worse than other four accepted dark matter profiles and only achieved a $\chi_{red}^2$ of 3.29.

A coreNFW halo is essentially a NFW halo which transforms an inner cusp into a finite central core \cite{read2016dark}. The change of inner cusp into a finite central core of this profile have brought improvement in the rotation curve fitting in the inner region of this galaxy. By referring to the inner region of this profile, the increasing dark matter velocity is less steep than the cuspy NFW profile. The steepness of dark matter velocity has made this profile to have good fit with total rotational velocity in the inner region of the galaxy. However, the increasing rate of dark matter velocity is found to be declining beyond 8 kpc radius, causing unsuitability of fitting with the continuously increasing total rotational velocity. This issue causes the coreNFW profile to be considered as not as good as Peudoisothermal profile and achieved a $\chi_{red}^2$ of 1.64.

The NFW profile is called ‘universal’ because it works for a large variety of halo masses, same shape of initial density fluctuation spectrum, from individual galaxies to the halos of galaxy clusters \cite{navarro1997universal} and this leads to the NFW profile that fits well to this galaxy as well. However, the NFW profile has the cuspy halo problem that increases steeply at small radii \cite{de2010core}. This problem can be seen in Figure 9, the inner region of the NFW profile is steeper than the coreNFW, Pseudoisothermal and Einasto profiles. From the inner region 0 kpc to 1 kpc, the steepness of the NFW profile causes the rotational velocity of halo, star and gas to mismatch with the total rotational velocity. This mismatch causes the fitting of NFW profile to be achieved $\chi_{red}^2$ of 0.49, which is slightly not as good as the case for the Einasto profile.

High-resolution N-body CDM simulations indicate that nonsingular three-parameter models such as the Einasto profile perform better than the singular two-parameter models such as Pseudoisothermal and NFW profiles, providing an excellent fit to a wide range of dark matter haloes \cite{retana2012analytical}. The Einasto profile involves a third parameter, n, the Einasto index, which describes the shape of the overall profile distribution, larger values of n results in steeper inner profiles and shallower outer profiles \cite{graham2006empirical}. In order to have the best-fitting with the total rotational velocity of this galaxy, the inner profile needs to be shallower and the outer profile needs to be steeper. In our case for this paper, the third parameter, n, acts a very important role to fulfill the criteria of the fitting, where we are able to use smaller values of n to make the inner profile shallower and the outer profile steeper (Figure 9). Furthermore, these three parameters model allows the profile to be tailored to each individual halo, thereby yielding improved fits \cite{navarro2004inner}. This can be seen in Figure 9 where each individual halo is well fitted with total rotational velocity. However, the addition of the third parameter and the profile tailored to each individual halo causes the Einasto profile to ultimately overfits the data with the value $\chi_{red}^2$ of 0.52.

Pseudoisothermal profile is a commonly used model for dark matter halo analysis and is often seen to better fit the galactic rotation curve than the NFW model \cite{de2001high,de2006high} and this works for this spiral galaxy as well. By looking at Figure 9, the overall Pseudoisothermal profile distribution fits very well and matches with Equation (1). In addition, the Pseudoisothermal rotation curve has a linear growth at the inner region then becomes flat at large radii \cite{de2008mass}. This trend can be seen on this galaxy too, where the rotational velocity of the Pseudoisothermal profile in the inner region below 1 kpc increases linearly and the rotational velocity becomes flat at large radii. The linearity at the inner region and the flatness at the large radii of the fitting characteristic are very suitable for this galaxy. This causes the Pseudoisothermal to achieve the best fitting among the nine dark matter profiles with $\chi_{red}^2$ of 1.25, which is the nearest to 1.

Based on the free parameters obtained as shown in Table 5, we calculated the total star mass and dark matter mass within radius 10 kpc for each dark matter profile as shown in Table 7. We obtained the star velocity from Equation (34) and subsequently calculated the star mass with Equation (36), while the mass of dark matter is calculated by applying the Equations (4), (10), (16), (19), (29) for Pseudoisothermal, NFW, Einasto, core-modified and coreNFW profiles, respectively.

\begin{table}
\centering
\begin{tabular}{|l|l|l|l|}
\hline
\begin{tabular}[c]{@{}l@{}} Dark Matter \\ Profile \end{tabular}    & Model      & Mass of star($M\odot$) & Mass of dark matter($M\odot$)    \\ \hline
\multirow{4}{*}{Cored} & Pseudoisothermal  & $(3.23\pm 7.66) \times 10^{10}$  & $(1.08\pm 0.0083) \times 10^{11}$ \\ \cline{2-4} 
& Einasto  & $(3.25\pm 3.67) \times 10^{10}$    & $(1.13\pm 0.00012) \times 10^{11}$ \\ \cline{2-4}
& core-modified  & $(3.11\pm 4.07) \times 10^{10}$  & $(1.01\pm 0.00035) \times 10^{11}$  \\ \cline{2-4} 
& coreNFW  & $(3.29\pm 7.61) \times 10^{10}$ & $(1.05\pm 0.0014) \times 10^{11}$  \\ \hline
\multirow{1}{*}{Cuspy} & NFW  & $(3.21\pm 1.66) \times 10^{10}$ & $(1.19\pm 0.10) \times 10^{11}$ \\ \hline 
\end{tabular}
\newline
\newline
Table 7: Total dark matter mass within radius 10 kpc for the dark matter profiles
\end{table}

The total dark matter mass for the five accepted dark matter profiles within radius 10 kpc in this galaxy is in the range from $(1.01\pm 0.00035) \times 10^{11} M\odot$ to $(1.19\pm 0.10) \times 10^{11} M\odot$. By taking consideration of the $\chi_{red}^2$, the Pseudoisothermal, Einasto, coreNFW and NFW profiles achieved the $\chi_{red}^2$ within the range of 0.40 $<$ $\chi_{red}^2$ $<$ 1.70. The mass of dark matter in this galaxy within the range of these four profiles is calculated to be from $(1.05\pm 0.0014) \times 10^{11} M\odot$ to $(1.19\pm 0.10) \times 10^{11} M\odot$.

\section{Conclusion}

For this galaxy, we analyzed the dark matter of the galaxy NGC 4321 by using the nine dark matter profiles. The nine dark matter profiles consist of seven core and two cuspy profiles. The core profiles are namely Pseudoisothermal, Burkert, Einasto, core-modified, DC14, coreNFW and Lucky13 profiles. The cuspy profiles are NFW and Moore profiles. We analyzed in detail how each dark matter profile performs in this galaxy. We rejected four dark matter profiles while five dark matter profiles are accepted for this galaxy. DC14 profile is rejected due to the stellar-to-halo mass ratio that is found to be out of the range of the profile condition. Lucky13 profile is rejected due to the unsuitability of the galaxy model setting of this new semi-empirical profile. The revised density law of Burkert profile causes the dark matter velocity to mismatch with regards to the total rotational velocity starting from the inner region of the galaxy. Further analysis showed that the Moore profile has the steepest rise in the inner region among the five dark matter profiles and this causes the dark matter velocity in the inner region to be higher than the total rotational velocity below the radius 2.0 kpc. These reasons suggest that these four dark matter profiles are not suitable for this galaxy.

For the remaining five accepted dark matter profiles, we analyzed the dark matter velocity together with the reduced chi-square test. The constant core of the core-modified profile causes the limited flexibility on fitting and hence producing a large value of $\chi_{red}^2$. The NFW profile cuspy halo has a weakness, where the inner region increases steeply at small radii, resulting in the total rotational velocity of halo, star and gas to be higher than the total rotational velocity by HI observed data at the inner region of this galaxy. The change of inner cusp into a finite central core of the coreNFW profile improved the NFW profile to become less steep in the inner region of the galaxy and this resulted in a better rotation curve fitting. The additional third parameter of the Einasto profile brings the flexibility of adjustment in the inner and outer radii of the galaxy and this enabled the Einasto profile to generate a better overall fitting. The Pseudoisothermal profile produced linear rotational velocity in the inner region and flat rotational velocity at the large radii of this galaxy. This fitting characteristic suits this galaxy very well by producing the best fitting among the nine dark matter profiles with a $\chi_{red}^2$ of 1.25.

Generally, the NFW, coreNFW, Einasto and Pseudoisothermal profiles achieve relatively good fittings and the mass of dark matter for these four profiles is calculated to be within the range from $(1.05\pm 0.0014) \times 10^{11} M\odot$ to $(1.19\pm 0.10) \times 10^{11} M\odot$. Ultimately, our results showed that the Pseudoisothermal profile achieved the best fitting among the nine dark matter profiles and the dark matter mass of this galaxy obtained through the use of this profile is calculated as $(1.08\pm 0.0083) \times 10^{11} M\odot$.

\begin{acknowledgements}
The authors would like to acknowledge the funding by the University of Malaya (FG033-017AFR).
\end{acknowledgements}

\clearpage

\newpage

%
%


%
%
\bibliographystyle{abnt-num}


\end{document}